\documentclass[useAMS,usenatbib]{mnras}
\usepackage{newtxtext,newtxmath}
\usepackage[T1]{fontenc}

\DeclareRobustCommand{\VAN}[3]{#2}
\let\VANthebibliography\thebibliography
\def\thebibliography{\DeclareRobustCommand{\VAN}[3]{##3}\VANthebibliography}

\usepackage{graphicx}	% Including figure files
\usepackage{amsmath}	% Advanced maths commands
\usepackage{listings}
\usepackage{siunitx}
\usepackage{pifont}
\usepackage{subcaption}
\usepackage{hyperref}
\usepackage{lipsum}
\usepackage{acro}
\usepackage{physics}
\usepackage{booktabs}  
\usepackage{placeins}  
\usepackage{pdflscape}
\usepackage{float} 
\usepackage{multirow}
\usepackage{multicol}
\usepackage[capitalize]{cleveref}
\title[The radial acceleration relation from MIGHTEE]{MIGHTEE-HI: The radial acceleration relation with resolved stellar mass measurements}

\author[V\u{a}r\u{a}\c{s}teanu et al.]{Andreea A. V\u{a}r\u{a}\c{s}teanu$^{1}$\thanks{email: \href{mailto:andreea.varasteanu@physics.ox.ac.uk} {andreea.varasteanu@physics.ox.ac.uk}}, Matt J.~Jarvis$^{1,2}$, Anastasia A.~Ponomareva$^{1,3}$, Harry Desmond$^{4}$, 
\and
Ian Heywood$^{1,5,6}$, 
Tariq Yasin$^{1}$, 
Natasha Maddox$^{7}$, 
Marcin Glowacki$^{8, 9}$, 
Michalina Maksymowicz-Maciata$^{7}$,
\and
Pavel E. Mancera Pi$\mathrm{\tilde{n}}$a$^{10}$,
Hengxing Pan$^{11}$
\\
$^{1}$Oxford Astrophysics, Denys Wilkinson Building, University of Oxford, Keble Road, Oxford, OX1 3RH, UK \\
$^{2}$Department of Physics and Astronomy, University of the Western Cape, Robert Sobukwe Road, 7535 Bellville, Cape Town, South Africa \\
$^{3}$Centre for Astrophysics Research, School of Physics, Astronomy and Mathematics, University of Hertfordshire, College Lane, Hatfield, AL10 9AB, UK \\
$^{4}$Institute of Cosmology \& Gravitation, University of Portsmouth, Dennis Sciama Building, Portsmouth, PO1 3FX, UK\\
$^{5}$Centre for Radio Astronomy Techniques and Technologies, Department of Physics and Electronics, Rhodes University, PO Box 94, Makhanda, 6140, \\ South Africa. \\
$^{6}$South African Radio Astronomy Observatory, 2 Fir Street, Black River Park, Observatory, Cape Town, 7925, South Africa. \\
$^{7}$School of Physics, H.H. Wills Physics Laboratory, Tyndall Avenue, University of Bristol, Bristol, BS8 1TL, UK. \\
$^{8}$Institute for Astronomy, University of Edinburgh, Royal Observatory, Edinburgh, EH9 3HJ, United Kingdom. \\
$^{9}$Inter-University Institute for Data Intensive Astronomy, Department of Astronomy, University of Cape Town, Cape Town, South Africa.\\
$^{10}$Leiden Observatory, Leiden University, P.O. Box 9513, 2300 RA, Leiden, The Netherlands.\\
$^{11}$National Astronomical Observatories, Chinese Academy of Sciences, Beijing 100101, People’s Republic of China.}
\date{Accepted XXX. Received YYY; in original form ZZZ}

\pubyear{\the\year{}}

\begin{document}
\label{firstpage}
\pagerange{\pageref{firstpage}--\pageref{lastpage}}
\maketitle

\begin{abstract}
The radial acceleration relation (RAR) is a fundamental relation linking baryonic and dark matter in galaxies by relating the observed acceleration derived from dynamics to the one estimated from the baryonic mass. This relation exhibits small scatter, thus providing key constraints for models of galaxy formation and evolution---allowing us to map the distribution of dark matter in galaxies---as well as models of modified dynamics. However, it has only been extensively studied in the very local Universe with largely heterogeneous samples.
We present a new measurement of the RAR, utilising a homogeneous sample of 19 H{\sc i}-selected galaxies out to $z=0.08$.
We introduce a novel approach of measuring resolved stellar masses using spectral energy distribution (SED) fitting across 10 photometric bands to determine the resolved mass-to-light ratio, which we show is essential for measuring the acceleration due to baryons in the low-acceleration regime.
%We find that the stellar mass-to-light ratio varies across galaxies and radially, favouring lower mass-to-light ratios compared to previous studies.
Our results reveal a tight RAR with a low-acceleration power-law slope of $\sim 0.5$, consistent with previous studies. Adopting a spatially varying mass-to-light ratio yields the tightest RAR with an intrinsic scatter of only $0.045 \pm 0.022$ dex, highlighting the importance of resolved stellar mass measurements in accurately characterising the gravitational contribution of the baryons in low-mass, gas-rich galaxies. We also find the first tentative evidence for redshift evolution in the acceleration scale, but more data will be required to confirm this. Adopting a more general MOND interpolating function, we find that our results ameliorate the tension between previous RAR analyses, the Solar System quadrupole and wide-binary test.
%These results highlight the importance of resolved stellar mass measurements in accurately characterizing the gravitational contribution of the baryons in low-mass, gas-rich galaxies.
\end{abstract}


%%%%%%%%%%%%%%%%%%%%%%%%%%%%%%%%%%%%%%%%%%%%%%%%%%

%%%%%%%%%%%%%%%%% BODY OF PAPER %%%%%%%%%%%%%%%%%%

\section{Introduction} 
\label{intro}
The missing gravity problem, highlighted by the flatness of rotation curves, arises from observations of galaxies that reveal a discrepancy between the visible matter (stars, gas and dust) and the dynamical one \citep{McGaugh_2004, vandenBergh2001}. % Trippe_2014}.
According to Newtonian dynamics, the rotational velocity of stars and gas in a galaxy should decrease with distance from the center. However, observations reveal that they remain fairly constant \citep{Rubin_1978, Bosma_1978}, implying the presence of an unseen mass component, known as dark matter, that dominates the outer regions of galaxies. 
This observation is a key motivation for the $\Lambda$CDM model, the current cosmological paradigm, in which galaxies form and evolve within massive dark-matter halos.
This framework successfully explains the large-scale structure in the Universe. However, it faces several challenges on smaller scales, particularly in relation to galaxy dynamics, where discrepancies arise \citep[see][for a review]{Bullock-LCDM-challenges}. Examples include the cusp-core problem \citep{1994_Moore, Flores_1994}, the missing satellites problem \citep{Moore_1999, Klypin_1999}, and the too-big-to-fail problem \citep{Boylan_Kolchin_2011, Boylan_Kolchin_2012, Garrison_2013, Papastergis_2017}.

Some of the most intriguing aspects of galaxy dynamics in this framework are the tight scaling relations between baryonic properties and the dynamics of galaxies. A well-known example is the Tully-Fisher Relation (TFR;~\citealt{TullyFisher1977}) that links the luminosity of a galaxy to the width of its global H{\sc i} profile, or its extended form, the baryonic Tully-Fisher Relation (bTFR), which links the baryonic mass (stars and gas) to the flat rotation velocity of the galaxy, spanning over 5 dex in baryonic mass \citep{McGaugh_2000, McGaugh_2012, Ponomareva2017, Lelli_2019, Ponomareva2021}.
%major tools to test galaxy formation and evolution models through their ability to reproduce the statistical properties such as slope scatter and zero point.

The radial acceleration relation (RAR) generalizes the bTFR by connecting the local observed dynamics of galaxies to the mass distribution within them across the full extent of galaxies. Previously termed the mass discrepancy--acceleration relation~\citep{Sanders_1990,McGaugh_2004}, the RAR as such was first reported by \cite{McGaughLelli2016} using the SPARC database (Spitzer Photometry and Accurate Rotation Curves), consisting of 175 late-type galaxies with accurate H{\sc i} rotation curves \citep{SPARC-database}. The RAR is a remarkably tight empirical correlation between the observed radial acceleration ($ \mathrm{g_{obs}}$), derived from rotation curves (thus including both baryonic and dark matter in the $\Lambda$CDM paradigm) and the one derived solely from the distribution of baryons ($\mathrm{g_{bar}}$).
The RAR suggests that the distribution of dark matter is closely tied to the distribution of visible matter, with minimal scatter, largely driven by observational uncertainties \citep{Lelli2017}. 

The high acceleration end of this relation corresponds to baryon-dominated regions in massive galaxies. Here, the RAR follows a one-to-one relation~\citep{SR-RAR}, which implies that the observed dynamics can be fully accounted for by baryons alone. However, at lower accelerations, below a certain characteristic acceleration scale of $\approx 10^{-10}$\,m~s$^{-2}$ \citep{McGaughLelli2016, Lelli2017}, the observed dynamics deviates significantly from what we infer from baryons alone, and this is generally attributed to the influence of dark matter. The inner parts of massive galaxies contribute to the high acceleration portion of the RAR, whereas the outer parts of high-mass galaxies and all regions of low-mass galaxies contribute to the lower acceleration regime. 
%This coupling between luminous and dark matter components---also referred to as the mass discrepancy-acceleration relation \citep{McGaugh_2004}---echoes "Renzo’s rule", which states that each feature in a galaxy’s luminosity profile corresponds to a feature in its rotation curve 
%\citep{sancisi2003visiblematterdark}.
Such an intimate baryon-dark matter coupling is not trivially expected in a $\Lambda$CDM cosmology, where galaxies form and evolve via stochastic processes (e.g, hot and cold mode gas accretion to dark matter halos, supernova and AGN feedback), which should naturally introduce substantial scatter, particularly at the low-mass end of the galaxy population, where dark matter dominates.

However, the RAR shows remarkably low scatter, potentially even consistent with zero \citep{Lelli2017}, making it the tightest known dynamical galaxy scaling relation \citep{Desmond_2023}. 
%This tightness is not only a striking empirical result, 
This makes it a powerful diagnostic tool to provide key insights into the elusive nature of dark matter, by enabling us to impose stringent observational constraints on galaxy formation models~\citep{Desmond_MDAR}. Any successful theoretical framework must be able to reproduce this tight relation across a wide range of galaxy types and masses. This stringent requirement has been used by some to argue for Modified Newtonian Dynamics (MOND;~\citealt{Milgrom_1983,Sanders_1990}) as an alternative to $\Lambda$CDM, which produces the relation more naturally (e.g.~\citealt{Lelli2017}).
% —whether based on $\Lambda$CDM or alternative theories such as MOND \citep{Milgrom_1983, Sanders_1990}—

%predict the RAR naturally 
 %Galaxies properites depend on the history of the dark matter halo they reside in, merger and mass accretion histories, and als on baryonic processes like supernovae and AGN feedback that can redistribute mass. 
% feedback induced cores by baryonic feedback to erase central cusps 
%Courteau 2007 - The success of a particular theory resides in its ability to predict the slope, scatter, and zero-point of any robust galaxy scaling relation
%models to accurately reproduce these relationships while accounting for complex feedback processes
%Regularity must somehow emerge from stochasticity 
The RAR has been studied extensively at redshift $z \approx 0$ for various galaxy samples, from rotationally-supported galaxies (spirals and irregulars) to local dwarf spheroidals and pressure-supported early-type galaxies (ETGs), even extending to weak lensing-based studies \citep{Lelli2017, Mistele_2024, Brouwer_2021}. This has also been extended to $z \approx 0.02$ with ultra-diffuse galaxies in clusters \citep{Freundlich_2022} and galaxy groups and clusters \citep{Tian_2020, Chan_2020, Tian_2024}.
%particularly intriguing that UDGs obey the same scaling relations as field spirals despite their very diﬀerent environments and formation scenarios - Freundlich 2022. low accelerations + strong external field - testing ground for MOND theories 
%While individual galaxies of different morphologies tend to follow the same RAR, systems like galaxy groups and clusters do not appear to lie on the same relation. 
It is important to note that accurately measuring $g_{\rm bar}$ and $g_{\rm obs}$ is dependent on knowledge of their true distance, inclination and mass-to-light ratio, but they are often considered nuisance parameters in RAR studies \citep[e.g.][]{Desmond_2023}.

Much of the work to measure and understand the RAR using the SPARC galaxies, have either fixed their mass-to-light ratios \citep{Lelli2017} or varied these "nuisance parameters" when fitting the RAR galaxy-by-galaxy with a functional form \citep{Li_2018, Chae_2020b, Chae_2021, Chae_2022}, finding a small intrinsic scatter ($< 0.1$\,dex).
%to find an rms scatter of 0.0057 dex from the residuals around their best-fit relation with fixed $a_{0} = 1.2$.
On the other hand, \cite{Desmond_2023} performed a full joint inference to map the degeneracies between all parameters and infer the acceleration scale with or without an external field effect (EFE), to find an intrinsic scatter of 0.034 $\pm$ 0.001 (statistical) $\pm$ 0.001 (systematic).
% and weak evidence for the EFE.
Furthermore, \cite{Stiskalek_2023} show that the RAR satisfies all postulated criteria for a unique fundamental 1-dimensional correlation that governs the radial dynamics in late-type galaxies, with all other dynamical scaling relations being nothing more than projections of the RAR.  
%Garaldi - dwarf spheroidals follow the same RAR as brighter galaxies, but with higher scatter

The observed RAR has motivated numerous efforts to reproduce it within the standard $\Lambda$CDM framework using both cosmological hydrodynamical simulations (e.g., \citealt{Wadsley2017, Ludlow2017, Tenneti2017-MasssiveBlackIIsimulations, Garaldi_2018, Dutton_2019, Mercado-RAR-Hooks}) and semi-analytic models (e.g., \citealt{Desmond_MDAR, Navarro_2017}).
% Navarro - RAR is just a reflection of the self-similar nature of cold DM haloes 
%Yet, the reproducibility of the RAR in cosmological hydrodynamical simulations and semi-analytic models in a standard $\Lambda$ CDM framework remains contentious.
%has become a central point of tension in efforts to reconcile observed galaxy dynamics with predictions from cosmological simulations \citep{Desmond_2016, Ludlow2017, Tenneti2017-MasssiveBlackIIsimulations}.
%This poses a significant challenge for theoretical models of galaxy formation within the $\Lambda$CDM framework, which must reproduce not only the average trend of the RAR but also its tightness across a broad range of galaxy masses and morphologies. %this tight empirical correlation across a wide range of galaxy types and masses.
%The RAR is thus, yet another useful tool for testing galaxy formation models in both a $\Lambda$CDM cosmology, as well as alternative theories of gravities, such as MOND \citep{Milgrom_1983, Sanders_1990};
%Thus, the tightness of the RAR is difficult to reproduce in cosmological simulations, and its status is under a lot of debate
While some studies have successfully recovered a tight relation similar to that observed (\citealt{Wadsley2017, Paranjape2021}), they often show significant differences in the inferred acceleration scale or intrinsic scatter compared to SPARC studies \citep{Ludlow2017, Tenneti2017-MasssiveBlackIIsimulations}. 
For instance, \cite{Wadsley2017} found agreement with the SPARC RAR, though their results were limited by sample size. Conversely, \cite{Ludlow2017} (EAGLE simulation) and \cite{Tenneti2017-MasssiveBlackIIsimulations} (MassiveBlack-II simulation) reported acceleration scales substantially larger than obtained with SPARC, and the latter also a power-law form that is not consistent with the observations. More recently, \cite{Paranjape2021} showed the RAR could emerge naturally from baryonic processes without fine-tuning, with predicted deviations from the MOND-inspired RAR at extremely low accelerations. In parallel, observational probes at large radii using weak lensing techniques in isolated Milky Way size galaxies also yielded mixed results \citep{Brouwer_2021}, due to the omission/inclusion of hot, ionized gas component in the baryonic budget. The agreement of the weak lensing of the RAR with the SPARC one was however much improved by the updated analysis of~\citet{Mistele_2024}.
Altogether, these results highlight the ongoing challenges in reproducing the observed RAR within $\Lambda$CDM.
% While the relation emerges naturally in alternative theories of gravity, such as Modified Newtonian Dynamics (MOND) \citep{Milgrom_1983, Sanders_1990}, its status in $\Lambda$CDM remains actively debated. % open question

To investigate the RAR, one needs photometric observations that trace the baryonic mass, coupled with spectroscopic information to determine the dynamical mass. 
Atomic hydrogen, emitted at 21\,cm, is particularly important for determining the rotation curves of galaxies because it is dynamically cold, with a low velocity dispersion compared to the overall rotational velocity and extends far beyond the stellar disc, out to the large radii where dark matter is expected to dominate.
Other emission lines such as CO \citep{Topal_2018} or H$\mathrm{\alpha}$ \citep{Tiley_2016, Ubler_2017} can be used at higher redshifts but they only trace the central mass distribution and do not extend beyond the optical disc. %tro trace the dark matter halo potential \citep{Frank_2016}.
%HI gas is particularly important for determining the rotation curves of galaxies because it dynamically cold, meaning it has a low velocity dispersion compared to the overall rotational velocity of  This makes it an excellent tracer of the gravitational potential, extending far beyond the stellar disc, out to large radii, where dark matter is expected to dominate.

The baryonic component of the RAR is typically constructed from the distribution of stars and H{\sc i} gas.
The conversion from light to mass is derived using stellar population synthesis models, and a fixed $\Upsilon_{*}$ value in the near-infrared wavelength range has been found to provide an effective dust-free tracer of the older stellar population. This is also less sensitive to variations in parameters like age and metallicity compared to the optical bands \citep{Bell2001, Meidt-0.6, Norris2014, Rock2015}. 
%Meidt - a single stellar population synthesis model 
% Note: \citep{Bell2001} Errors in the dust-reddening estimates do not strongly a†ect the Ðnal derived stellar masses of a stellar population.
For example, many studies use a constant mass-to-light ratio using the Spitzer 3.6\,$\mu$m band, ranging from $\Upsilon_{*} = 0.42$ \citep{SchombertMcGaugh2014} to $\Upsilon_{*} = 0.6$ \citep{Meidt-0.6}. 
\cite{Lelli_2016} adopted a value of $\Upsilon_{*} = 0.7$ for the bulge and $\Upsilon_{*} = 0.5$ for the disc component using Spitzer 3.6\,$\mu$m photometry for the SPARC database
% (Spitzer Photometry and Accurate Rotation Curves) database of 175 late-type galaxies at $z \sim 0$, to study the RAR 
\citep{SPARC-database}.

However, one persistent source of uncertainty in constructing the baryonic component of the RAR is the assumption of a constant mass-to-light ratio, $\Upsilon_{\star}$. 
%An alternative approach to mitigating this uncertainty is to move to higher redshifts and measure the radial variations, removing the need to treat it as a nuisance parameter \citep{Desmond_2023}. 
In this paper, we aim to address this in a homogeneous way with a sample of H{\sc i}-selected galaxies that have excellent multi-wavelength ancillary data from which we can measure accurate $\mathrm{\Upsilon_{\star}}$ ratios with spatially resolved SED fitting and probe the low acceleration regime of the RAR. %which is often more difficult to constrain due to observational limitations
Combined with H{\sc i} kinematics, these observations enable us to construct both the baryonic (stars and gas) and the dynamical components required to probe the RAR at low acceleration beyond the very local Universe. 

This paper is organized as follows.
In Section~\ref{data} we outline the MIGHTEE-HI data we use to form a sample of 19 galaxies with which to investigate the RAR. In Section~\ref{methods} we describe the photometric measurements to determine the radial stellar-mass surface density and the mass-to-light ratios of our galaxies along with the measured rotation curves and the surface mass density of H{\sc i}. In Section~\ref{results} we present our results and in Section~\ref{summary} we discuss our results and summarise our conclusions. Throughout
the paper we assume $\mathrm {\Lambda CDM}$ cosmology with $H_0$ = 70 $\mathrm{km s^{-1}}$ $\mathrm{Mpc^{-1}}$, $\Omega_{m}$ = 0.3 and $\Omega_{\Lambda}$ = 0.7. Unless otherwise stated, all logarithms are base 10. 

\section{Data}
\label{data}
\subsection{MIGHTEE-HI}
\label{MIGHTEE-HI}
The MeerKAT International
GigaHertz Tiered Extragalactic Exploration survey \cite[MIGHTEE; ][]{Jarvis2016}, is a medium-deep, medium-wide survey providing simultaneous radio continuum \citep{Heywood2021,Hale2025}, spectral line \citep{MIGHTEE-DR1} and polarisation observations \citep{Taylor2024}. 
MIGHTEE covers an area over 20 deg$^2$ over four well-known extragalactic deep fields: COSMOS, XMM-LSS, Extended Chandra Deep Field South (ECDFS) and ELAIS-S1. 

The MIGHTEE-H{\sc i} emission project represents one of the first deep, blind, %untargeted
medium-wide interferometric surveys for neutral hydrogen. The Early Science spectral line products are described in detail in \cite{Maddox2021}. 
The MIGHTEE Early Science data has led to various publications and scientific results, including \cite{Ponomareva2021}, which is particularly relevant to this paper. This study defined a sample of 67 galaxies from the MIGHTEE Early Science data to measure H{\sc i} rotation curves, analyze their kinematics, and investigate the bTFR over the past billion years.

Building on these results, this work makes use of the first major data release (DR1) of the MIGHTEE-HI deep spectral line observations in the COSMOS field \citep{MIGHTEE-DR1}, which provides eight times better spectral resolution compared to the Early Science data.
The DR1 imaging products were created from 94.2 hours of on-target observations using 15 pointings. 
 The HI imaging products achieve an angular resolution of 12" at z = 0. The spectral line observations cover two interference-free regions, namely L1 ($0.23 < \mathrm{z_{HI}} < 0.48$) and L2 ($0 < \mathrm{z_{HI}} < 0.1$), of MeerKAT’s L-band (856-1712 MHz) with a spectral resolution of 26 kHz (equivalent to 5.5 km s$^{-1}$ at z = 0) for L2, and 104.5 kHz for L1. Both the L1 (960-1150 MHz) and L2 (1290-1520 MHz) sub-bands were also imaged with three different resolution settings. A detailed description of the DR1 data is available in \citep{MIGHTEE-DR1}.
%the ES data.
%in L2 band
The DR1 imaging products from MIGHTEE-H{\sc i}, with the higher spectral resolution, lay the foundation for more accurate H{\sc i} kinematics required for this paper.  %The sample used in this paper comes from L2 only. 

\subsection{Optical and Near-infrared data}
In this work, we leverage the wealth of deep multi-wavelength data over the COSMOS field, which are crucial for stellar mass measurements to complement the H{\sc i} observations.
Optical $u$-band photometry is sourced from CFHT \citep{CFHTLS}, while \textit{$GRIZ$} photometry is provided by the HyperSuprimeCam Subaru Strategic Program (HSC; \cite{Aihara2018, Aihara2019}). For the near-infrared \textit{$YJHK_{s}$} photometry, we rely on the UltraVISTA \citep{UVista} Data Release 6 imaging. The 5$\sigma$ depth of these data are in the range 25-27\,mag (AB) for a 2~arcsec aperture \citep[see Table~1 in][for more information]{Adams2023}, thus they are significantly deeper than the data usually available for photometric measurements of relatively low-redshift galaxies.
We note that a further advantage of using these data,  is that the imaging all has approximately the same seeing ($\sim 0.8$\,arcsec), mitigating the need for significant aperture corrections between bands. This is also the reason that we do not use the Spitzer 3.6\,$
\mu$m band in this study, as the resolution is a factor of $\sim 3$ poorer than the ground-based optical and near-infrared data that we use.

\subsection{Sample selection}
Based on the MIGHTEE-DR1,  we select all galaxies
with inclinations greater than $20^\circ$, as measured from the H{\sc i} moment-0 maps to ensure reliable kinematic measurements. % since corrections to the rotational velocities through the sin($i$) factor become large at smaller inclinations. 
We also
require that they are extended across at least three resolution elements along their major axes in the MIGHTEE data. Further details of the selection and automated kinematic modeling can be found in \cite{Ponomareva2021}. %\ty{what about removal of disturbed galaxies etc.? are there no quality flags similar to SPARC?}

This results in a sample of 19 galaxies to study the radial acceleration relation, focusing on those with high-quality, resolved rotation curves, up to $z = 0.08$ (Ponomareva et al 2025, submitted). % as the rotation curves represent a direct tracer of the centripetal acceleration.%
This purely H{\sc i}-selected dataset, complemented by the detailed photometry across 10 optical and near-infrared bands described above, is thus ideal for exploring the radial acceleration relation from the perspective of a sample selected on the H{\sc i} gas mass. Such H{\sc i} selection predominantly results in a sample of low-mass disc galaxies with  relatively low stellar mass, and as such enable us to probe predominantly the low-acceleration part of the RAR.
% but do not facilitate investigating the higher accelerations.}

\section{Methods}
\label{methods}
\subsection{Photometry}
\label{photometry}
We extract resolved optical and near-infrared photometry of the low-redshift H{\sc i} galaxies in the COSMOS field using the {\sc photutils} package \citep{larry_bradley_2024_10967176}.

\subsubsection{Isophotal fitting}
Cutouts were created for all galaxy images, centering each on the target galaxy, followed by background subtraction. Source detection was performed using image segmentation via the {\sc photutils} detection tool, identifying all sources with at least 10 connected pixels above a 2$\sigma$ threshold. %above the background noise level in our images.
To isolate the target galaxy, we applied two masking steps: a manual masking of foreground stars that overlap with the galaxy of interest if applicable, and an automatic masking of unrelated sources (galaxies, stars) in the image using the segmentation map. We then modified the segmentation map to retain only the galaxy and background pixels, setting all other sources to zero, yielding an image free of flux contamination from nearby objects. 
% follows the iterative method described by \citep{Jedrzejewski1987}
Elliptical isophotes were then fitted to the surface brightness distribution of each galaxy using {\sc photutils}, following \cite{Jedrzejewski1987}. %The isophotal fitting process follows the iterative method described by \citep{Jedrzejewski1987}. 
Accurate masking is crucial to ensure reliable photometric measurements and relatively straightforward for smooth, regular galaxies, however it can become problematic in late-type systems where clumps of star formation in the disc are hard to distinguish from foreground stars, for example. 
% becomes more challenging in late-type systems, where clumpy star formation in the disc can be difficult to distinguish from unrelated foreground sources.
To mitigate these effects, we employed sigma clipping during the isophote fitting to exclude deviant points, improving the stability and accuracy of the fits in low surface brightness regions or in galaxies with prominent non-elliptical features such as spiral arms.
%to clean deviant points and mitigate the effect of bright clumpy star-forming regions or masked regions within the galaxy, thus enhancing stability and accuracy in regions with low surface brightness or non-elliptical features like spiral arms, foreground stars, etc.
The fitting process begins with an initial elliptical isophote, defined by estimated values for the center ($x$, $y$), ellipticity, and position angle of the galaxy. The image is then sampled along this elliptical path to generate a 1D intensity profile as a function of angle, from which the mean intensity and pixel count within the isophote are measured. The algorithm proceeds by incrementally adjusting the semi-major axis, using the parameters of the previous fitted ellipse as the starting point for each subsequent fit.
%Intensity errors are derived from the RMS scatter of the pixel values along each ellipse. 
The isophotal analysis returns, for each fitted ellipse, the mean intensity along the elliptical path in surface brightness units, ellipticity, and position angle (PA), along with their associated uncertainties. The surface brightness profiles are then extracted from the mean intensity values, excluding masked regions.
 %The intensity errors are derived from the RMS scatter along the ellipse.  or median intensity values 
%\begin{figure}
%\centering
%\includegraphics[width = 0.8\linewidth]{Figures/Ellipticity_variation_J100211.2+020118_R.pdf}
%\caption{Example of a radial variation of the axis ratio $\mathrm{b/a}$ for one representative galaxy in our sample. The ellipicities vary significantly in the first 80 pixels, with a significant jump, followed by a steep decline near the center, and a flattening towards the outskirts. This is largely due to the change in ellipticity and position angle for the fitted isophotes (see Figure~\ref{fig:ellipse-fitting}).}
%\label{eps-variation}
%\end{figure}

It is important to point out that when fitting elliptical isophotes,  the ellipticity and position angle of the isophotes can vary significantly with radius and across different photometric bands, due to different galaxy components.
% (see Fig \ref{eps-variation}). %such as bars or spiral arms for instance 
While this variability can be useful for detailed studies of galaxy structures like bars and spiral arms that leave distinct signatures in the radial surface brightness profile of galaxies, it can result in inconsistently sampling the same physical regions of a galaxy. This consistency is crucial for later deriving stellar masses and surface mass densities using SED fitting.

Since {\sc Photutils} is optimized to fit elliptical isophotes, particularly in early-type galaxies, attempting to simultaneously fix all the isophote parameters caused the fitting algorithm to fail. To address this, we adopted an alternative method, that consisted of three main steps. 
First, we used the \texttt{fit image} task and allowed the ellipticity and position angle to vary freely, keeping only the center fixed. We chose the HSC $G$ band as a reference due to its depth, with a 5$\sigma$ AB limiting magnitude of 27.4 mag \citep{Aihara_2022}. The $G$-band effectively captures key structural features and extends far enough into the  outer disc of galaxies for a robust comparison of the H{\sc i} and stellar disc morphologies. %making it a good tracer of the HI disc geometry as well. 
We also verified our results using the $R$-band, our second deepest band with a 5$\sigma$ AB limiting magnitude of 27.1 mag, traditionally used to derive optical inclinations, as it is more sensitive to the older stellar population that traces stellar mass. The inferred axis ratios are consistent between $G$- and $R$-bands within uncertainties.
Second, we used the outermost isophote from the $G$-band fit to define a fixed geometry—fixing the center, ellipticity, and position angle-which was visually inspected to ensure it provided a good overall description of the galaxy morphology. 
Finally, based on this geometry, we extracted surface brightness profiles in all optical and near-infrared bands via the {\sc Ellipse Sample} method. %to extract intensities along these predefined elliptical paths. %based on a fixed geometry.
This ensured consistent sampling of physical regions across bands and avoided the convergence issues associated with trying to fit all parameters at once. 
%that arise trying to fit all parameters at once.
%We therefore implemented this approach using the G-band geometry, though we also tested the R-band as well.
%The procedure consisted of three main steps.
%First, we fit elliptical isophotes in the $G$-band with a fixed center, whilst allowing ellipticity and position angle to vary freely for the initial fit.
%Second, we defined the geometry with fixed ellipticity and position angle with the outermost isophote found in step one chosen (and visually checked to ensure they provided a good overall description of the galaxy morphology). 
%And finally, we sampled intensities along these fixed elliptical paths using the predefined $G$-band geometry (with fixed center, position angle, and ellipticity). 
%This allowed us to extract intensities from all optical and near-infrared bands along the same set of elliptical paths, thus ensuring consistency in sampled regions across bands, and avoiding the issues that arise trying to fit all parameters at once. %at the same radii
\begin{figure}
    \centering
    \begin{subfigure}[b]{0.48\linewidth}
        \centering
        \includegraphics[width=\linewidth]{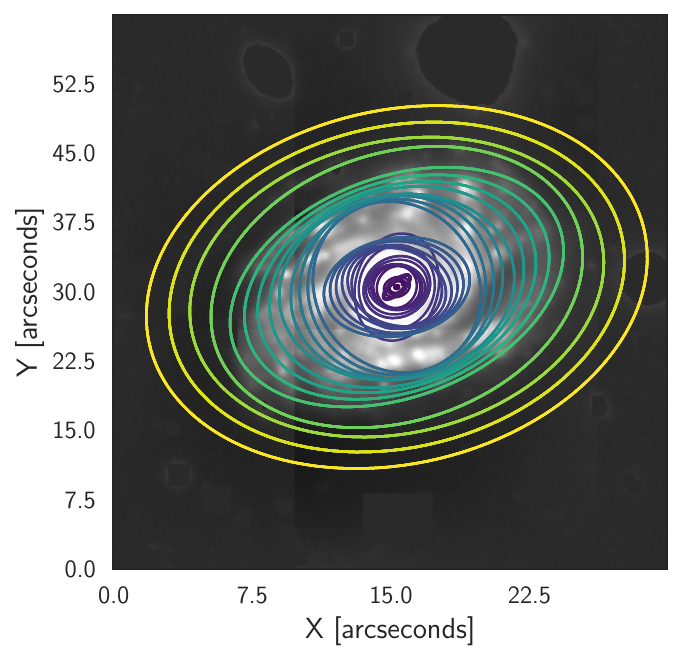}
    \end{subfigure}
    \hfill
     \begin{subfigure}[b]{0.48\linewidth}
        \centering
        \includegraphics[width=\linewidth]{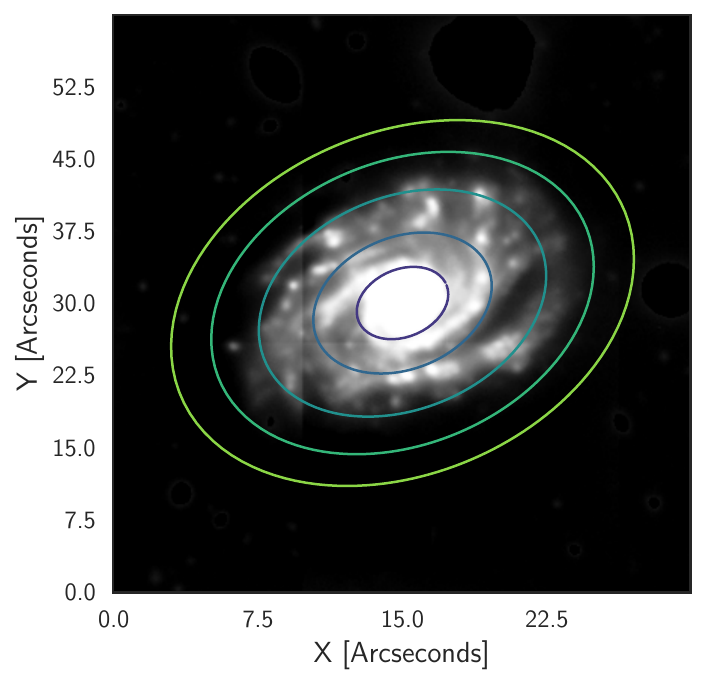}
    \end{subfigure}
    \caption{\textit{Left}: Elliptical isophotes on the $G$-band image with varying position angle and ellipticity for one of our galaxies. \textit{Right}: Elliptical apertures that we use for the analysis on the same image placed every 5~arcsecs with fixed center, position angle and ellipticity.}
    \label{fig:ellipse-fitting}
\end{figure}
An example of using the method outlined above can be seen in Figure~\ref{fig:ellipse-fitting} for a representative galaxy in our sample. 
We then computed the uncertainties in both position angle and axis ratio by summing in quadrature the {\sc Photutils} error of the outermost isophote and the standard deviation from multiple measurements.

\subsubsection{Correcting for inclination}
\label{inclinations}
%Inclinations are crucial for both photometry and kinematics.
The inclinations of galaxies play a key role in this study, as they affect both the surface brightness profiles and the derivation of rotational velocities from H{\sc i} kinematics. Typically, inclinations can be derived from both optical and H{\sc i} data \citep{Verheijen_2001}. In general, the H{\sc i} disc is much thinner than the stellar disc and its intrinsic thickness can be neglected but due to flaring, it can become significantly thick in the outer parts (\citealt{Bacchini_2019, Mancera_2025}). 
Since our H{\sc i} data is only marginally resolved, we choose to adopt the optical inclinations, as our optical data is sufficiently deep to reliably trace the underlying stellar disc. The inclinations ($i$) were derived from the axis ratio using the standard relation:
\begin{equation} 
\cos^2(i) = \frac{(b/a)^2 - q_0^2}{1 - q_0^2},
\end{equation} where $b$ and $a$ are the semi-major and semi-minor axes in $G$-band, and $q_0$ represents the intrinsic axis ratio of the disc, typically between 0 (thin disc) and 0.4 \citep{Fouque1990}. In our analysis, we adopt the thin disc assumption with $q_0$ = 0. We verified that assuming different $q_0$ values (thin disc with $q_0 = 0$, or following standard prescriptions with $q_0 = 0.2$) had a negligible impact on the derived inclinations, with variations well within the uncertainties. 
%confirmed that the difference in inclination with or without assuming a thickness for the disc $q_0$ is negligible and within the uncertainties.
 The specific intensity (flux per unit area) varies with inclination, such that in highly inclined galaxies the surface brightness increases as the projected area of the galaxy is decreased from its face-on value. To obtain accurate measurements, it is therefore essential to correct for this projection effect by adjusting the observed surface brightness ($\mu_{\rm obs}$) to its face-on surface brightness value ($\mu_0$) at each radius using: 
%thus corrected the observed surface brightness profiles for inclination using:
\begin{equation}\label{eq-inclination-correction}
    \mu_{0} = \mu_{\rm obs}- 2.5 \log_{10}\mathrm{b/a}   .
\end{equation}

We also applied Galactic extinction corrections using Schlegel dust extinction maps \citep{Schlegel} that employ a $R_V$ = 3.1 extinction curve, using the appropriate extinction coefficients corresponding to the different photometric bands. In the near-infrared, these were taken from the VISTA technical information. For the optical bands, however, we utilized the Python package {\sc extinction} that is commonly used to calculate the Galactic extinction and reddening at a given wavelength based on the \cite{Fitzpatrick99} model, given a certain $R_V$. 
%Following this, the redenning corrections, $A_\lambda$, were found to be higher in the optical bands compared to the near-infrared ones, 
The average extinction in $G$ and $R$ bands is $\approx 0.07$ mag and only $0.006$ for the $K_{s}$-band. This reflects the benefits of using the data across the COSMOS field 
%at low galactic latitude, where dust attenuation is negligible.
%The average extinction in the G and R bands was found to be $0.07$ mag, and only $0.006$ mag in the K band. This highlights the advantage of using data across the COSMOS field to investigate the RAR, 
as it lies at high Galactic latitude and therefore suffers minimal galactic extinction.
%, compared to the majoroity SPARC galaxies. 
%corrective term assuming an optically thin disc?? see Trujilo 
%= 2.5C log (a/b); C = 0 for optical thick disc, C = 1 for transparent disc 
%The total (apparent) magnitude of the galaxy is the surface brightness integrated over the entire galaxy. 
The surface brightness was measured for each galaxy over a range of elliptical isophotes  and the total integrated flux in each photometric band was obtained by summing the pixel values within the largest fitted elliptical isophote.
To measure the uncertainties on the fluxes, circular apertures with areas equal to that of the fitted ellipse at a specific semi-major axis were randomly placed within a larger cutout image, ensuring they were placed outside other bright sources or galaxies. A Gaussian was fitted to the histogram of flux values measured in these apertures, and the standard deviation of the fitted Gaussian was used as an error estimate for the flux within the specified aperture. This method ensures that the calculated errors reflect the noise characteristic of the specific aperture size. 

\subsubsection{SED fitting}
The total fluxes in all bands were used to measure the stellar masses through spectral energy distribution (SED) fitting for each galaxy. To achieve this, we employed the {\sc BAGPIPES} code (Bayesian Analysis of Galaxies for Physical Inference and Parameter Estimation; \citealt{Bagpipes}). {\sc BAGPIPES} contains several star formation history (SFH) models, including: exponential, delayed exponential, log-normal and double power law and uses a Chabrier Initial Mass Function \citep[IMF; ][]{Chabrier2003} with Stellar Population Synthesis models based on \cite{BruzualCharlot}, with three dust extinction models: Calzetti \citep{Calzetti2000}, Cardelli Milky Way dust law \citep{Cardelli_1989} and \citep{CharlotFall2000}. It works on the assumption of energy balance where the strength of dust emission is equal to the amount of energy removed from the UV-optical by the dust attenuation. 

We tested several parametric star formation history models, however the choice of SFH did not affect the final stellar mass results. In fact, for most of the MIGHTEE-HI galaxies, no star formation history (SFH) model is preferred over the other  \citep{Tudorache_2024}. 
Therefore, for the rest of this analysis, we fit all our photometry with the delayed exponential model. We model the dust attenuation using the Calzetti law by allowing the extinction to vary between $A_{V} = 0 - 2$, whilst keeping the redshift fixed as all of our galaxies have precise redshifts from their H{\sc i} emission line. We note that the choice of extinction law does not affect our results.
We employ the stellar grids that contain remnants, i.e., the stellar mass included in white dwarfs, neutron stars, etc., but exclude gas lost by stellar winds or supernovae. For the rest of this paper, we quote stellar masses and surface mass densities including remnants, if not specified otherwise.\footnote{We tested both scenarios, both with remnants and without, and it does not significantly impact our final results of stellar masses, stellar surface densities or mass-to-light ratios.}
\begin{figure}
        \centering
        \includegraphics[width=\linewidth]{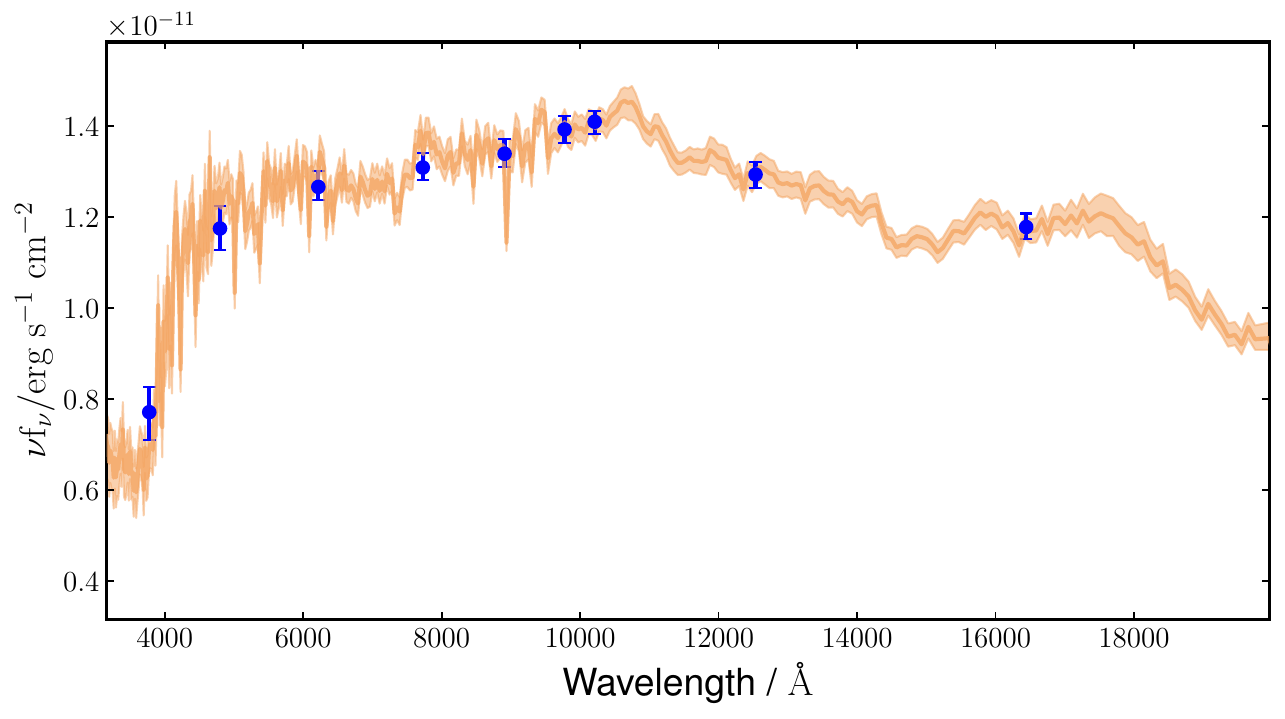}
        %15_fit.pdf}
    \caption{An example of SED fit with Bagpipes across the 10 photometric bands, from HSC \textit{ugrizy} and VISTA \textit{YJHKs}. The input photometry is given by the blue points and the orange line is the posterior SED, both presented in the rest-frame.}
    \label{fig:SED-fit}
\end{figure}

We obtain the total stellar masses, $M_\star$ for all our galaxies using the photometry across the 10 photometric bands. Figure~\ref{fig:SED-fit} illustrates an example of an SED fit obtained for a galaxy in our sample. The uncertainties on the stellar masses are based on the posterior distributions from {\sc Bagpipes} and do not take into account systematic uncertainties that arise from the choice of IMF or star-formation history for example. 
The near-infrared photometry is particularly useful to robustly estimate the total stellar masses, as it is minimally affected by dust attenuation and recent star formation episodes. The total stellar masses for our samples can be found in Table~\ref{tab:total_mass_to_light_ratios}.

\subsubsection{Resolved stellar mass surface densities}
After deriving the stellar masses, we compute the resolved stellar mass surface densities for the galaxies in our sample. To achieve this, we first fit their surface brightness profiles with an analytical model and then use these fits to convert the surface brightness into stellar mass surface density as a function of radius, $\Sigma_{*}(r)$.

Typically, surface brightness profiles of galaxies are well characterized by a Sersic profile \citep{Sersic_1968}:
\begin{equation} 
       I(R) = I_e \exp \left\{ -b_n \left[ \left( \frac{R}{R_e} \right)^{1/n} - 1 \right] \right\},
\label{sersic-function} 
\end{equation} 
where $I_e$ is the intensity at the half-light radius, $R_e$ and $b_n$ is a constant that depends on the Sersic index $n$, with $n$ describing the light profile or how concentrated the light is (with brighter and more massive galaxies having larger $n$ values \citep{Capaccioli1993}). The constant $b_n$ can be approximated by 1.992$n$ - 0.3271 \citep[e.g.][]{Capaccioli_1989, Prugniel1997}.
For some galaxies in our sample that lack distinct bulges or contain pseudo-bulges, we evaluate whether this single Sersic profile more accurately describes the galaxy compared to a double Sersic model, containing both bulge and disc components. The latter is described by the following functional form:
\begin{equation} 
    I = I_{e_1} \exp \left[-b_{n_1} \left(\frac{R}{R_{e_1}}\right)^{1/n_{1}} - 1\right] + I_{e_2} \exp \left[-b_{n_2} \left(\frac{R}{R_{e_2}}\right)^{1/n_{2}} - 1\right] .
\label{double-sersic} 
\end{equation}
To achieve this, we use {\sc Nautilus} \citep{nautilus}, a nesting sampling algorithm  that relies on Bayesian inference to estimate the best model and map the posterior distribution of the parameters, for the bulge $I_{e_1}$, $R_{e_1}$, $n_{1}$ and disc components $I_{e_2}$, $R_{e_2}$ and $n_{2}$. 
We set uniform priors on the parameters of the Sersic function to cover the range in intensity and radius of our data points and set $0.5 < n < 4$. 
We use the Bayesian evidence ($Z$) to compare the single and double Sersic models, which represents the probability of the observed data given the model. Given two different models, the Bayes factor $B$ is defined by: 
\begin{equation}
    \log_{10} B = \log_{10} Z(d|M_{D}) - \log_{10} Z (d | M_{S}),
\end{equation} where $M_{S}$ and $M_{D}$ are the double and single Sersic model that we are comparing. 
For all galaxies in our sample, the logarithm of the Bayes factor was $>9$ indicating that the double Sersic model decisively fits the data better.
We thus proceed to fit the surface brightness profiles with a double Sersic. 

\begin{figure}
    \centering
    \includegraphics[width=0.98\linewidth]{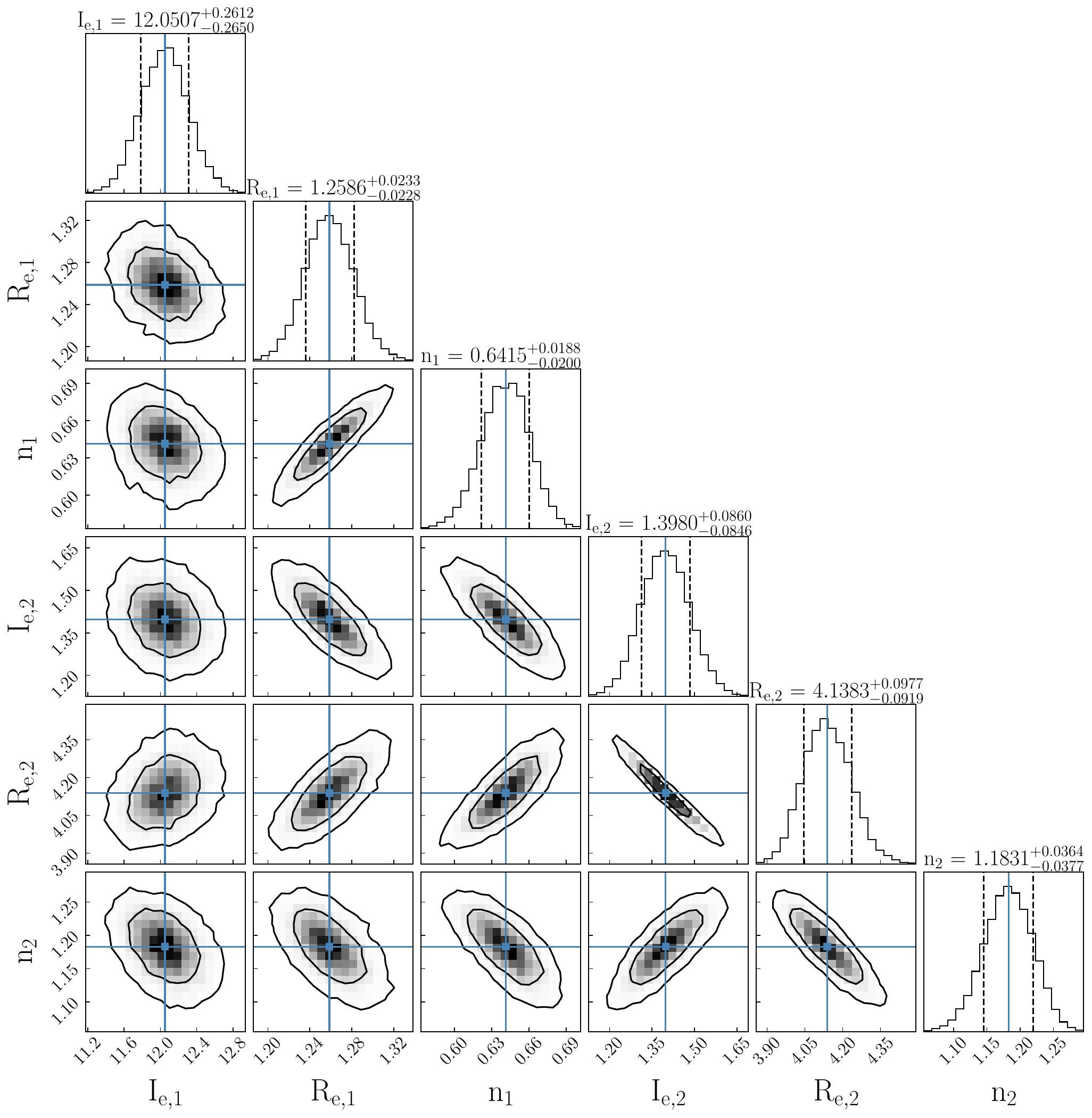}
    \caption{Example of posterior distribution for the set of 6 parameters describing the double Sersic function. The intensities at the half-light radius are given in units of $\mathrm{\mu Jy/arcsecond^2}$ and the radii are expressed in arcseconds. The blue lines represent the maximum likelihood estimates for the parameters, whilst the dashed lines show the 16th and 84th percentiles. Contours indicate the 68\% and 95\% confidence regions for the 2D posteriors.}
    \label{fig:Nautilus-corner-plot}
\end{figure}

\begin{figure}
    \centering
    \includegraphics[width =\linewidth]{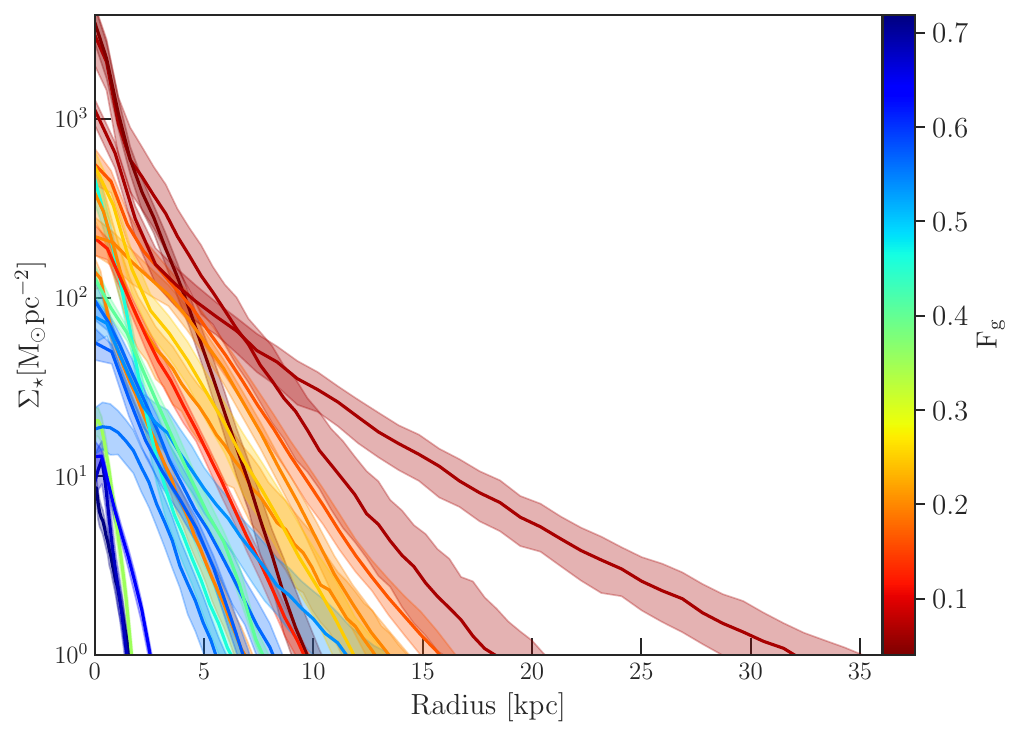}
    \caption{Resolved stellar surface mass densities for our sample of galaxies, colour coded by their gas fraction, $\mathrm{F_{g} = M_{HI}/M_{bar}}$.}
    \label{fig:RAR-stars-surf-dens}
\end{figure}

To recover the intrinsic, face-on surface brightness values, we convolve the model double Sersic profile with the PSF to match our data, and run {\sc Nautilus} to derive the best fit values, as shown in Figure~\ref{fig:Nautilus-corner-plot}. We fit this double Sersic model independently for each filter for every galaxy and calculate the surface brightness at fixed radii, employing a radial spacing of 0.6~arcseconds. We use these surface brightness profiles in {\sc Bagpipes} to obtain the inclination corrected radial stellar mass surface densities. We visually inspect the SED fits and the corner plots at every radius to ensure that the surface mass density $\Sigma_{*}$ is robustly constrained. 
Figure~\ref{fig:RAR-stars-surf-dens} presents all stellar mass surface densities obtained for our sample---high-mass galaxies have shallower declines in their surface brightness profiles, indicative of stellar contribution at all radii, whereas low-mass, gas-dominated galaxies show steeper declines.

\subsubsection{Mass-to-light ratio variations}
\label{ml-variations}
One of the most fundamental properties that can offer insights into the formation and evolution of a galaxy is its stellar mass, in addition to the luminosity produced by its stellar populations. The stellar mass-to-light ratio ($\mathrm{\Upsilon}_{\star}$) enables a direct translation between photometry and dynamics. This plays a key role in rotation curve decomposition and mass modeling of disc galaxies, where it represents the major source of uncertainty and thus, the assumed mass-to-light ratio directly affects the inferred distribution of baryons and dark matter.
%The mass-to-light ratio represents the major source of uncertainty in studies of mass modelling.  distribution of dark matter and baryons in
%galaxies
Typically, the stellar mass-to-light ratio in different photometric bands is known to depend on stellar age, metallicity, IMF, and dust extinction and can be derived from stellar population models \citep{BruzualCharlot, Bell_2003}. 
%These models integrate an assumed IMF with star formation and chemical enrichment histories, combined with stellar isochrones, to estimate the mean \mathrm{M/L} ratio 
Mass-to-light ratios are highly wavelength-dependent, especially in bluer filters, where there is substantial scatter.
However, incorporating the near-infrared has been found to yield more robust stellar mass estimates, as the stellar mass-to-light ratios are less sensitive to age compared to UV or optical data \citep{Into_2013, Meidt-0.6, McGaugh2015}.
%Nonetheless, intermediate-age stars such as thermally pulsating asympttotic giatn granch (TP_AGB) can significantly enhance the NIR fluxes, complicating the picture
A constant stellar mass-to-light ratio has been commonly adopted in studies of galaxy dynamics to disentangle the contributions of dark matter and baryons to the total mass distribution. However, the systematic biases introduced by this assumption have only been quantified in a few studies, potentially affecting the accuracy of derived mass profiles and scaling relations \citep{Ponomareva2017,Liang_2024}. % they use early type galaxies here thought in TNG100 hydrodynamic simulation and they find that the ML ratio is generally overestimated and dm fraction underestimated by 20% on average - maybe not relevant but the idea behind

In this section, we aim to investigate mass-to-light ratio $\mathrm{\Upsilon_{\star}}$ variations as a function of radius for all galaxies in our sample and compute the total $\mathrm{\Upsilon_{\star}}$ for our galaxies. For this purpose, we use the near-infrared VISTA $K_{s}$-band, as near-infrared bands are less affected by dust and the light is emitted predominantly by the older stellar population that forms the bulk of the stellar mass \citep{Sorce2013}, as discussed previously. Additionally, at a wavelength of 2.2\,$\mu$m, it is close to the commonly employed 3.6\,$\mu$m band for similar studies \citep[e.g.][]{Verheijen_2001}. %\footnote{We note that we do not use the 3.6\,$\mu$m data due to the significantly poorer spatial resolution compared to our ground-based imaging}

\begin{figure}
    \centering
    \includegraphics[width = \linewidth]{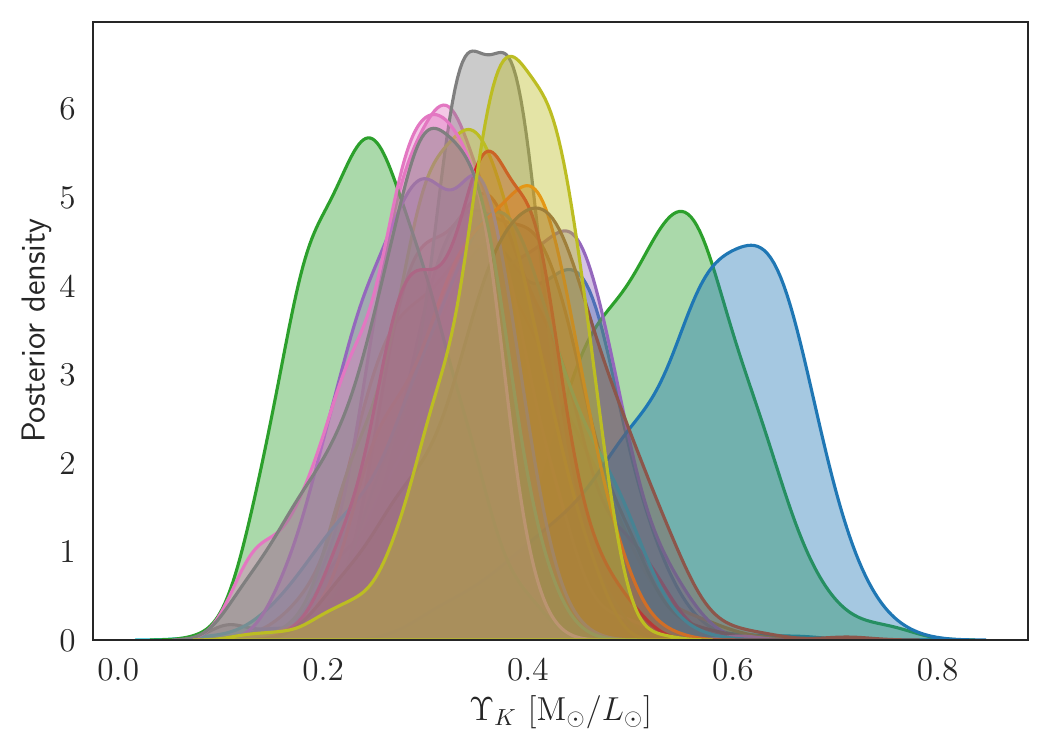}
    %ML_histogram_remnants.pdf}
    %\linewidth]{Figures/ML_histogram_remnants.pdf}
    \caption{Distribution of mass-to-light ratios in the $K_s$-band ($\mathrm{\Upsilon_{\star}}$) from the mass-to-light ratio posterior samples derived from SED fitting with {\sc Bagpipes}.}
    \label{fig:ml-histogram-remnants}
\end{figure}

%For the total mass-to-light ratios, the total K band luminosity was calculated as:
%using the distance modulus 
%\begin{equation}
%\text{L} = 10^{0.4 \times (M_{\odot,K} - M_K)},
%\end{equation} with $M_{K}$ obtained using the distance modulus and the galaxy's K band integrated magnitude.
%shows the distribution of mass-to-light ratios in the K band, with a median value of 0.33 and a standard deviation of 0.1.
%This calculation was performed for all galaxies in the sample, and these results, 
%The measured $K_{s}-$band luminosities along with the total stellar masses are presented in Table~\ref{tab:total_mass_to_light_ratios} and 
The distribution of mass-to-light ratio in the $K_{s}-$band for our galaxies is shown in Figure~\ref{fig:ml-histogram-remnants}. We find that the total mass-to-light ratios obtained from SED fitting cover a wide range of values from 0.24 to 0.57, with a median of 0.35. These findings align well with those of \citet{Ponomareva2017}, who employed a similar methodology, as well as with results from the DiskMass Survey \citep{Martinsson_2013}. However, these values are in tension with recent studies such as \cite{Marasco_2025}, who used 2MASS $K_s$-band data and found a median $\mathrm{\Upsilon_{\star}} = 0.7$, with significantly higher scatter. This discrepancy is likely due to their sample being a subsample of the more heterogeneous SPARC sample, which has a significantly larger fraction of high-stellar-mass galaxies that tend to exhibit higher mass-to-light ratios. 

\FloatBarrier
\vspace{0.5 cm}
\begin{table}
\centering
\resizebox{0.48\textwidth}{!}{%
\begin{tabular}{lrrll}
\toprule
Galaxy & $z$ & $i_{\rm opt}$ [$^\circ$] & 
$\log_{10}(M_\star / M_\odot)$ & 
$\Upsilon_{\star}\ [\mathrm{M_\odot /L_\odot}]$ \\
\midrule
J095846.8+022051 & 0.00577 & 70 & $7.48 \pm 0.08$ & $0.33 \pm 0.08$ \\
J095927.9+020025 & 0.01297 & 41 & $7.65 \pm 0.10$ & $0.35 \pm 0.09$ \\
J100005.8+015440 & 0.00622 & 66 & $7.82 \pm 0.06$ & $0.53 \pm 0.08$ \\
J095904.3+021516 & 0.02458 & 50 & $7.98 \pm 0.08$ & $0.36 \pm 0.08$ \\
J100211.2+020118 & 0.02134 & 41 & $9.18 \pm 0.09$ & $0.38 \pm 0.08$ \\
J100009.3+024247 & 0.03267 & 56 & $9.49 \pm 0.09$ & $0.37 \pm 0.07$ \\
J100115.2+021823 & 0.02845 & 73 & $8.79 \pm 0.09$ & $0.32 \pm 0.06$ \\
J095720.6+015507 & 0.03205 & 48 & $10.10 \pm 0.07$ & $0.36 \pm 0.06$ \\
J100143.2+024109 & 0.04699 & 45 & $9.43 \pm 0.09$ & $0.34 \pm 0.06$ \\
J100259.0+022035 & 0.04426 & 43 & $10.92 \pm 0.10$ & $0.37 \pm 0.10$ \\
J100055.2+022344 & 0.04426 & 42 & $10.53 \pm 0.08$ & $0.57 \pm 0.10$ \\
J095923.2+024137 & 0.04764 & 61 & $9.75 \pm 0.10$ & $0.37 \pm 0.08$ \\
J100236.5+014836 & 0.04554 & 44 & $9.47 \pm 0.12$ & $0.24 \pm 0.07$ \\
J100117.1+020337 & 0.06155 & 47 & $9.50 \pm 0.09$ & $0.34 \pm 0.07$ \\
J100003.9+015253 & 0.06521 & 60 & $9.19 \pm 0.09$ & $0.30 \pm 0.07$ \\
J095755.9+022608 & 0.07125 & 76 & $10.08 \pm 0.10$ & $0.39 \pm 0.09$ \\
J100217.9+015124 & 0.06238 & 49 & $10.31 \pm 0.12$ & $0.29 \pm 0.07$ \\
J100103.7+023053 & 0.07193 & 39 & $10.09 \pm 0.10$ & $0.30 \pm 0.08$ \\
J095907.8+024213 & 0.07908 & 46 & $10.81 \pm 0.07$ & $0.38 \pm 0.06$ \\
\bottomrule
\end{tabular}%
}
\caption{Redshifts, optical inclinations, stellar masses, and total mass-to-light ratios in $K_{s}-$band for the galaxies in our sample. Stellar masses are derived from SED fitting using a delayed exponential star formation history \citep{Mobasher2015}, with uncertainties reflecting the posterior distributions obtained from {\sc Bagpipes}.}
\label{tab:total_mass_to_light_ratios}
\end{table}

\begin{figure}
    \centering
    \includegraphics[width = \linewidth]{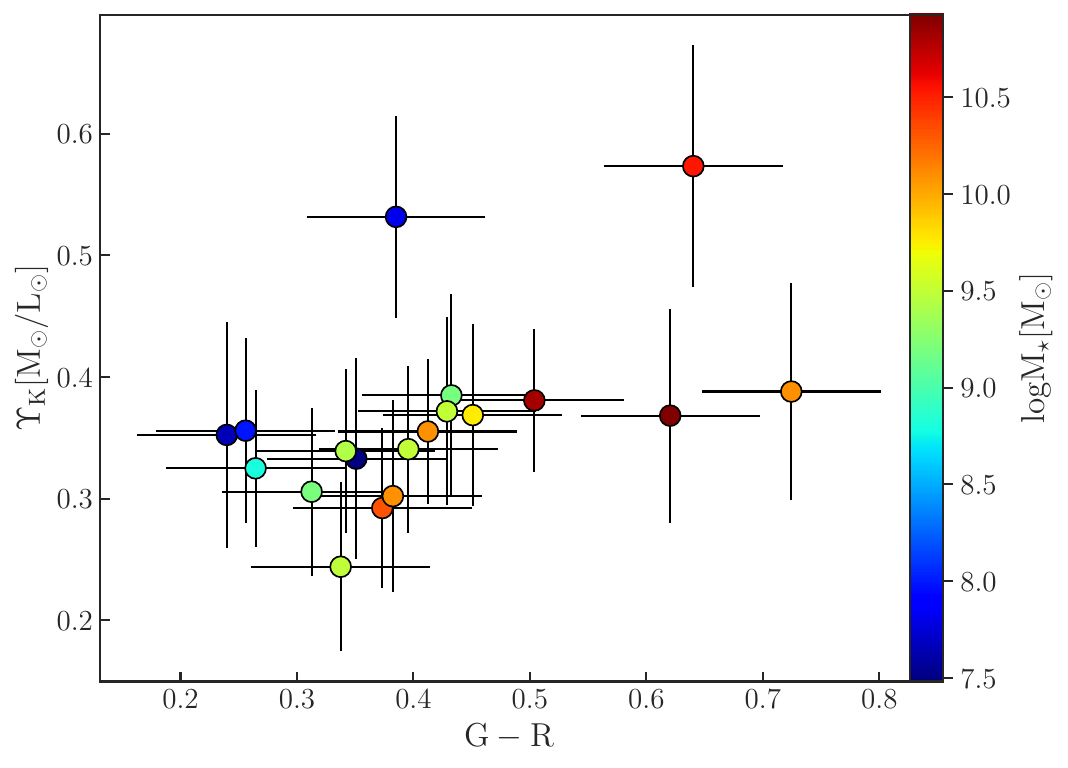}
    \caption{Mass-to-light ratio in $K_s$-band for our galaxies as a function of $G-R$ colour, colour coded by stellar mass.}
    \label{fig:ml-colour-Klum}
\end{figure}
%{\it left}: {\it right}:
In Figure \ref{fig:ml-colour-Klum} we show the mass-to-light ratio in relation to the $G - R$ colour. This shows a clear correlation, with a Pearson coefficient of 0.66. This trend is particularly relevant for our sample of late-type galaxies: lower $G - R$ values correspond to bluer, star-forming, disc-dominated systems that exhibit lower mass-to-light ratios, whereas redder $G - R$ values are associated with more bulge-dominated spirals and higher mass-to-light ratios.

To determine the radially-varying mass-to-light ratios, we use the ratio of stellar surface mass density, obtained from SED fitting, to the $K_s$-band surface brightness, both inclination corrected and derived from the double Sersic profiles.

%$\mathrm{I (L_\odot /pc^2)}$ is simply the double Sersic profile computed at the same radii as the stellar surface mass density. 
%The distance $d$ was computed using Hubble's Law at the galaxy's redshift by employing astropy's cosmology sub-package that computes the luminosity distance of a galaxy given its redshift \citep{Astropy_2018}. 

\begin{figure}
    \centering
    \includegraphics[width=\linewidth]{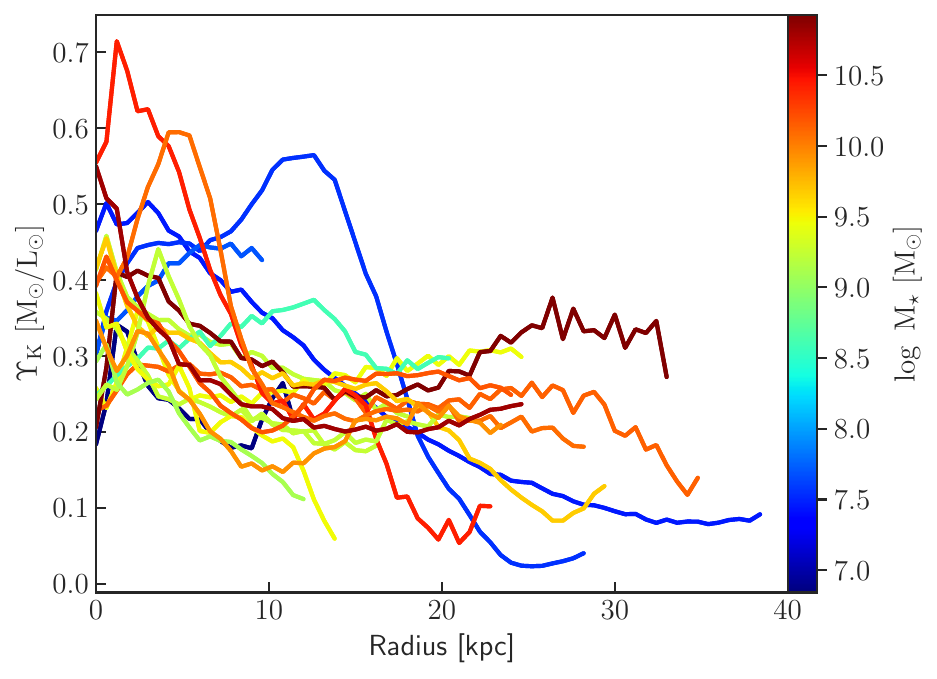}
    \caption{Mass-to-light ratio variations in $K_s$-band as a function of radius for all galaxies in our sample, colour coded by stellar mass. Unsurprisingly, the mass-to-light ratios are typically higher in the center, followed by a decrease and flattening at large radii, where the surface brightness exponentially decreases.}
    \label{fig:ml-ratio-variations}
\end{figure}

In Figure~\ref{fig:ml-ratio-variations} we show the radial variations in the mass-to-light ratio for all galaxies in our sample, colour-coded by stellar mass. The figure reveals a clear trend where the mass-to-light ratio is higher in the central regions and flattens beyond a certain radius.
These variations of $\Upsilon_{*}$ in the $K_{s}$ band both as a function of radius and colour, suggest that a constant $\Upsilon_{*}$ is not appropriate to describe the complicated physics of disc galaxies when converting light to stellar mass. 
%The gradients in the mass-to-light ratios are indicative of differences in the IMF and star formation histories. 

\subsection{H{\sc i} surface density profiles and rotation curves}
\label{HI-kinematics}
After deriving the stellar contribution, we must now compute the gas surface mass densities to assess their contribution to the baryonic component of the RAR. %and derive rotation curves to obtain the 

Due to the faintness of the H{\sc i} line, resolved observations of galaxies in H{\sc i} become increasingly challenging at higher redshifts. Kinematic modeling requires H{\sc i} observations with high spatial and velocity resolution, combined with a high signal-to-noise ratio (SNR), to accurately derive rotation curves. However, the new generation of large-scale H{\sc i} surveys are transforming this landscape, coupled with new software developments that enable kinematic modeling even for marginally resolved galaxies.  

These include {\sc 3D Barolo} \citep{DiTeodoro2015} and {\sc Tirific} \citep{Tirific}, that can effectively determine the underlying kinematics in galaxies with as few as three resolution elements along the major axis, provided that the signal-to-noise ratio (SNR) is greater than 2-3 \citep{DiTeodoro2015, Mancera_Pina_2020}.

In this work, we use the software {\sc 3D Barolo} (3D Based analysis of Rotating Objects from Line Observations; \citealt{DiTeodoro2015}), to derive rotation curves and radial H{\sc i} surface density profiles. {\sc 3D Barolo} automatically fits a 3-dimensional tilted-ring model to emission-line data cubes. 
We use the 3D fit task that works with a masked data cube to exclude noisy pixels. The source mask is generated using MASK=SMOOTH with a signal-to-noise ratio threshold of 3 (SNRCUT=3). Additionally, we set NORM=AZIM for an azimuthal normalization of the moment-0 map to obtain the average gas surface density $\mathrm{\Sigma_{gas}}$ for each ring in the model. Throughout the analysis, we assume a razor-thin H{\sc i} disc. %as any effect of disc thickness is dominated by the size of the beam. 
To account for beam smearing, {\sc 3D Barolo} convolves the model with the beam. The software also includes a built-in source finding algorithm that finds a galaxy in the cube and reliably estimates the galaxy's center and systemic velocity during the initial run.
% mitigates the effect of beam smearing by convolving the model with the beam. 
 %Although {\sc 3D Barolo} works very well with poorly resolved data, %However, despite its reliability in estimating the kinematic center and systemic velocity, 
 
Despite this however, {\sc 3D Barolo} is sensitive to initial parameter guesses, particularly the inclination angle, which is more challenging to constrain in lower-resolution data (see also Fig 2 in \citealt{Ponomareva2021}). Therefore, we input the inclination based on optical photometry, as discussed in section \ref{inclinations}, allowing it to vary within its uncertainties, while keeping other parameters free.

The fitting process consists of two steps: an initial fitting, where the rotation velocity, velocity dispersion, inclination and position angle are all free parameters to fit, and a second step, where only the rotation velocity and velocity dispersion are fitted, while the inclination and position angle of the rings are fixed. 
%To obtain the rotation curve, we input the initial inclination estimates and fit the kinematic parameters, such as $v_{rot}$, with the disc geometry. 
We assume a radial separation equal to half the beam size, i.e., two rotation curve points per beam. The uncertainties on the rotation velocities are estimated using {\sc 3D Barolo}'s default Monte Carlo error calculation method, implemented through the built-in function FLAGERRORS.

\begin{figure}
    \centering
    \includegraphics[width =\linewidth]{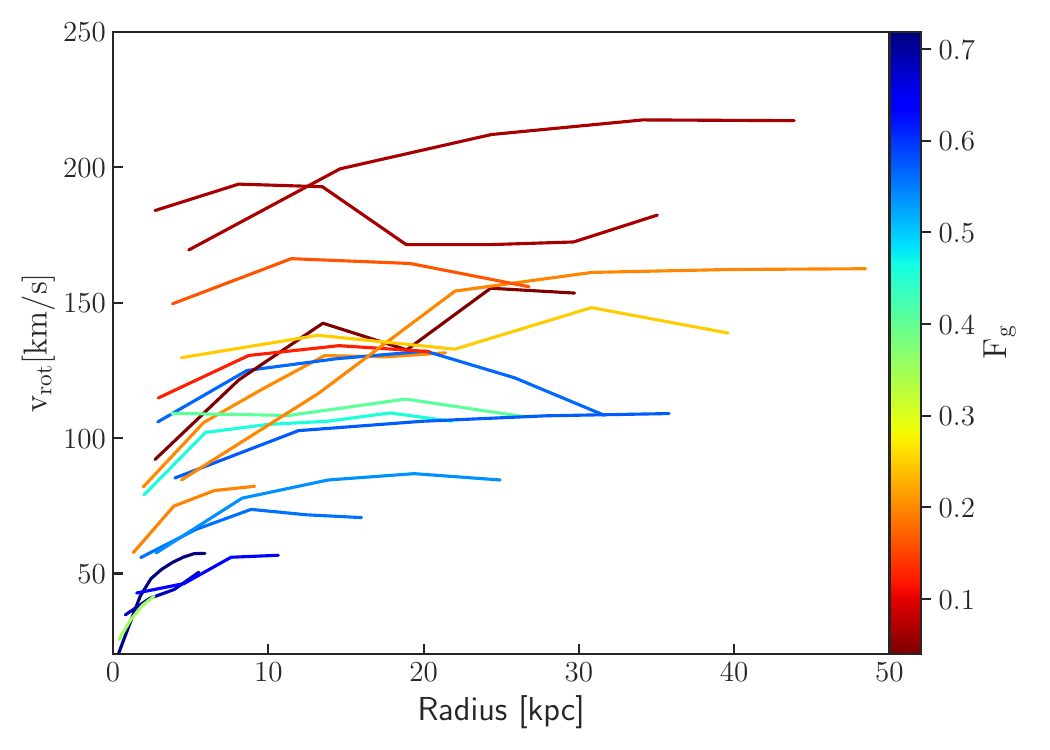}
    \caption{Rotation curves for our sample of galaxies, colour coded by their gas fraction, $\mathrm{F_{g} = M_{HI}/M_{bar}}$.}%with the shading regions indicating the uncertainties }
    \label{fig:rotation-curves}
\end{figure}
The rotation curves for our sample are presented in Figure~\ref{fig:rotation-curves} (see also Ponomareva et al 2025, submitted).
%which are in good agreement with other more standard methods like the difference between the receding and approaching halves of the disc \citep{Barolo}
%Assuming that the HI emission is optically thin

 After obtaining the rotation velocities, we also derive the azimuthally averaged H{\sc i} surface mass densities at each radius, corrected for inclination. %Assuming a razor-thin HI disc, the relation between $\Sigms_\text{int}$ and $\Sigma_{obs}$ is 
 This allows us to quantify the baryonic contribution of the gas component to the total radial acceleration.
 The mean H{\sc i} flux density is obtained at each radius from the input data cubes, corrected to face-on values, and automatically converted to physical units of $\mathrm{M_\odot /pc^{2}}$
 \citep{Meyer_2017} in {\sc 3D Barolo}.
 In disc galaxies, the dominant baryonic mass components are stars and atomic gas, and the contribution from molecular gas is typically small \citep{Catinella_2018}. However, we should still account for these components insofar as it is feasible. We therefore apply a correction factor to the face-on H{\sc i} surface densities to account for contributions from helium, metals, and molecular gas. 
%Since molecular hydrogen is less accessible, the molecular gas mass is usuallty traced by the next available molecule, CO, however CO is rarely detected in low mass galaxies Leroy 2008, and on top of that, the conversion factors from CO flux to H2 flux and mass are quite uncertain. 
%This ensures an accurate estimate of the total gas contribution to the baryonic radial acceleration. 
The total gas mass, corrected for the hydrogen fraction \citep{McGaugh_2020}, can be expressed as:
\begin{equation}
\mathrm{M_g} = X^{-1} ({M_{\rm HI} + M_{\rm H_2}}), 
\end{equation} where ${M_{\rm HI}}$ and $M_{\rm H_2}$ represent atomic and molecular hydrogen, respectively. The total hydrogen fraction $X$ is related is then related to the stellar mass by

\begin{equation} X(M_\star) = 0.75 - 38.2 \left( \frac{M_\star}{1.5 \times 10^{24}~\mathrm{M}_{\odot}} \right)^{0.22}.
\end{equation}

%Typically, the variations in the hydrogen fractions are minuscule \citep{McGaugh_2020}.
As we do not have a direct measurement of molecular gas for our sample of galaxies, we also use the stellar mass-molecular gas relation found by \cite{McGaugh_2020}, who have shown that the molecular gas mass is approximately 7 percent of the stellar mass of a galaxy. In their work, they combined two known scaling relations between the molecular gas mass and star formation rate (see Equation (1) in \citealt{McGaugh2015}) and between the stellar mass and star formation rate \citep{McGaugh2017}, such that the molecular gas mass is given by:
\begin{equation}
\log_{10} ({M_{\rm H_{2}}}) = \log_{10}(M_\star) - 1.16. 
\end{equation} 

The molecular gas mainly contributes to the baryonic radial acceleration in the central regions of galaxies \citep{Young_1991, Leroy_2008, Saintonge_2016, Saintonge_2022}. 
As a result, not including it in the baryonic budget can lead to an underestimation of $g_\mathrm{{bar}}$ in the inner regions (higher acceleration). 
 \begin{figure}
    \centering
    \includegraphics[width = \linewidth]{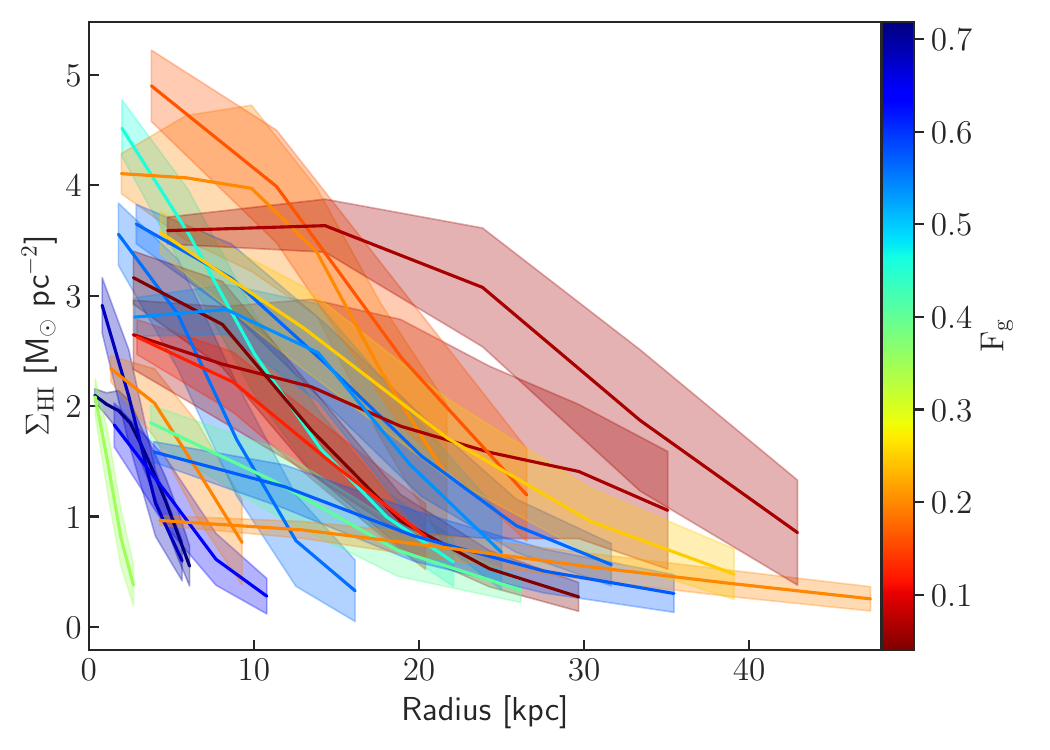}
    \caption{All HI radial surface densities derived using {\sc 3D Barolo}, colour coded by their gas fraction, $\mathrm{F_{g} = M_{HI}/M_{bar}}$.}
    \label{fig:RAR-HI-surf-dens}%redoing this with gas fractions, same for stellar surface densities
\end{figure}

We apply these correction factors to the face-on H{\sc i} surface densities, which are presented in Fig \ref{fig:RAR-HI-surf-dens}, to determine the total gas surface density as a function of galaxy radius.

\subsection{Baryonic circular velocities}
\label{circular-velocities}

We use the stellar and gas surface mass density profiles and numerically solve Poisson's equation to compute the circular velocity contribution of each baryonic component to the total rotation curves,
%\begin{equation} \nabla^2 \Phi = 4 \pi G \rho.\label{eq:Poisson} \end{equation}
where the total baryonic circular velocity is given by the sum in quadrature of the individual contributions from the gas and stars.

%\begin{equation} v_{\rm bar}^2 = v_{\rm stars}^2 + v_{\rm gas}^2. \end{equation}
%}\footnote{https://gitlab.com/iogiul/galpynamics/} 
 We use the {\sc Galpynamics} software \citep{Iorio2018}, which numerically solves Poisson’s equation under the assumption of vertical hydrostatic equilibrium \citep{Cuddeford1993OnTP}. The software takes as input the mass surface density profiles and returns the circular velocity at each radius. 

For the stellar component, we use the surface mass densities derived from SED fitting and model them with a fourth-degree poly-exponential function, following \citealt{Bacchini_2019} and \citealt{Mancera_Pina_2022_flaring}:
\begin{equation} \Sigma_{\star}(R) = \Sigma_{0} \exp\left(-\frac{R}{R_d}\right) \left(c_1 R + c_2 R^{2} + c_3 R^{3} + c_4 R^{4} \right), 
\end{equation} where $\Sigma_{0}$ is the central surface mass density, $R_d$ is the disc scale length, and $c_1, c_2, c_3, c_4$ are the coefficients of the polynomial.
This functional form is flexible enough to capture the rich structure of stellar profiles, including any dips and peaks, and any deviations from a simple exponential.
In addition to the radial distribution, the vertical mass distribution of the stellar disc must also be specified. Several functional forms have been used in the literature, such as $\mathrm{sech}^2$, $\mathrm{sech}$, and exponential profiles \citep{van_der_Kruit_1988, Bershady_2010}. We test both a $\mathrm{sech}^2$ and an exponential vertical profile. The $\mathrm{sech}^2$ profile, often used in galactic disc models, arises from the assumption of isothermal vertical equilibrium and has a scale height $h_z$ related to the disc scale length by $h_z = 0.1 R_d$ \citep{van_der_Kruit_2011}. For the exponential case, we adopt a constant scale thickness of $z_d = 0.196 \: R_{d}^{0.633}$, following the empirical relation from \cite{Bershady_2010} based on photometry of edge-on galaxies. 
We find that the choice of vertical stellar distribution has a minimal impact on the resulting RAR. Therefore, for simplicity, we adopt an exponential stellar vertical profile with a fixed scale height in all subsequent analysis and present our results based on this configuration.
%Observation in the near-infrared have indicated that the exponential form might likely be the the best functional form Aoki et al 1991

\begin{figure}
    \centering
    \includegraphics[width=\linewidth]{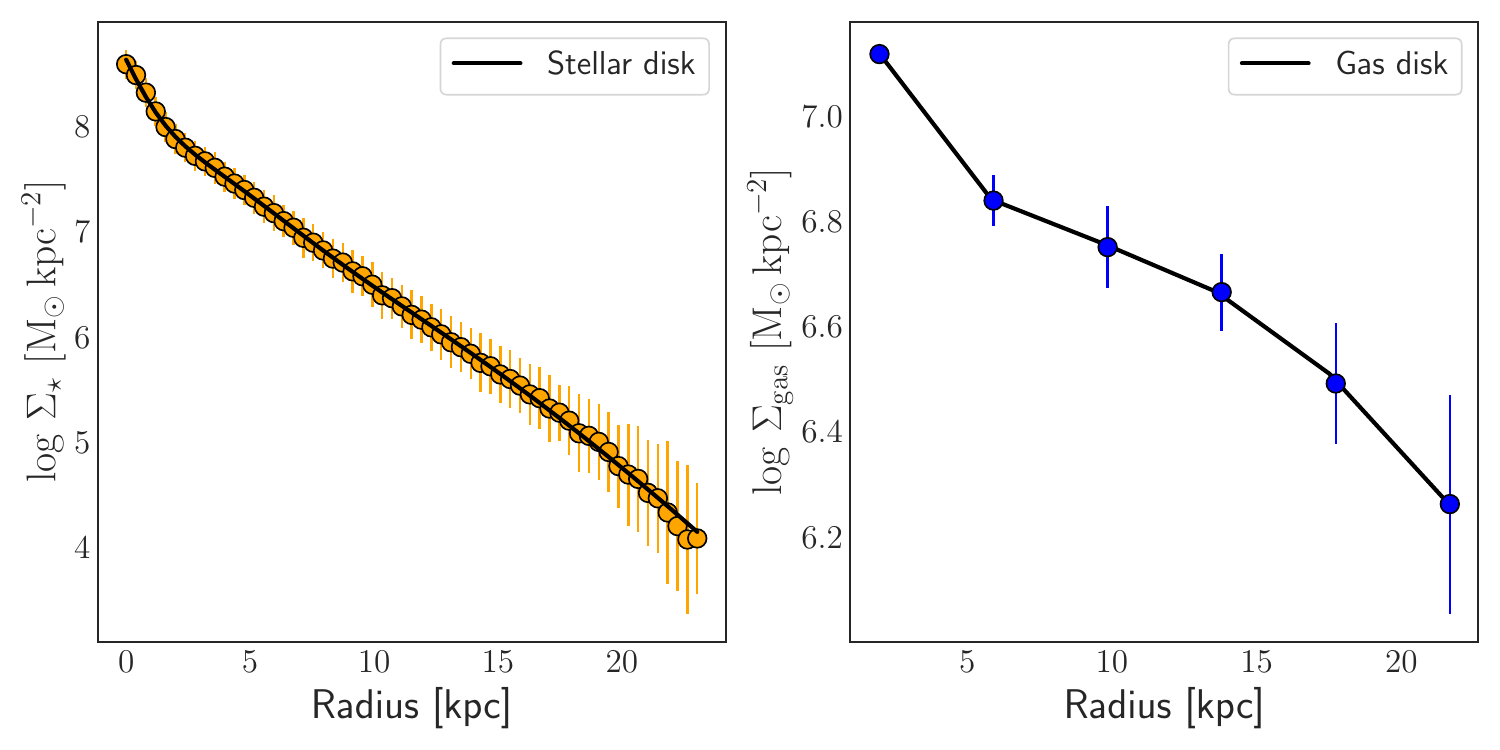}
    \caption{Radial stellar and gas surface density profiles for a representative high mass galaxy. The observed data points are shown alongside the solid lines, which represent poly-exponential fits derived using {\sc Galpynamics}. Error bars smaller than the size of the markers are not displayed.}
    \label{fig:baryonic-surf-dens-fit}
\end{figure}

For the gas component, we assume a razor-thin disc in the modeling. This assumption is justified given that all galaxies in our sample are marginally resolved in H{\sc i}, and that, in the mass regime of our galaxies, differences between thin and realistic flared gas discs have a negligible effect on the gravitational potential (see \citealt{Mancera_Pina_2022_flaring}). The radial surface-density profiles are modeled either with a simple exponential or, for more complex cases, a third-order poly-exponential function. %For the gas disc, given that all galaxies in our sample are marginally resolved, we use a poly-exponential thin disc model. For galaxies with simpler exponential profiles, we apply an exponential disc model, while for more complex gas surface-density profiles, we use a third-order poly-exponential function. %The values are corrected for metals and molecular gas as discussed above. 
These functional forms for both stars and gas, effectively model the variety of observed radial surface mass density profiles of the galaxies in our sample. An example of poly-exponential fits to the surface density profiles of stars and gas is shown in Fig \ref{fig:baryonic-surf-dens-fit} for a representative galaxy in our sample. 

To estimate the uncertainties on the circular velocities, we resample the radial stellar and gas surface densities used in {\sc Galpynamics} based on their associated errors. Assuming a Gaussian error distribution, we generate 100 realizations of the stellar and gas disc models by drawing new values for the surface densities ($\Sigma_\star$, $\Sigma_{gas}$) and scale lengths ($R_d$) from distributions centered on their best-fit values with standard deviations set by their respective uncertainties. For each realization, {\sc Galpynamics} numerically solves for the circular velocity of the stellar and gas components. The final uncertainties on the star and gas circular velocities are then derived from the standard deviation of the resulting velocity distributions for all resampled models.
Note that we do not correct for pressure support as this is negligible for the high rotational velocities of our sample \citep{Iorio_2017, Pavel_2021}. As such, the rotation velocity of the gas is a direct tracer of the gravitational potential.
%as the random gas motions for our sample of galaxies are negligible  

\begin{figure}
    \centering
    \begin{subfigure}[b]{0.48\linewidth}
        \centering
        \includegraphics[width=\linewidth]{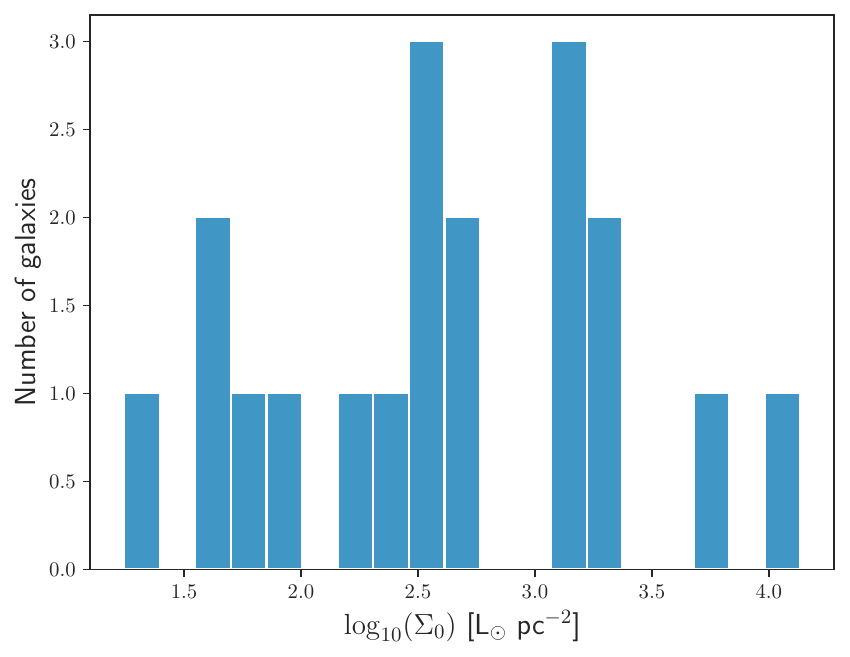}
    \end{subfigure}
    \hfill
     \begin{subfigure}[b]{0.48\linewidth}
        \centering
        \includegraphics[width=\linewidth]{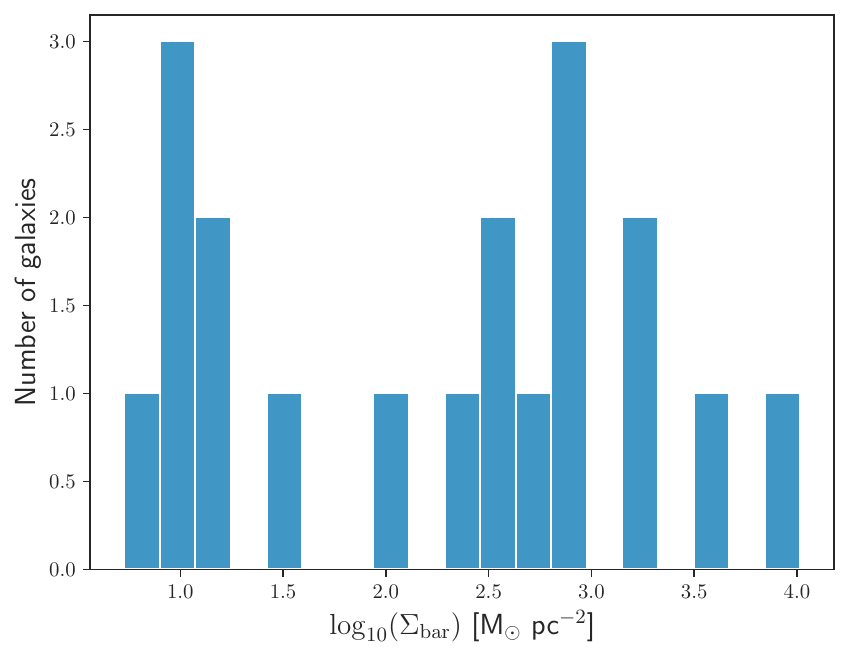}
    \end{subfigure}
   \caption{The central surface brightness $\Sigma_{0}$ and equivalent surface density distribution ($\Sigma_{\rm bar} = \frac{3}{4} \frac{M_{\rm bar}}{R_{\rm bar}^{2}}$, where $M_{\rm bar}$ is the baryonic mass (stars and gas), and $R_{\rm bar}$ is the radius where the baryonic circular velocity $\mathrm{v_{bar}}$ is maximum) histograms for our sample of 19 H{\sc i} galaxies.}
\label{fig:overall_properties_histograms}
\end{figure}

\begin{figure}
    \centering
    \includegraphics[width=0.5\textwidth]{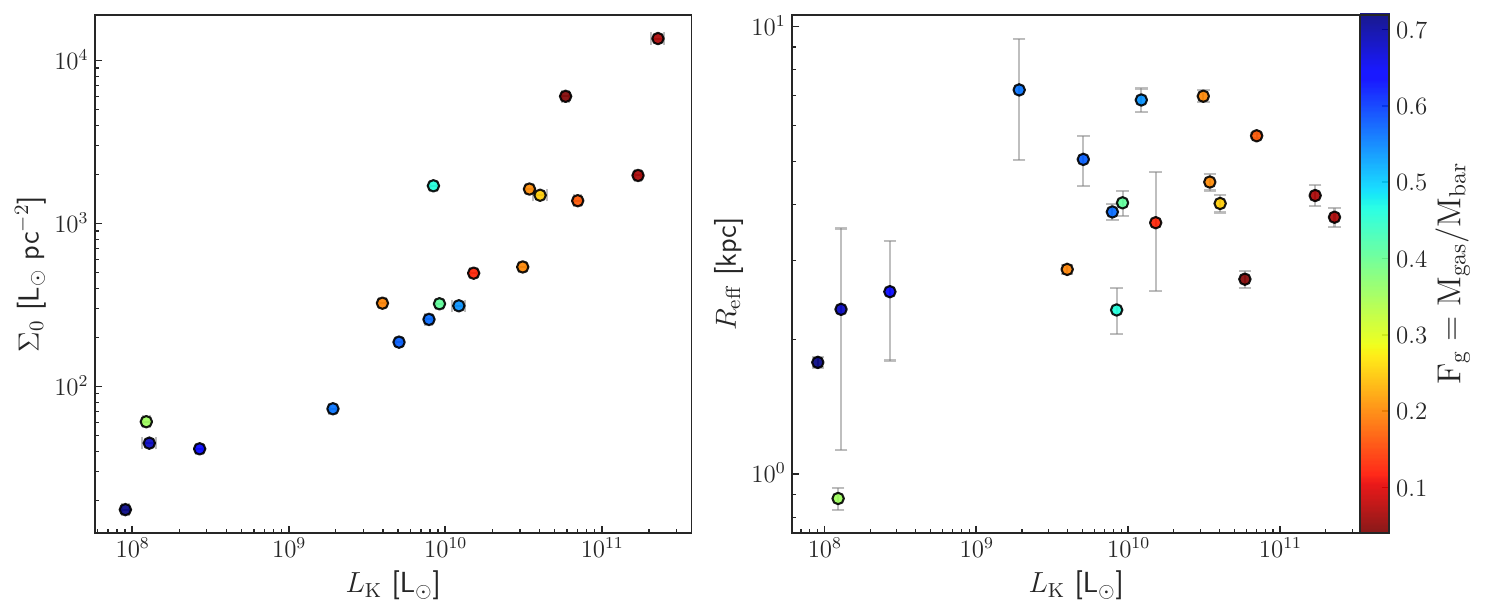}
    \caption{Luminosity in K band versus central surface brightness (left) and effective radius (right) for our sample of galaxies. Error bars smaller than the size of the markers are not displayed.}
    \label{fig:Lk_vs_SB0_Reff}
\end{figure}

\begin{figure}
    \centering
    \includegraphics[width=0.5\textwidth]{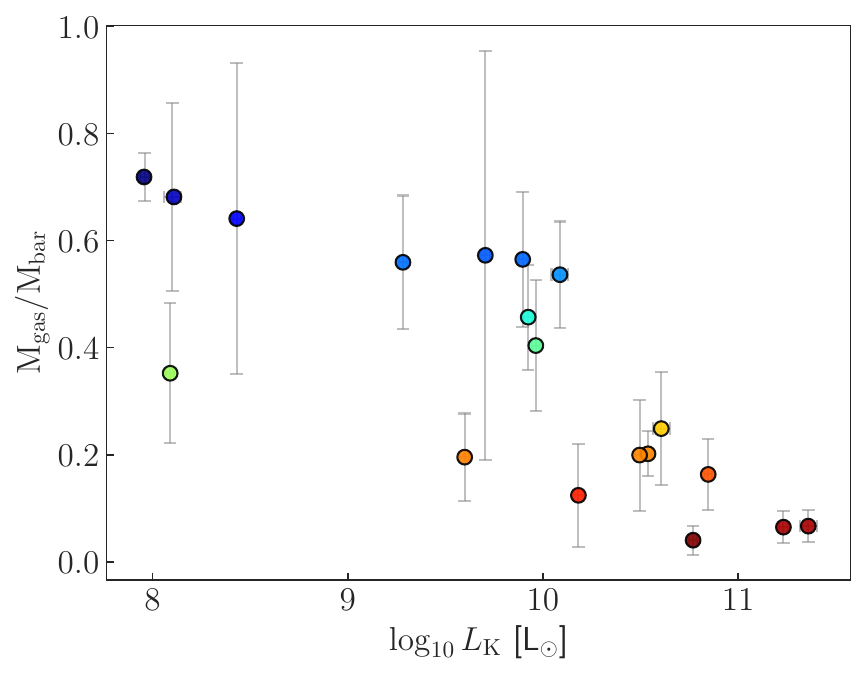}
    \caption{The gas fraction ${F_{g} = M_{\rm gas}/M_{\rm bar}}$, as a function of total $K$-band luminosity for our sample of galaxies. The gas fraction anticorrelates with L$_{K}$. Error bars smaller than the size of the markers are not displayed.}
    \label{fig:gas_fraction_vs_luminosity}
\end{figure}

The overall properties of our galaxies are shown in the histograms in Figure~\ref{fig:overall_properties_histograms}.
Figures ~\ref{fig:Lk_vs_SB0_Reff} and ~\ref{fig:gas_fraction_vs_luminosity} show the central surface brightness, effective radius, and gas fraction as a function of total K band luminosity for our sample.

% Our sample spans a wide range of luminosities, from (~10$^{8}$ to ~10$^{11.5}$ L$_\odot$), effective radii and gas content, as shown in Figure ~\ref{fig:Lk_vs_SB0_Reff}.

\section{Results}
\label{results}
The contributions from stars and gas are added together, and the derivative of the potential gives the required baryonic acceleration:

\begin{equation}
   g_\mathrm{bar}(r) = -\frac{\partial \Phi}{\partial r} = \frac{v_\mathrm{bar}^{2}(r)}{r},
   \label{eq:g_bar_eq}
\end{equation}
where $g_{\rm bar}$ is the acceleration due to the baryonic mass, $v_{\rm bar}$ is the circular velocity resulting from the baryonic mass, both determined at radius $r$. 

The total centripetal acceleration, derived from rotation curves, is given by: 
\begin{equation}
    g_\mathrm{obs}(r) = \frac{v_\mathrm{rot}^2(r)}{r},
    \label{eq:g_obs_eq}
\end{equation}
where $v_{\rm rot}(r)$ is the velocity measured from the rotation curve at radius $r$.

\begin{figure*}
    \centering
    \includegraphics[width=\linewidth]{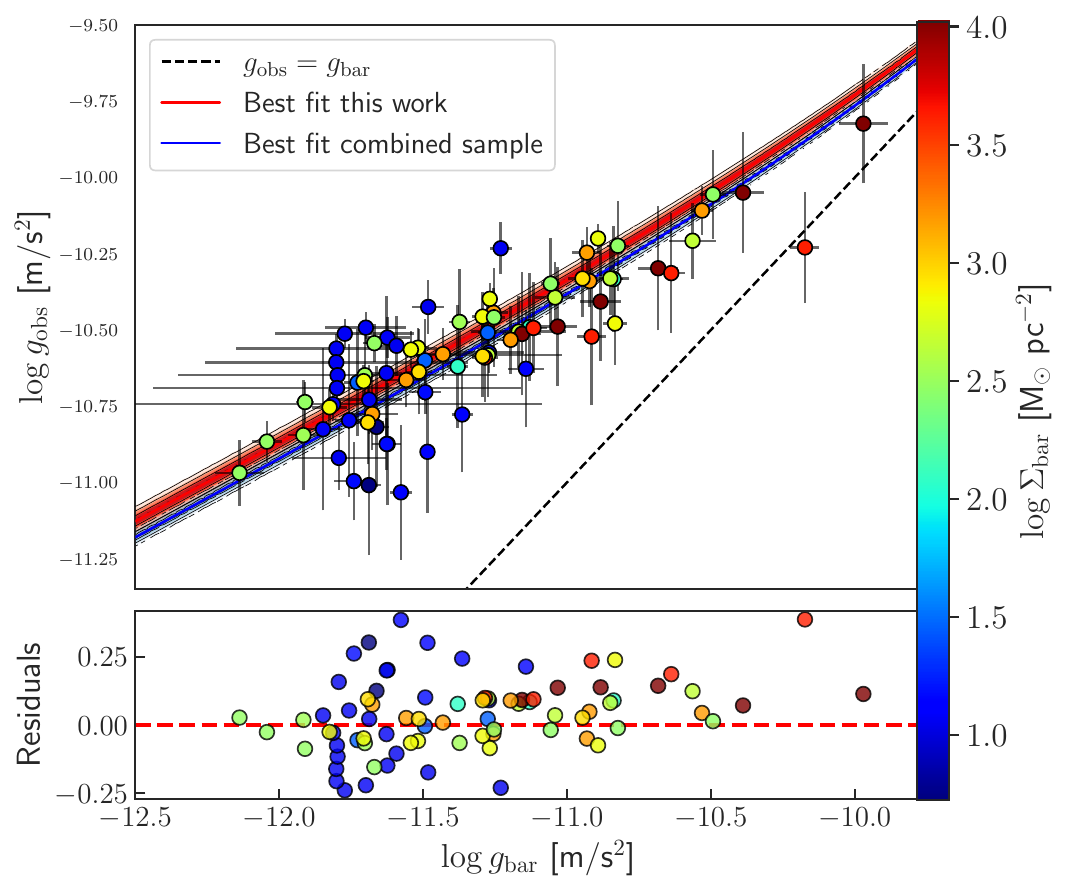}
    \caption{{\it Top}: The RAR for our sample of 19 late type galaxies, colour coded by their equivalent baryonic surface density. 
    The red shaded regions represent the full posterior predictive distribution from the MOND inspired fit to our sample alone, shown at 1, 2 and 3 $\sigma$ confidence intervals, with the solid red line shown as the best fit. Similarly, the blue shaded regions and blue solid line represent the full posterior predictive distribution and best fit for the combined dataset (our sample + high acceleration portion of the SPARC sample), not shown here.
    {\it Bottom}: The residuals (model - data) around the best-fit model to our sample alone.}
    \label{fig:RAR}
\end{figure*}

The baryonic and dynamical acceleration components allow us to analyze the radial acceleration relation at all radii. 
Due to the angular resolution of our H{\sc i} data, we discard the innermost points, which correspond to less than 5 arcseconds in radius. The radial acceleration relation for our sample, derived using the varying mass-to-light ratio, is shown in Figure~\ref{fig:RAR}, colour coded by the equivalent baryonic surface density. This quantity was introduced by \cite{McGaugh_2005} and is given by $\Sigma_{\rm bar} = \frac{3}{4} \frac{M_{\rm bar}}{R_{\rm bar}^{2}}$, where $M_{\rm bar}$ is the baryonic mass (stars and gas), and $R_{\rm bar}$ is the radius where the baryonic circular velocity $\mathrm{v_{bar}}$ is maximum.

\subsection{Fitting the RAR}
\label{rar-fits}
We model the RAR using two functional forms: the MOND-inspired relation from \cite{McGaughLelli2016} (``RAR'' or ``McGaugh--Lelli--Schombert'' interpolating function) and a general double power law proposed by \cite{Lelli2017}. These were both fit using the Python package {\sc Roxy} \citep{roxy}. This implements the  ``Marginalised Normal Regression (MNR)'' method, which fits a function to data accounting for uncertainties in both $x$ and $y$, intrinsic scatter in the relation and unknown ``true'' $x$ values through the use of a Gaussian hyperprior with inferred mean $\mu_\text{gauss}$ and standard deviation $w_\text{gauss}$. We apply MNR to the accelerations in base 10 logarathmic space, so that both $\mu_\text{gauss}$ and $w_\text{gauss}$ are expressed in dex. Extensive mock tests showed that MNR, unlike most others methods employed in the literature, is unbiased~\citep{roxy}. The likelihood is sampled using the No U-Turn Sampler (NUTS) method of Hamiltonian Monte Carlo. 

The MOND-inspired functional form \citep{McGaughLelli2016} is described by the following equation:
\begin{equation}
    \mathrm{g_{obs}} = \mathrm{F (g_{bar})} = \mathrm{\frac{g_{bar}}{1 - e^{-\sqrt{g_{bar}/a_{0}}}}},
    \label{eq:RAR-eq-mond}
\end{equation} where $a_{0}$ represents the acceleration scale.

We adopted uniform priors for $\log_{10}(a_0)$ between $-15$ and $5$, and for the intrinsic scatter between $0$ and $3$ dex. %$[0, 3]$
Running {\sc Roxy} with 700 warm-up steps and 5000 samples (ample to ensure convergence), we recovered a best-fit value of the acceleration scale $a_0 = (1.69 \pm 0.13) \times 10^{-10}\ \mathrm{ms^{-2}}$ and an intrinsic scatter of $0.045 \pm 0.022$ dex. This is significantly smaller than the 0.12 dex total scatter reported by \cite{Lelli2017} for SPARC galaxies, and consistent with more recent estimates of the intrinsic scatter \citep{Li_2018, Chae_2021, Chae_2022,  Desmond_2023}, although we note the small dynamic range in $g_{\rm bar}$ for our sample.

To overcome some of the problems associated with the relatively small dynamic range, we also perform a joint fit using our sample combined with the SPARC RAR data at high accelerations (above our highest value for $\log_{10} g_{\mathrm{bar}}$) 
for which we impose a quality cut to include only galaxies with inclinations greater than 30 degrees. The high-acceleration SPARC data are less likely to have a significant departure from their assumed mass-to-light ratio of $0.5$ at 3.6\,$\mu$m, as the high-acceleration points are largely derived from the inner parts of galaxies where the stellar populations are generally older and more homogeneous compared to the star-forming discs that dominate the low-acceleration part of the RAR.
Running {\sc Roxy} on this combined dataset yields a best-fit acceleration scale of $a_0 = (1.32 \pm 0.13) \times 10^{-10}\ \mathrm{ms^{-2}}$ and an intrinsic scatter of $0.064 \pm 0.007$ dex. 
This decrease in $a_0$ is expected, as the combined sample spans a wider range of baryonic accelerations than our dataset alone. 
In particular, the SPARC galaxies extend further into the high-acceleration regime, where the RAR follows the 1:1 line. By contrast, our galaxies are limited to $g_{\rm bar} \lesssim 10^{-10}~\mathrm{ms^{-2}}$ at the high acceleration end, primarily populating the lower acceleration regions, where the data points begin to peel away from the 1 to 1 line. The primary reason for this is our selection bias toward low-mass, gas-rich galaxies.
%due to our selection of low-mass, gas-rich systems at higher redshift.

In Figure~\ref{fig:RAR} we present the posterior predictive RAR relations with 1, 2, and 3$\sigma$ confidence intervals from our fitted model, shown as red shaded regions around the best fit relation. For comparison, we also overlay the best-fit RAR from the combined dataset in blue with the full posterior predictive.
%which includes both our sample and the SPARC data at high accelerations. 
Notably, the combined best-fit line does not align with the center of the posterior of the fit to our data alone, but instead lies outside the 2$\sigma$ confidence region. This suggests that our data alone favors a different acceleration scale than the combined sample, which is dominated by the high-mass SPARC galaxies. %80 compared to 670 in SPARC high acceleration

The deviation of our dataset from the combined best-fit RAR trend could stem from selection biases, as our sample is dominated by low-mass, gas-rich galaxies with lower baryonic accelerations. Furthermore, differences in the assumed mass-to-light ratios used to derive $g_{\rm bar}$ are likely to also play a significant role at these low accelerations, where the mass-to-light ratios in the outer regions of galaxies tend to be lower than in the central regions (Figure~\ref{fig:ml-ratio-variations}).
However, the very low intrinsic scatter in our RAR supports the conclusions of \cite{Lelli2017, Desmond_2023, Stiskalek_2023}, that the RAR is a fundamental relation.
%\%begin{equation}
%\log g_{\mathrm{obs}} = \log \hat{g}_{y} + (\alpha - \beta) \log \left(1 + \frac{g_{\mathrm{bar}}}{\hat{g}_{x}} \right) + \beta \log \left( \frac{g_{\mathrm{bar}}}{\hat{g}_{x}} \right)
%\end{equation}
%explored an

\begin{figure*}
        \centering
        \includegraphics[width=\linewidth]{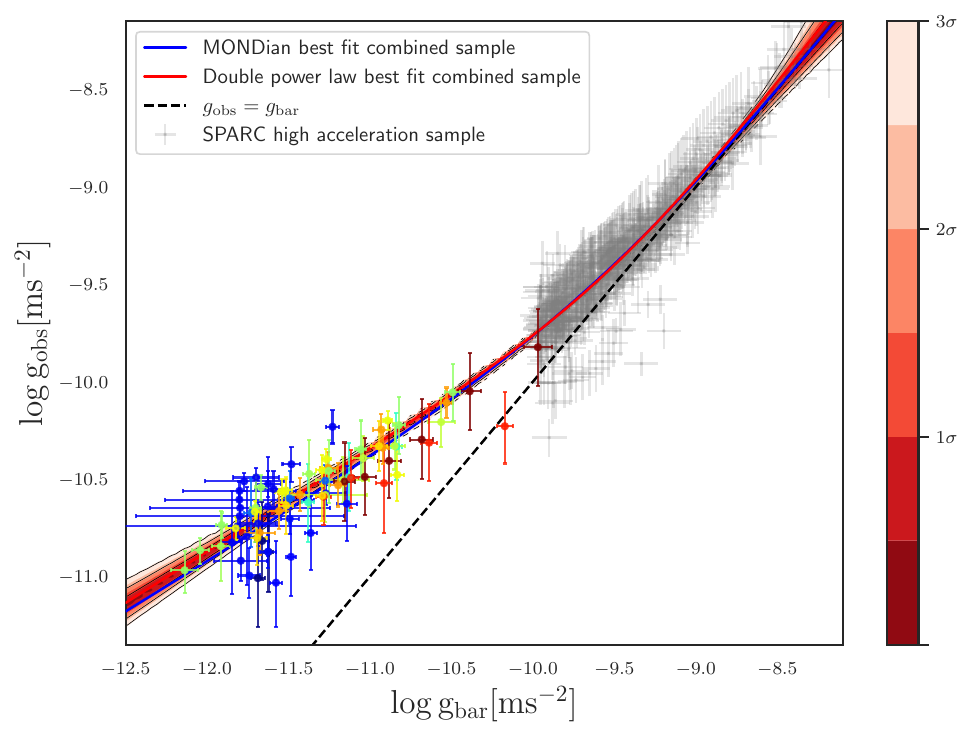}
    \caption{The posterior predictive plot for the combined sample (this work colour-coded by equivalent baryonic surface density + SPARC data in faded grey) at 1, 2 and 3 $\sigma$ confidence intervals from the double power law fit, with the MONDian best fit line in blue. Note that only the SPARC points with greater $g_\text{bar}$ than our entire sample are included in our combined sample and shown here.}\label{fig:RAR_dblplaw_combined_posterior_predictive}
\end{figure*}

The alternative functional form for the RAR: a general double power-law model, as proposed by \cite{Lelli2017}, is given by:
%\begin{equation}
%g_{\mathrm{obs}} = \hat{g}_{obs} \left(1 + \frac{g_{\mathrm{bar}}} {\hat{g}_{bar}} \right)^{ (\alpha - \beta)} \left( \frac{g_{\mathrm{bar}}}{\hat{g}_{bar}} \right)^ {\beta}, 
%\end{equation}

\begin{equation}
y = \hat{y}\left(1 + \frac{x} {\hat{x}} \right)^{ (\alpha - \beta)} \left( \frac{{x}}{\hat{x}} \right)^ {\beta}, 
\end{equation}
where $\alpha$ and $\beta$ are the high and low acceleration slopes for $\mathrm{x} \gg \mathrm{\hat{x}}$ and for $\mathrm{x} \ll \mathrm{\hat{x}}$, respectively.
 In the case of our dataset alone, the limited sample size results in poor constraints on several parameters,  particularly the high-acceleration slope  $\alpha$ and $\mathrm{\log \hat{x}}$.
 %Nevertheless, fitting the DPL allows us to compare its relative performance against the simpler,
 %Due to the limited sample size of our data and the larger number of parameters to constrain ($\mathrm{g_{bar}}$, $\mathrm{g_{obs}}$, and $\beta$), some of the parameters are not well constrained. 
 To quantitatively assess the quality of both models, we computed the Bayesian Information Criterion (BIC): 
 \begin{equation} {\rm BIC }= -2 \ln \hat{L} + k \ln (n), 
 \end{equation} where $\ln \hat{L}$ is the log-likelihood of the best-fit model, $k$ is the number of free parameters, and $n$ is the number of data points. For our sample, the MOND-inspired fit is clearly favored with a $\Delta \mathrm{BIC} = 10.6$, owing to its simpler form and fewer parameters to fit (1 compared to 4).
 %smaller parameter space.
However, when we repeat the analysis on the combined dataset (our sample + the high acceleration SPARC galaxies), the situation changes: the larger dynamic range in accelerations probed and improved statistics enable better parameter constraints, and the double power-law fit is preferred over the MOND-inspired model, with a $\Delta \mathrm{BIC} = 11.5$. 
This suggest that the more flexible double power-law form better accommodates the full diversity of baryonic accelerations. % Interestingly, for both our sample and the combined dataset, the slope of the double power-law fit in the low acceleration regime ($\beta$) is consistent with MOND predictions; however, this does not rule out consistency with $\Lambda CDM$. 
The posterior predictive RAR relation for this case is shown in Figure~\ref{fig:RAR_dblplaw_combined_posterior_predictive}, and the best-fit parameters are displayed in Table~\ref{best-fit-params-roxy-dblplaw}.
%2776 data points compared to (-11.35, -9.5)
%with the best fit line for the double power law in green - I changed it back to red
\begin{table*}
    \centering
    \caption{Best-fit parameters for the double power-law functional form of the RAR, derived using {\sc Roxy} and including the combined sample (with SPARC high acceleration data points).}
    \begin{tabular}{lcccccc} 
        \hline
         & $\hat{x}$ [$10^{-10}$ m s$^{-2}$]  & $\hat{y}$ [$10^{-10}$ m s$^{-2}$]  & $\alpha$ & $\beta$ & $\sigma_\text{int}$ [dex] & Sample size \\
        \hline
         
         This work  & $15.6 \pm 18.5 $ & $4.5\pm  2.8$ & $2.13 \pm 1.42$ & $0.46 \pm 0.04 $ & $0.046 \pm 0.022 $ & 80 \\
        
        Combined sample & $3.5 \pm 2.3$ & $2.9 \pm 1.3$ & $1.10 \pm 0.16$ & $0.52 \pm 0.03$ & $0.062 \pm 0.007 $  & 670 \\
        \hline 
    \end{tabular}
    %\end{adjustbox}
    \label{best-fit-params-roxy-dblplaw}
\end{table*}

\begin{table*}
    \centering
    \caption{Best-fit acceleration scale ($a_0$) and intrinsic scatter ($\sigma_{\rm int}$) of the RAR under different $\Upsilon_\star$ assumptions, obtained using {\sc Roxy}.}
    \begin{tabular}{l|cc|ccc}
        \hline
        Sample & \multicolumn{2}{c|}{Fixed $\delta$ = 1} & \multicolumn{2}{c}{Varying $\delta $} \\
        & \( a_{0} \) [$10^{-10}$ m s$^{-2}$] & $\sigma_{\rm int}$ [dex] & \( a_{0} \) [$10^{-10}$ m s$^{-2}$] & $\delta$ & $\sigma_{\rm int}$ [dex]  \\
        \hline
        
        Varying $\Upsilon_{\rm{K}}^{\rm{fiducial}}$ & $1.69 \pm 0.13$ & $0.045 \pm 0.022$ & $2.00 \pm 0.15$ & $3.94 \pm 1.4$ & 0.043 $\pm$ 0.021 \\
        Varying $\Upsilon_{\rm{K}}^{\rm{no \ mol}}$ & $2.06 \pm 0.15$ & $0.038 \pm 0.021$ & $2.38 \pm 0.17$ & $ 3.69\pm 1.27 $ & 0.037 $\pm$ 0.021 \\
        Fixed $\Upsilon_{\rm{K}} = 0.6$   & $1.08 \pm 0.09$ & $0.06 \pm 0.02$ & $1.48 \pm 0.10$ & $4.85 \pm 1.21$ & $0.032 \pm 0.020$ \\
        Radial average $\Upsilon_{\rm{K}}$ & $1.47 \pm 0.13$ & $0.09 \pm 0.02$ & $1.91 \pm 0.15$ & $4.22 \pm 1.23$ & $0.068 \pm 0.020$  \\
        SPARC RAR & $1.15 \pm 0.02$ & $0.082 \pm 0.003$ & $0.96 \pm 0.05$ & $0.84 \pm 0.03$  & $0.081 \pm 0.003$\\
        Low acceleration SPARC & $1.16 \pm 0.02$ & $0.087 \pm 0.003$ & 0.85 $\pm$ 0.06 &0.75 $\pm$ 0.03 & 0.084 $\pm$ 0.003  \\
        MIGHTEE + high acceleration SPARC & $1.32 \pm 0.13$ & $0.064 \pm 0.007$ & $1.76 \pm 0.15$ & $1.29 \pm 0.11$ & $0.061 \pm 0.007$\\
       % MIGHTEE $\Upsilon_{K} = 0.6$ + high acceleration SPARC & $1.15 \pm 0.02$ & $0.082 \pm 0.003$ & 1.27 $\pm$ 0.12 & 1.12 $\pm$ 0.09 & 0.049 $\pm$ 0.010 \\
        \hline
    \end{tabular}
    \label{tab:ml-assumptions-constraints-table}
\end{table*}

\subsection{Effect of the mass-to-light ratio}
\begin{figure*}
    \centering
    \includegraphics[width = \textwidth]{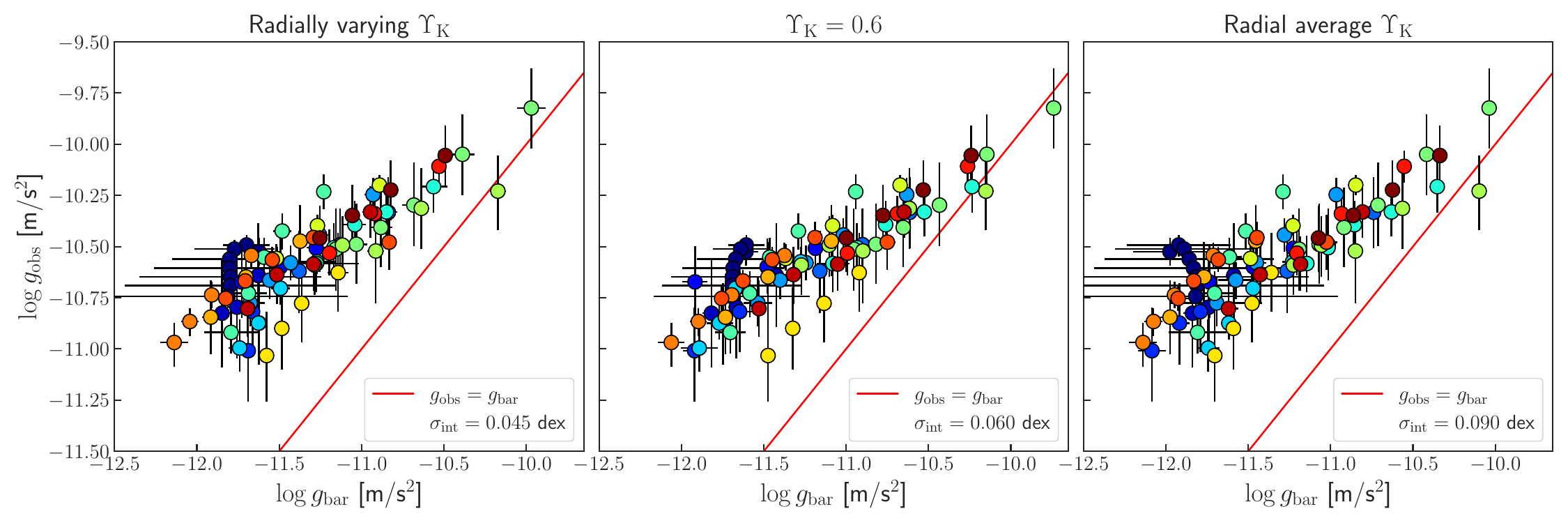}
    \caption{From left to right: the RAR obtained based on different assumptions of the mass-to-light ratio. First panel corresponds to the varying mass-to-light ratio case, the second to a constant $\Upsilon_{\star}$ ratio across galaxies and all radii and the third one to a different $\mathrm{\Upsilon_{K}}$ between galaxies from the average value obtained from the resolved SEDs (see Appendix \ref{appendix:resolved_SEDS} for the resolved SEDs across galaxies' radii). }
    \label{fig:RAR_comp_different_ML}
\end{figure*}

As noted previously, past studies of the RAR have adopted a single mass-to-light ratio for galaxies spanning a range in mass and morphologies.
In this section we therefore examine the impact of assuming a single, constant mass-to-light ratio for our sample, as well as a constant ratio for each individual galaxy within our sample (not including SPARC data in both cases). For the latter case, we adopt the average radial $K_s$-band mass-to-light ratio from the resolved SED fits.
We perform the same fits to the RAR using equation \ref{eq:RAR-eq-mond} for these two cases and provide the results in Table~\ref{tab:ml-assumptions-constraints-table}. The RARs obtained based on these different $\Upsilon_{\star}$ assumptions are presented in Fig~\ref{fig:RAR_comp_different_ML}. 

In both cases, we recover an overall trend similar to that of the scenario of radially varying mass-to-light ratios.
%Although assuming a single mass-to-light ratio value is an oversimplification, 
For the more massive galaxies where the average mass-to-light ratio approaches the standard value of 0.6 in the $K_s$-band (equivalent to $\approx 0.5$ at 3.6\,$\mu$m) in the central regions, this assumption has little effect, and the data points remain largely unchanged. However, at larger radii (low $g_{\rm bar}$), the points shift to higher accelerations. 
Under this simplification, the intrinsic scatter increases by 0.01\,dex and the acceleration scale changes considerably from 1.69 to 1.08, deviating by nearly 4$\sigma$ from our best-fit $a_0$ obtained in the radially varying scenario.  
 
Assuming a constant $\mathrm{\Upsilon_{K}}$ across galaxy radii but allowing it to vary between galaxies has a comparable impact on the RAR. This similarity arises because the average integrated mass-to-light ratio for our sample is 0.36. For high-mass galaxies, an increase in $\mathrm{\Upsilon_{K}}$ has a minimal effect at low accelerations, as the data points shift rightward but remain aligned with the overall trend. In the inner regions, assuming a constant $\mathrm{\Upsilon_{K}}$ does not significantly change the results since the assumed value is already close to the true $\mathrm{\Upsilon_{K}}$ in the center. For lower-mass galaxies, while the data points similarly shift to the right, the general RAR trend is still preserved, but the intrinsic scatter increases to 0.09 $\pm$ 0.02 dex. 

In addition, we find that assumptions about the molecular gas content also have a non-negligible impact on the inferred RAR parameters.
When molecular gas corrections are omitted, the intrinsic scatter does not decrease significantly  ($\sigma_{\rm int} = 0.038 \pm 0.021$\,dex); however, the best-fit acceleration scale changes by nearly $2\sigma$ to a value of $a_0 = 2.06 \pm 0.15$\,m~s$^{-2}$ from the value obtained when molecular gas is included. %\ty{I think it decreases insignificantly is the correct thing to describe here - else the first sentence of the next paragraph seems odd.}%care must be therfore taken with molecular gas, as its effect is pronounced particularly in the central regions of low-mass galaxies, which correspond to the low-acceleration regime of the RAR, where the relation begins to deviate from the 1 to 1 line

These results demonstrate that adopting a more sophisticated approach outperforms simpler assumptions, as it leads to significantly lower intrinsic scatter in the RAR. This underscores the importance of accounting for radial variations in mass-to-light ratios and the contribution from all components in the baryonic surface mass budget, particularly at the low-acceleration end, where such effects are most pronounced and predictions from MOND and $\Lambda$CDM are likely to be the most divergent. % care needs to be taken
% as central regions of low mass galaxies correspond to the outer regions of high mass systems, where the RAR is expected to deviate from the 1 to 1 line
%such effects can substantially affect the inferred relation. % 
Neglecting these variations can introduce biases in the inferred radial baryonic acceleration, ultimately affecting the shape and tightness of the relation.   
Given that the RAR is one of the tightest known dynamical scaling relations, its reproducibility serves as a stringent test for galaxy formation and evolution models, as they must be able to reproduce both its shape and remarkably small scatter (as well as its further ``fundamental'' features identified in~\citealt{Stiskalek_2023}). Accurately modeling $\Upsilon_\star$ is therefore essential for placing meaningful constraints on the dynamical mass distribution of disc galaxies and the physical processes that determine galaxy dynamics.

\subsection{Redshift evolution}

\begin{figure}
    \centering
    \includegraphics[width = 0.5\textwidth]{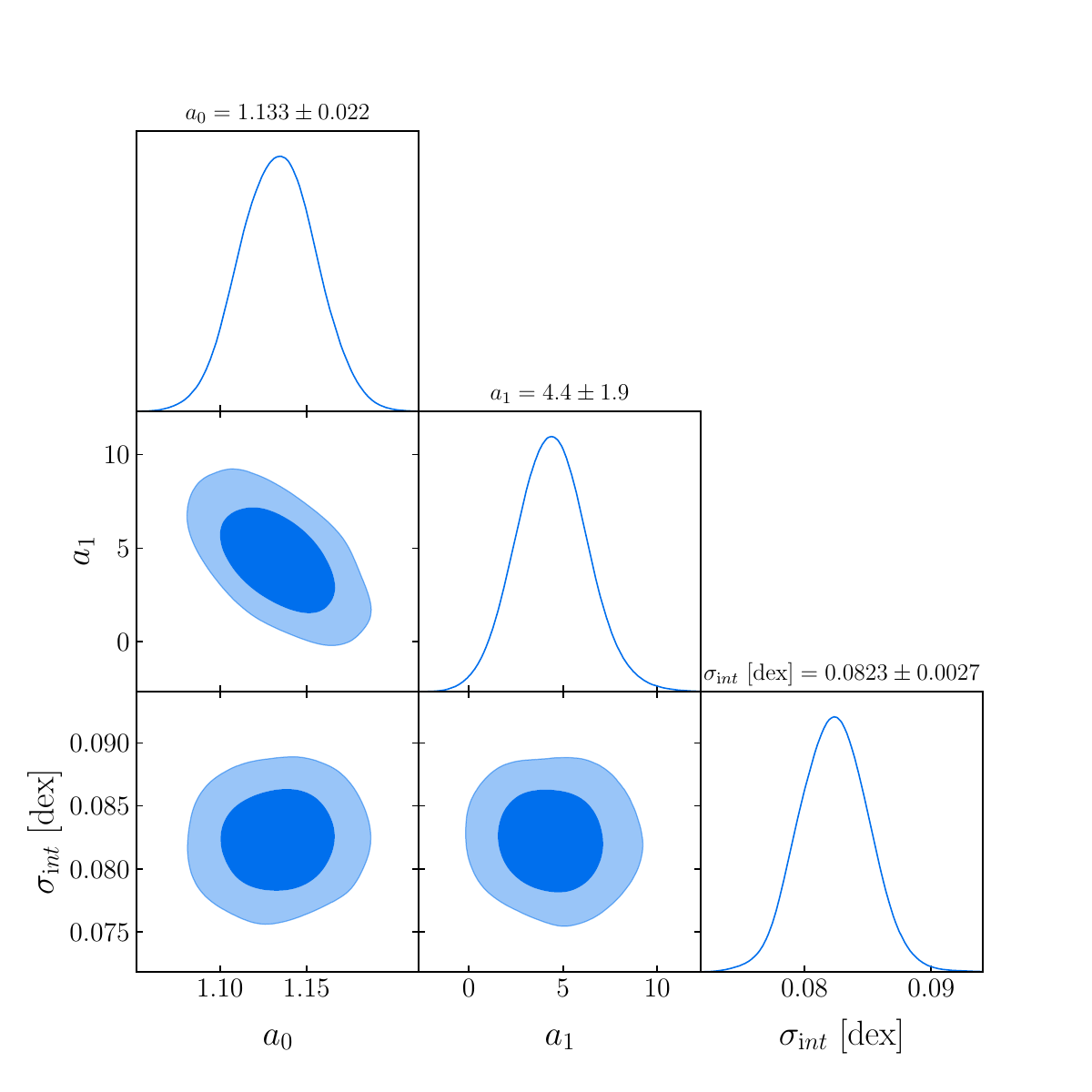}
    \caption{The corner plot showing the posterior distribution of the parameters from the redshift dependent fit to the RAR. The 2D posterior contours correspond to 68\% and 95\% confidence levels.} %
    \label{fig:RAR_corner_z}
\end{figure}

Our sample, although small, extends to significantly higher redshifts than previous samples used to investigate the RAR. In this section, we therefore investigate whether there is any significant evidence for evolution in the RAR with redshift. 
To investigate potential redshift evolution in the RAR, we extend our model for the combined (with SPARC) dataset by introducing a redshift-dependent acceleration scale of the form:
\begin{equation}
    a(z) = a_0 + a_1 \times z,
    \label{eq:a_z}
\end{equation}
where the $a_1$ term captures the evolution with redshift, $z$. We refit the RAR using this model in {\sc Roxy}, and show the resulting posterior distribution of the parameters in Figure~\ref{fig:RAR_corner_z}.

We find a tentative evidence for evolution with $a_1$ = 4.47 $\pm 1.88 \times 10^{-10}$ ms$^{-2}$, corresponding to a 2.4 $\sigma$ evidence that the acceleration scale ($a_0$) increases with redshift.

%To illustrate this redshift dependence, we show the redshift-dependent best fit for the combined sample in ~\ref{fig:RAR_fits_z}, with galaxies colour-coded by their redshift.
To explore whether this redshift evolution of the RAR could be linked to cosmological expansion (as suggested by~\citealt{Milgrom_2009}), we compare our redshift dependent acceleration scale $a_1$ to the redshift-dependence of the Hubble parameter. To do so we 
%We also investigate a potential connection between the acceleration scale $a_0$ and the Hubble constant $H_0$ by 
compute $H(z)$ assuming flat $\Lambda$CDM with $\Omega_\text{m}=0.3$ at the redshifts of our galaxies. Fitting a straight line to these points, we find d$H(z)$/d$z = 1.08 \times 10^{-18}$s$^{-1}$. To make this comparison independent of the proportionality constant, we compute the ratio $a_1 / a_0$ from our fit and compare it with (d$H(z)$/d$z)/H_0$, which is $\sim$ 0.47. We find that this value is consistent with our inferred ratio $a_1/a_0$ = $3.96 \pm 1.72$ at the 2$\sigma$ level, suggesting that our results are consistent with
$a_0(z) \propto H(z)$.
 
\begin{figure}
    \centering
    \includegraphics[width = 0.5\textwidth]{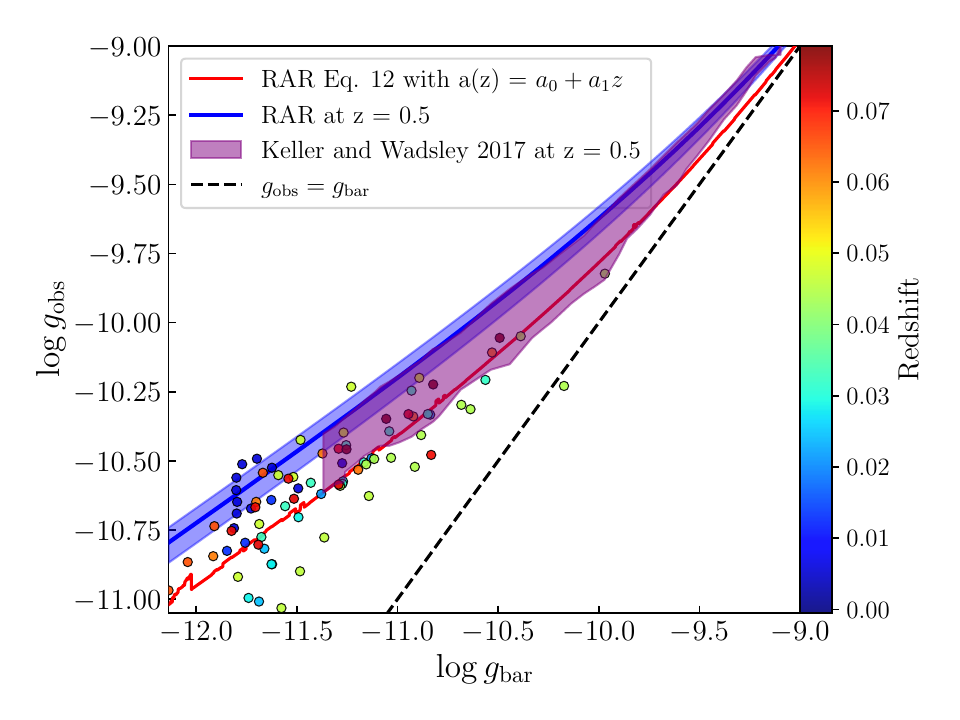}
  \caption{The extrapolated RAR at $z = 0.5$ for our sample with a varying mass-to-light ratio (colour-coded by their redshift) combined with SPARC (not shown), using a redshift dependent acceleration scale given by Eq.~\ref{eq:a_z}. The red line shows the best-fit $z=0$ RAR from the redshift dependent model, while the blue line corresponds to our extrapolated RAR at $z = 0.5$. The purple shaded region shows the result from the \citet{Wadsley2017} simulations at $z = 0.5$.}
    \label{fig:RAR_z_05}
\end{figure}

In Figure~\ref{fig:RAR_z_05}, we then compare our RAR at $z = 0.5$, as shown by the blue line, with the $\Lambda$CDM expectation from the hydrodynamical simulation of \cite{Wadsley2017}, denoted by the purple shaded region. 
We find that our RAR at $z = 0.5$ tracks the simulation prediction reasonably well across most of the $g_{\rm bar}$ acceleration range. We also performed a similar comparison at redshift $z = 1$ and find that our predicted RAR is again consistent within the uncertainties with the \cite{Wadsley2017} results. We caution however that this is an extreme extrapolation of a linear relation from our relatively low-redshift sample.
% Given the constraints on our $a_1$ however, an extrapolation at higher redshifts than this is not sensible.} %closely

Such evolution should also be evident in the baryonic Tully-Fisher relation (bTFr), due to its close relation to the RAR.
\cite{Ponomareva2021}, using a sample of 67 galaxies at $0<z<0.08$ from the MIGHTEE survey,  found no evidence for an evolution in the bTFr relation. 
\cite{Gogate2023}, using a sample of H{\sc i} selected galaxies in a cluster environment at $z\sim 0.2$, also found no evidence in the evolution in the normalisation of the bTFr. 
At first glance these both appear to be evidence against the evolution in the RAR that we find. However, we note that the analysis in \cite{Ponomareva2021} did not use the SPARC data to anchor the $z\sim 0$ relation, but just used the MIGHTEE data alone to investigate the evolution, resulting in much larger uncertainties which are formally consistent with our marginal evidence for evolution in the RAR. The lack of any evidence for evolution in the sample of \cite{Gogate2023} is also statistically consistent with our results as the redshift baseline is still relatively small, alongside the relatively large uncertainties on the bTFr measurement.
More recently, \cite{Jarvis2025} measured the bTFr to $z\sim 0.4$ using a sample of H{\sc i}-detected galaxies within the MIGHTEE survey. They find tentative evidence that the measured bTFr for the galaxies in their sample tend to have higher velocities for a given baryonic mass compared to the $z \sim 0$ relation. This would be in line with the form of the evolution that we find. %\hd{But $a_0$ goes into the normalisation of the BTFR in MOND, not its slope}. 
However, as explained in \cite{Jarvis2025}, such an offset in the bTFr could be accounted for with a slight overabundance of molecular gas in the total baryonic mass, which is not fully accounted for. Moreover, their sample size is small (11 objects) and is dominated by galaxies with high-stellar mass $M_\star >10^{9.8}$\,M$_{\odot}$ and high baryonic mass $M_{\rm bar} >10^{10.5}$\,M$_{\odot}$, thus it is difficult to compare consistently with our results using lower mass galaxies.

A thorough investigation of this redshift dependence will require a consistent analysis of the radially varying mass-to-light ratio for all galaxies, and preferentially for those galaxies to also be selected in a consistent way. 
With more data in the future (from e.g. MIGHTEE and Looking at the Distant Universe with the MeerKAT Array \citep[LADUMA; ][]{Blyth2016}, and eventually the Square Kilometre Array (SKA)), especially at higher redshift, this may provide a novel test of $\Lambda$CDM and MOND~\citep{Wadsley2017, zdep}. 
%We find good agreement between our relation and the simulation-based prediction across most of the $g_{bar}$ range, especially [-9.5, -11] with a slight 
%The blue line shows our best-fit RAR at this redshift, derived using Eq~\ref{eq:a_z}, while the purple shaded region corresponds to the 

%~\ref{fig:RAR_fits_z} shows that our results are consistent with the $\Lambda CDM$expectation from \cite{} at z = 0.5 and 1. 

%Our analysis provides the first observational constraint on $a_1$ parameter, serving as a preliminary step towards probing potential changes in the RAR across cosmic time with future deeper surveys.

\subsection{The shape of the MOND interpolating function}
So far we have fitted only a single MOND interpolating function (IF), the ``RAR'' or ``MLS'' IF of Eq.~\ref{eq:RAR-eq-mond}. We consider here three more general functional forms
% To explore more general forms, we consider here three more  functional forms of IFs,
given by ~\cite{Famaey_McGaugh_2012}: 
%the ``$\delta$-family'' of IFs, given by~\citep{delta_fam}:
\begin{subequations}
\begin{equation}
    g_{\mathrm{obs}} = g_{\mathrm{bar}} \left[ 1 - e^{ -\left( \frac{g_{\mathrm{bar}}}{a_0} \right)^{\delta/2} } \right]^{-1/\delta} 
    \label{eq:delta_family}
\end{equation}

\begin{equation}
g_{\mathrm{obs}} = g_{\mathrm{bar}} \left( \frac{1 + \sqrt{1 + 4 \left(\frac{g_{\mathrm{bar}}}{a_0}\right)^{-n}}}{2} \right)^{\frac{1}{n}} 
\label{eq:n_family}
\end{equation}

\begin{equation}
  {g_{\mathrm{obs}} = g_{\mathrm{bar}} \left[ \left(1 - e^{ -\left( \frac{g_{\mathrm{bar}}}{a_0} \right)^{\gamma/2} }\right)^{-1/\gamma} + \left(1 - \frac{1}{\gamma}\right) e^{ -\left( \frac{g_{\mathrm{bar}}}{a_0} \right)^{\gamma/2}} \right]}
    \label{eq:gamma_family}
\end{equation}

\label{eq:RAR_IF_shape}
\end{subequations}
 
These ``$\delta$-'', ``$n$-'' and ``$\gamma$-families'' contain an extra shape parameter ($n, \delta, \gamma$) that describes the sharpness of transition from the Newtonian (where $g_\mathrm{bar}$ $\gg$ a$_0$) to the deep-MOND (where $g_\mathrm{bar}$ $\ll$ a$_0$) regime. This reduces to the RAR IF at $\delta$ = $\gamma$ = 1. 

\begin{figure}
    \centering
    \includegraphics[width=0.5\textwidth]{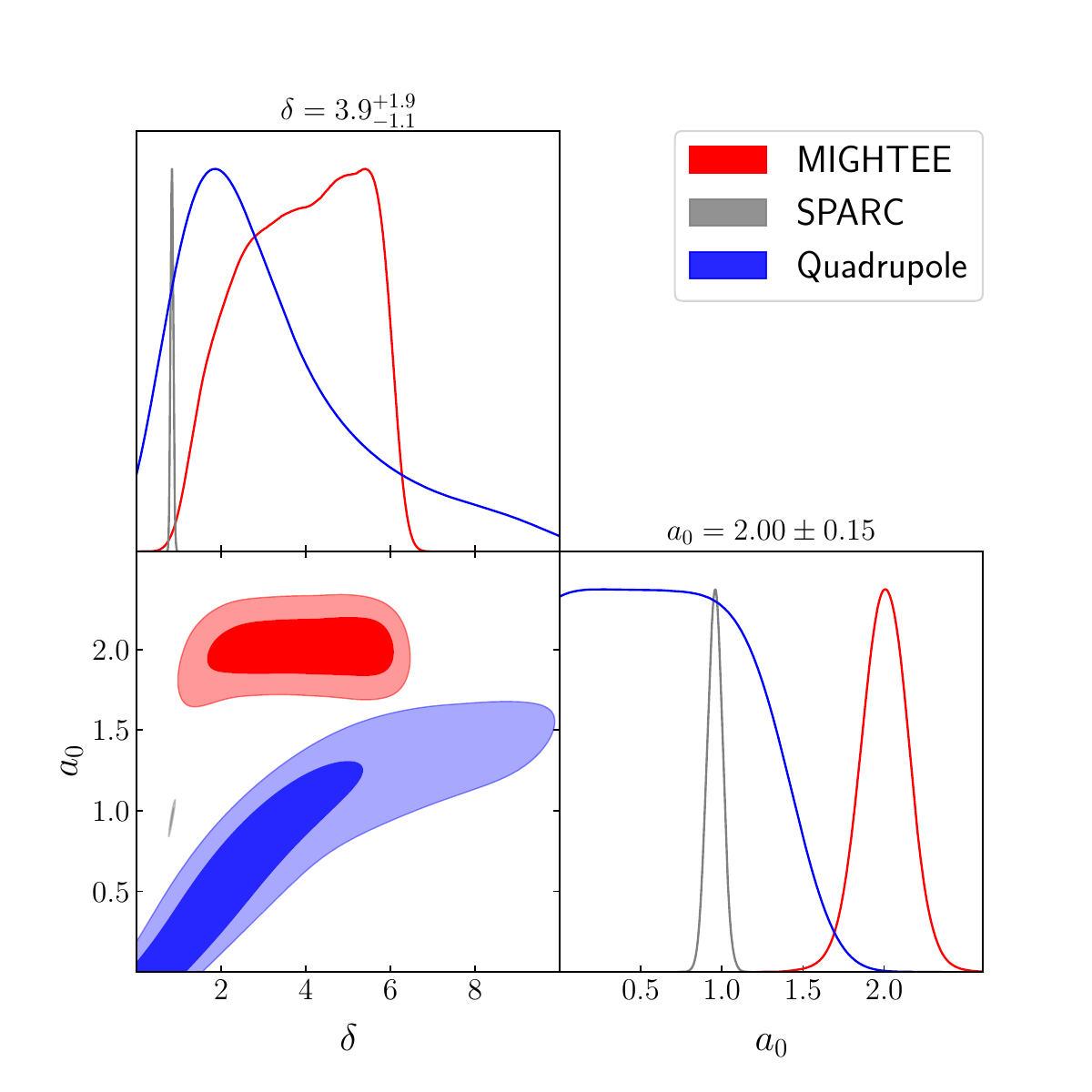}
    \caption{Posterior distributions for $a_0$ and $\delta$ from our sample and SPARC using the shape dependent IF described by  Eq.~\ref{eq:delta_family} and the Solar System quadrupole~\citep{Harry_SS_quadrupole}.}
    %\hd{Recommend cutting delta off at 10 (which is probably what I did in producing that chain) and adding some smoothing for the quadrupole posterior.}}}
    \label{fig:RAR_shape_constraints}
\end{figure}

%We show the more general 
%this more general function
First, we show the constraints using the $\delta$ family from our sample only, the full SPARC sample and the Solar System quadrupole (from \citealt{Harry_SS_quadrupole}) in Figure~\ref{fig:RAR_shape_constraints}. We adopt uniform priors for the acceleration scale between $0.001$ and 10, and for the shape $\delta$ between 0 and 30.  
%\hd{State the priors used}
Interestingly, our sample prefers a significantly sharper transition compared to SPARC, with a best-fit value for $\delta$ of 3.90 $\pm$ 1.39, and a higher acceleration scale with a slightly reduced intrinsic scatter, whereas SPARC alone prefers $\delta \approx 1$~\citep{Harry_SS_quadrupole}.

This is an important issue because the constraint from the SPARC RAR is in $\sim$9$\sigma$ tension with the inferred value of $a_0$ and $\delta$ from the quadrupole of the Solar System's gravitational potential, induced by the external field of the Milky Way and measured by the Cassini spacecraft's tracking of the ephemerides of Saturn~\citep{Harry_SS_quadrupole}. In contrast, the constraints from our sample alone are fully consistent with this measurement, and also with the null detection of a MONDian signal in the Wide Binary Test conducted with \textit{Gaia} data~\citep{Banik_WBT}. This is because $\delta \gtrsim 2-3$ makes the Solar System and Solar neighbourhood, at $\sim$1.8$\:a_0$, almost completely Newtonian. This might suggest systematics in the SPARC data resulting in an underestimate of $\delta$, which would put MOND on a much stronger footing by removing this critical source of inconsistency. In this regard it is interesting to note that excluding the SPARC galaxies with bulges results in a best-fit $\delta\approx2-2.5$, compatible with our sample (see table 2 of~\citealt{Harry_SS_quadrupole}).
% By itself, this result is completely consistent with the Cassini constraints and Wide Binary Test (WBT). 
%Our RAR contraints on the shape parameter are in excellent agreement with the Cassini ones, 

\subsection{Comparison with SPARC}\label{sec:sparc_comp}
%Fixing our sample to a constant mass-to-light ratio $\Upsilon_{\star} = 0.6$ reduces the acceleration scale and increases the shape parameter, with a reduced intrinsic scatter. 
%So far, we have discusses only the $\delta$ family,  The first one is given by the low acceleration SPARC data points, with  $\log_{10}(g_{\rm bar} /$ m~s$^{-2}) < -10$, matching the acceleration range probed by our galaxies. For the second one, we combined our constant mass-to-light ratio sample (with $\Upsilon_{\star}$ = 0.6) with the high acceleration portion of the SPARC dataset to create a joint sample. The results from these fits, including constraints on the acceleration scale and intrinsic scatter, are presented in Table~\ref{tab:ml-assumptions-constraints-table}.}
% within the acceleration range of our dataset
%incorporate

\begin{table*}
    \centering
    \begin{tabular}{lccccc}
    \hline
         Sample & MIGHTEE $\mathrm{\Upsilon_{\star}}$ & SPARC $\mathrm{\Upsilon_{\star}}$ & $a_0$ & Shape $\delta$ & $\sigma_\text{int}$ \\
        \hline
       MIGHTEE & Varying $\Upsilon_{\rm{K}}$& - & 2.00 $\pm$ 0.15 & 3.94$\pm$ 1.4 & 0.043 $\pm$ 0.021  \\
       
        MIGHTEE & $\Upsilon_{\rm{K}} = 0.6 $ & - & 1.48 $\pm$ 0.095 & 4.85 $\pm$ 1.21 & 0.032 $\pm$ 0.020 \\
        \hline
        
        SPARC & - & $\Upsilon_{\rm{\star, disc}}$ = 0.5, $\Upsilon_{\rm{\star, bulge}}$ = 0.7 & 0.96 $\pm$ 0.05 & 0.84 $\pm$ 0.03 & 0.081 $\pm$ 0.003 \\
        
         SPARC discs & - & $\Upsilon_{\rm{\star, disc}}$ = 0.27 & 0.22 $\pm$ 0.06 & 0.34 $\pm$ 0.02 & 0.102 $\pm$ 0.004 \\
        
        Low acceleration SPARC & - &$\Upsilon_{\rm{\star, disc}}$ = 0.5, $\Upsilon_{\rm{\star, bulge}}$ = 0.7 & 0.85 $\pm$ 0.06 &0.75 $\pm$ 0.03 & 0.084 $\pm$ 0.003 \\
        
    Low acceleration SPARC discs & - &$\Upsilon_{\rm{\star, disc}}$ = 0.27 & 0.20 $\pm$ 0.06 & 0.32 $\pm$ 0.02 & 0.103 $\pm$ 0.004 \\
   \hline
       MIGHTEE + high acceleration SPARC &  Varying $\Upsilon_{\rm{K}}$ & $\Upsilon_{\rm{\star, disc}}$ = 0.5, $\Upsilon_{\rm{\star, bulge}}$ = 0.7 &  1.76 $\pm$ 0.15 & 1.29 $\pm$ 0.11 & $0.061 \pm 0.006$\\
    
    MIGHTEE + high acceleration SPARC &  $\Upsilon_{\rm{K}} = 0.6$ & $\Upsilon_{\rm{\star, disc}}$ = 0.5, $\Upsilon_{\rm{\star, bulge}}$ = 0.7 & 1.27 $\pm$ 0.12 & 1.12 $\pm$ 0.09 & 0.049 $\pm$ 0.010 \\
    
    MIGHTEE + high acceleration SPARC discs &  Varying $\Upsilon_{\rm{K}}$ & $\Upsilon_{\rm{\star, disc}}$ = 0.27 & 0.81 $\pm$ 0.21 & 0.48 $\pm$ 0.05 & 0.028 $\pm$ 0.017 \\
         
    \hline
    \end{tabular}
    \caption{Constraints on the RAR parameters $a_0$ (acceleration scale), $\sigma_\text{int}$ (intrinsic scatter) and sharpness of transition (Shape) for the more general $\delta$-family of interpolating functions. We show various data combinations and mass-to-light models to investigate the cause of the differences between the constraints from the MIGHTEE and SPARC samples (see Sec.~\ref{sec:sparc_comp}).}
    \label{tab:delta-constraints-table}
\end{table*}

For completeness we next consider fitting all interpolation function (IF) families to different samples, aimed at checking consistency across different acceleration regimes and datasets.
In addition to the original samples, we also consider our MIGHTEE data with a constant mass-to-light ratio, $\Upsilon_{\rm{K}}$ = 0.6, as well as the low acceleration subset of SPARC, with  $\log_{10}(g_{\rm bar} /$ m~s$^{-2}) < -10$, which corresponds to the acceleration range probed by our galaxies. 
%to faciliate a more direct comparison
%we then select the pure discs in the SPARC dataset
%We also apply a reduced mass-to-light ratio for the SPARC pure disc galaxies, similar to the average value preferred by our sample (equivalent to $\Upsilon_{\star}$ = 0.27 in 3.6 $\mu$m) and
%repeat the analysis for all these different samples (low acceleration SPARC discs, all SPARC discs, and a combined sample from our varying mass-to-light ratio case with high acceleration SPARC discs data points).

To further refine the comparison between our sample and SPARC, we isolate pure disc galaxies within SPARC and apply a reduced mass-to-light ratio  $\Upsilon^{\rm{K}}_\star = 0.35$—matching the average value preferred by our own sample—equivalent to $\Upsilon^{3.6}_{\star} = 0.27$ in Spitzer 3.6 $\mu$m band. We then fit this adjusted SPARC discs subset and its corresponding low acceleration portion (low acceleration SPARC discs, as per Table~\ref{tab:delta-constraints-table}). %data points.
% to assess the 
Moreover, we repeat the analysis on two additional joint datasets: the first one--our MIGHTEE sample with a constant $\Upsilon_\star$ combined with the high acceleration portion of SPARC data, and the second one--our varying $\Upsilon_\star$ MIGHTEE sample with the high acceleration data points of the SPARC discs with reduced mass-to-light ratio. 

The results for the $\delta$ family for all these cases are shown in Table~\ref{tab:delta-constraints-table}, and the constraints from the $n$ and $\gamma$ families are presented in Table~\ref{tab:n-gamma-constraints-table}. 
These clearly show that our MIGHTEE sample alone prefers a higher acceleration scale, steeper shape parameter $\gtrsim 2$ and smaller intrinsic scatter than either the SPARC dataset (or any subset of it), or any of variant of the combined datasets. 
%In contrast, 
The SPARC data prefers $\delta\approx1$ in all cases, which is not consistent with the Solar System tests mentioned above in a MONDian interpretation. 
%Combining SPARC with our constant mass-to-light ratio sample however, leads to an increase in the shape parameter. 
%The constraints on the shape parameter from our constant mass-to-light ratio sample and the combined one in this case are consistent within 3$\sigma$ for all three IF families.
%as shown by the blue and red contours.%~\ref{fig:RAR_ML_06_delta_corner}, ~\ref{fig:RAR_ML_06_n_corner} and ~\ref{fig:RAR_ML_06_gamma_corner}.
Reducing the SPARC mass-to-light ratio to match our sample does not reconcile the discrepancy in the inferred parameters $(a_0, \delta, \sigma_{\rm int})$ between SPARC and our data—or with Solar System constraints. In fact, it leads to even lower inferred values for the acceleration scale $a_0$ and shape parameter $\delta$ and increases the intrinsic scatter to $\sim$ 0.1 dex. 
Likewise, combining MIGHTEE (regardless of the $\Upsilon_{\star}$ model)
%(with or without a varying $\Upsilon_{\star}$)
with the full SPARC sample or pure SPARC discs fails to alleviate the tension. 
% This suggests that the discrepancy likely arises from radial variations in the mass-to-light ratio—accounted for in our sample but absent in SPARC—and underscores the importance of modeling such variations, particularly in the low-acceleration regime.
Neither adjusting the global $\Upsilon_\star$ in SPARC nor restricting the analysis to its low-acceleration regime brings the SPARC-derived constraints closer to those from MIGHTEE. Similarly, modifying our own sample (by adopting a fixed mass-to-light ratio of 0.6) does not significantly alter the shape parameter $\delta$. These findings suggest that the discrepancy is not solely due to the mass-to-light ratios (or their radial variations), but is intrinsic to the data used. It may be because the
% also the differences in the homogeneity of the analysis. The
MIGHTEE data is a homogeneous sample, whereas SPARC comprises a heterogeneous compilation of galaxies.
%This highlight the critical role of both accounting for radial variations in the mass-to-light ratio and homogeneous / consistent data analysis 
%and consistency
%the principal driver of this is the homogeneous analysis of MIGHTEE, compared to the heterogeneous SPARC dataset, highlighting the importance of a homgeneous analysis, whilst accounting for radial mass-to-light ratio variations. 

\section{Summary and conclusions} 
\label{summary}
%Provide the first measurement of the Radial Acceleration Relationship in the baryon-dominated inner regions of disc galaxies probed by H$\alpha$ rotation curves and extend the RAR studies beyond the Local Universe
In this paper, we perform an analysis of the radial acceleration relation in 19 rotationally supported galaxies by leveraging the homogeneous sample of H{\sc i}-selected galaxies from the MIGHTEE-H{\sc i} survey, with spatially resolved H{\sc i} kinematics and homogeneously analyzed photometry from SED fitting across 10 optical and near-infrared bands. The aim of this study was to investigate the RAR using a novel approach---resolved stellar mass modeling from spectral energy distribution fitting, extending to a low-mass, homogeneously selected H{\sc i}-galaxies, whilst increasing the redshift range to $z \sim 0.08$.

We perform resolved SED fitting in the optical and near-infrared to measure the radial mass-to-light ratio variations across our galaxies. Using these variations, we derive the stellar surface densities, which, combined with the gas surface densities, allow us to determine the radial acceleration due to baryons. 
Our results reveal a tight RAR with an acceleration scale $a_{0} = 1.69 \pm 0.13 \times 10^{-10}$~m~s$^{-2}$, higher than previously reported for the SPARC galaxies alone, and with an intrinsic scatter of $\sigma_{\rm int} = 0.045 \pm 0.022$ dex. 
%This seems to be consistent with the idea of increasing acceleration scale with redshift evolution, as reported by several authors in simulations \citep{Wadsley2017, Mercado-RAR-Hooks}.
We also explore the impact of adopting a constant mass-to-light ratio versus one that varies across the galaxy sample compared to our initial approach of varying mass-to-light ratio. We find that adopting a spatially varying mass-to-light ratio yields the tightest RAR, suggesting that at least within the low-acceleration regime, this becomes increasingly important.
We also combine our sample with the high acceleration portion of the SPARC RAR data and fit the same functional form, yielding a lower acceleration scale of $a_0$ = 1.32 $\pm$ 0.13 \,$\mathrm{m~s^{-2}}$. However, the combined best-fit RAR is in tension with the trend preferred by our data alone at the 2$\sigma$ level. This is not unsurprising, as this tension highlights the critical role of adopting an appropriate mass-to-light ratio, particularly in the low acceleration regime. Different assumptions on the $\mathrm{\Upsilon_{\star}}$ ratio can lead to significant discrepancies in the inferred RAR. 
This is particularly relevant in the context of constraining dark-matter halo profiles---where an incorrect $\mathrm{\Upsilon_{\star}}$ assumption could systematically bias the interpretation of the scaling relation and lead to discrepancies in the inferred dark matter distributions--or in testing theories of modified gravity such as MOND.
% , or dark matter. 
The comparison of different $\mathrm{\Upsilon_{\star}}$ prescriptions underscores that using a spatially resolved approach provides the tightest correlation with the smallest intrinsic scatter in the RAR.
We also fit a double power law to our data. 
Although our limited sample size does not justify a more complicated parameterisation, combining our data with the SPARC high acceleration sample, we find that such a parameterisation is preferred. 
%The joint dataset clearly favors the double power-law model over the simpler MOND-inspired form, as indicated by the high $\Delta \mathrm{BIC}$ value. 
This suggests that the more flexible functional form better accommodates the diversity of baryonic accelerations in such a large dataset. Interestingly, for both our sample alone and the combined dataset, the slope of the double power-law fit in the low acceleration regime of $\beta\sim 0.5$ is consistent with MOND predictions; however, this does not rule out consistency with $\Lambda$CDM. 

We also perform the first measurement of the potential evolution in the RAR with redshift. Although our sample size is small, we find tentative evidence for redshift evolution at the 2.4$\sigma$ level. However, such an evolution would need to be confirmed with a much larger sample, preferably extending to higher redshifts, but analysed in a consistent manner following the work presented here. 

In addition, we consider a generalised interpolating function of the $\delta$ family. We find a sharper transition from the Newtonian to the deep MOND regime, compared to previous SPARC studies, with a shape $\delta$ of 3.90 $\pm$ 1.39, and a higher acceleration scale with a slightly reduced intrinsic scatter. Interestingly, this has much greater consistency with constraints on MOND from the Solar System quadrupole \citep{Harry_SS_quadrupole} and wide binary test \citep{Banik_WBT}.
We further examine the $n$ and $\gamma$ IF families with similar results. In each case we
% fit two additional MOND interpolating functions and again
find that using a varying mass-to-light ratio results in a higher $a_0$ than when one adopts a single value of $\mathrm{\Upsilon_{\star}} = 0.6$. This again emphasises the need for accurate measurements of the mass-to-light ratio, particularly in the low-acceleration regime in order to probe MONDian theories.

The uniqueness of this study resides in the fact that we probe the low acceleration regime, which is often more challenging due to observational limitations, by using a homogeneously analyzed sample. Unlike previous studies, which primarily focused on local galaxies, our method also extends the analysis to higher-redshift using a multi-wavelength approach, providing a more comprehensive view of the RAR across cosmic time. This is particularly relevant for future H{\sc i} surveys with the SKA, where improved spatial resolution, better sensitivity and sky coverage will allow for even deeper investigations into the disc-halo relation towards higher redshifts. Furthermore, incorporating other emission line tracers, such as $\mathrm{CO}$, ionized gas H$\alpha$, and stellar kinematics, can aid in constraining the rotation curves in the inner parts of disc galaxies, whilst also providing a direct measurement of their contribution to the baryonic mass, where accurately mapping the dark matter and baryonic distributions are crucial in mitigating the disc-halo degeneracy. % Resolved studies of Dark Matter in disc galaxies with ionised gas kinematics or 

Ultimately, our work emphasizes the necessity of accurately determining the mass-to-light ratios using deep multi-wavelength observations for
% as a promising way forward in
constraining the dark matter and baryonic distributions, expanding towards homogeneously selected samples beyond the local Universe. This not only enables tests of cosmic evolution but also helps mitigate systematics that affect local samples, such as peculiar velocities and uncertainties in distance estimates. %This comprehensive approach will provide new insights into the coupling between baryonic and dark matter.
Future advancements, including higher-resolution H{\sc i} data
%for the outer parts
and combined optical emission line tracers (CO, $\mathrm{H\alpha}$), will be crucial in refining our understanding of galaxy formation and evolution and constraining the dark matter properties of galaxies. 

\section*{Data availability}
The MIGHTEE-H{\sc i} spectral cubes are available from \href{https://archive-gw-1.kat.ac.za/public/repository/10.48479/jkc0-g916/index.html}{https://doi.org/10.48479/jkc0-g916}  \citep{MIGHTEE-DR1}. The optical and near-infrared data used to measure the stellar surface brightness properties of the galaxies are all in the public domain. Other data underlying the article are available on request to the corresponding author.

\section*{Acknowledgments}
AV, MJJ, IH and TY acknowledge the support of a UKRI Frontiers Research Grant [EP/X026639/1], which was selected by the European Research Council. MJJ, AAP and IH also acknowledge support from the STFC consolidated grants [ST/S000488/1] and [ST/W000903/1] and the Oxford Hintze Centre for Astrophysical Surveys which is funded through generous support from the Hintze Family Charitable Foundation. HD is supported by a Royal Society University Research Fellowship (grant no. 211046).
MG is supported through UK STFC Grant ST/Y001117/1. MG acknowledges support from the Inter-University Institute for Data Intensive Astronomy (IDIA). IDIA is a partnership of the University of Cape Town, the University of Pretoria and the University of the Western Cape. PEMP acknowledges funding from the Dutch Research Council (NWO) through the Veni grant VI.Veni.222.364. For the purpose of open access, the author has applied a Creative Commons Attribution (CC BY) licence to any Author Accepted Manuscript version arising from this submission. We thank Adam Carnall for providing the stellar grids with remnants and Deaglan Bartlett for assistance with {\sc Roxy}. %Michalina Maksymowicz-Maciata and Hengxing Pan for comments on the manuscript.%and Pavel E. Mancera Pi$\mathrm{\tilde{n}}$a
This work is based on observations made with the MeerKAT telescope.
This research made use of {\sc Photutils}, an Astropy package for
detection and photometry of astronomical sources \citep{larry_bradley_2024_10967176}, {\sc Roxy} \citep{roxy}, {\sc Scipy} \citep{Scipy_2020}, {\sc fgivenx} \citep{fgivenx}, {\sc numpyro} \citep{numpyro1, numpyro2}.
%%%%%%%%%%%%%%%%%%%%%%%%%%%%%%%%%%%%%%%%%%%%%%%%%%



\bibliographystyle{mnras}
\bibliography{ref.bib} 
%\begin{thebibliography}{99}
%\end{thebibliography}

%%%%%%%%%%%%%%%%%%%%%%%%%%%%%%%%%%%%%%%%%%%%%%%%%%

%%%%%%%%%%%%%%%%% APPENDICES %%%%%%%%%%%%%%%%%%%%%

\appendix
%\section{Sample Data Table}
%\input{gbar_gobs_table}
\label{appendix}

\section{Radial mass to light ratio variations}
\label{appendix:resolved_SEDS}
Each figure below shows, on the left, the G-band image of the galaxy with ellipses spaced every 5 arcseconds. On the right, the top panel displays the stellar surface mass density profile inferred from resolved SED fitting (in blue), with the $K_s$-band surface brightness profiles overlaid in red for comparison. The bottom panel shows the corresponding radial variations in the $K_s$-band mass-to-light ratio, $\Upsilon_{K}$.

\begin{figure*}
   \begin{subfigure}[t]{0.48\textwidth}
      \centering
      \includegraphics[height=7cm, keepaspectratio]{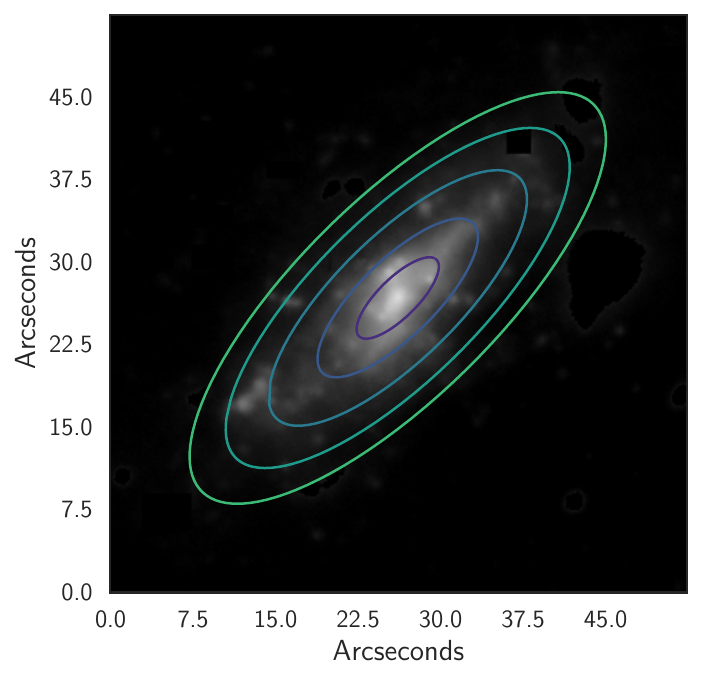}
   \end{subfigure}
   \hfill
   \begin{subfigure}[t]{0.48\textwidth}
      \centering
      \includegraphics[height=7cm, keepaspectratio]{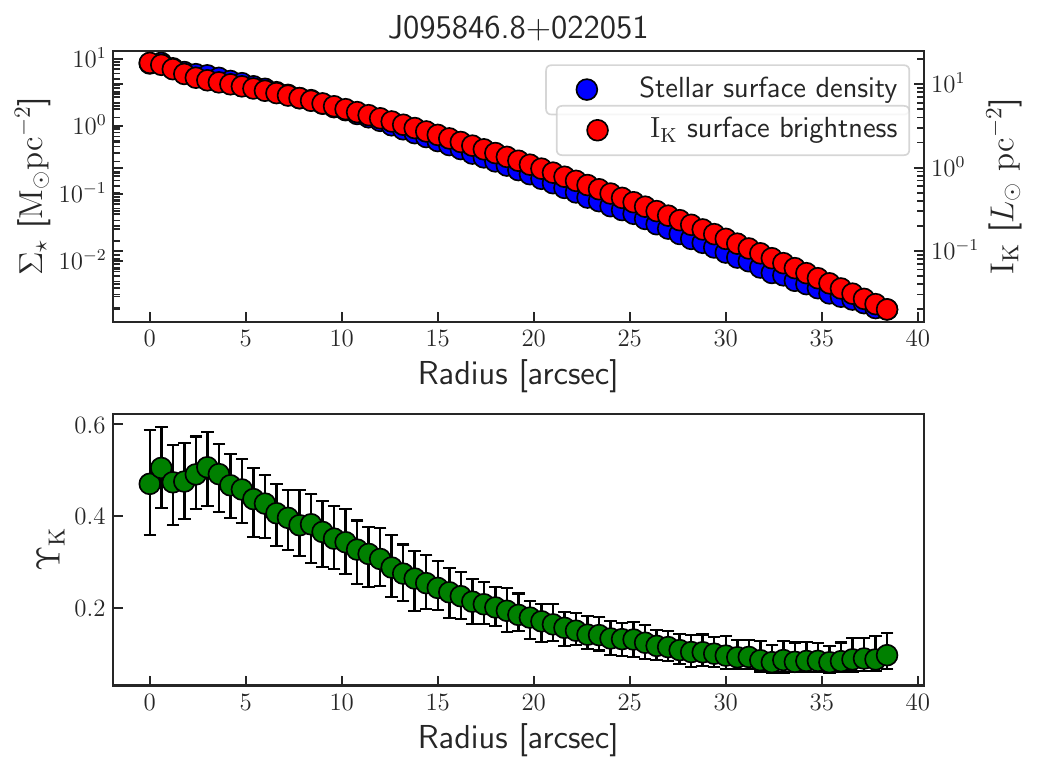}
   \end{subfigure}
   \caption{J095846.8+022051}
\end{figure*}

\begin{figure*}
   \begin{subfigure}[b]{0.48\textwidth}
   \centering
      \includegraphics[height=7cm, keepaspectratio]{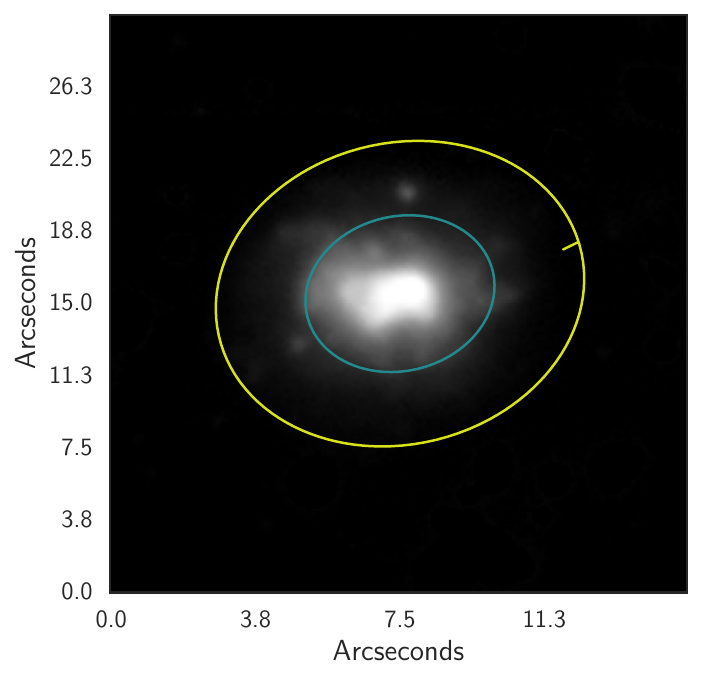}
    \end{subfigure}
    \hfill
    \begin{subfigure}[b]{0.48\textwidth}
      \centering
      \includegraphics[height=7cm, keepaspectratio]{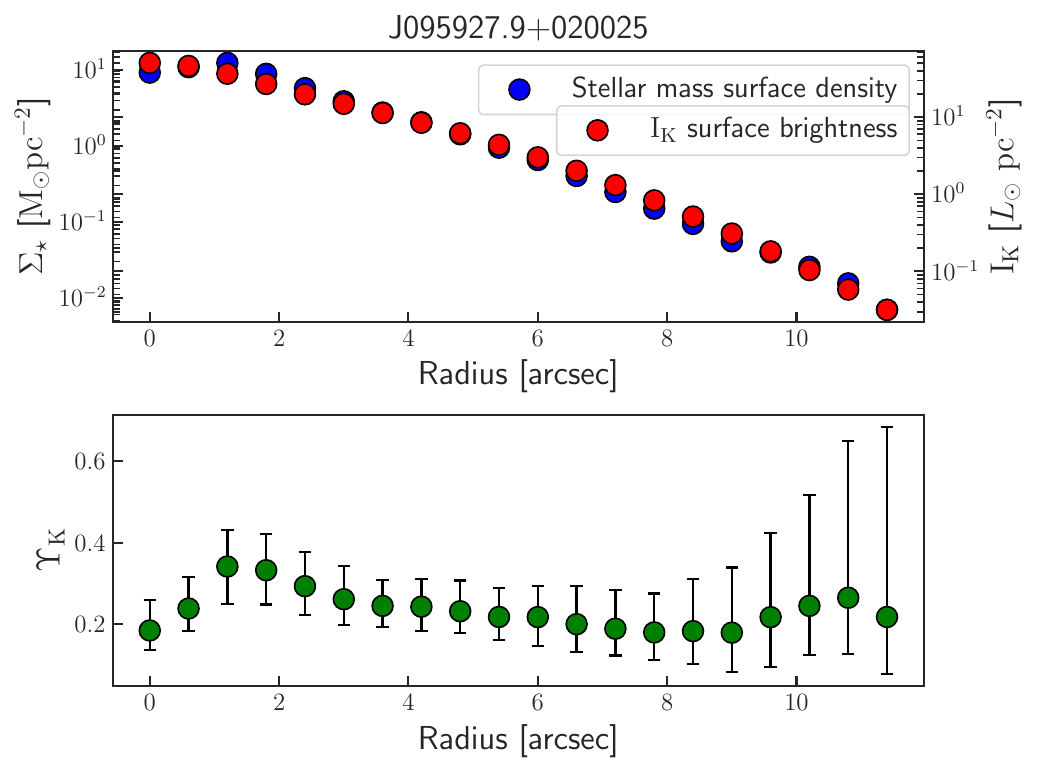}
    \end{subfigure}
    \caption{J095927.9+020025}
\end{figure*}

\begin{figure*}
   \begin{subfigure}[b]{0.48\textwidth}
      \includegraphics[height=7cm, keepaspectratio]{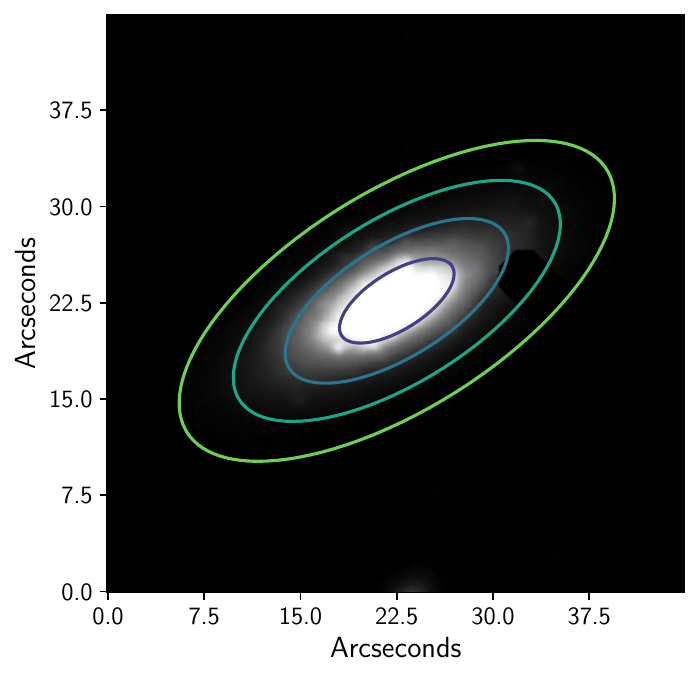}
    \end{subfigure}
    \hfill
    \begin{subfigure}[b]{0.48\textwidth}
      \centering
      \includegraphics[height=7cm, keepaspectratio]{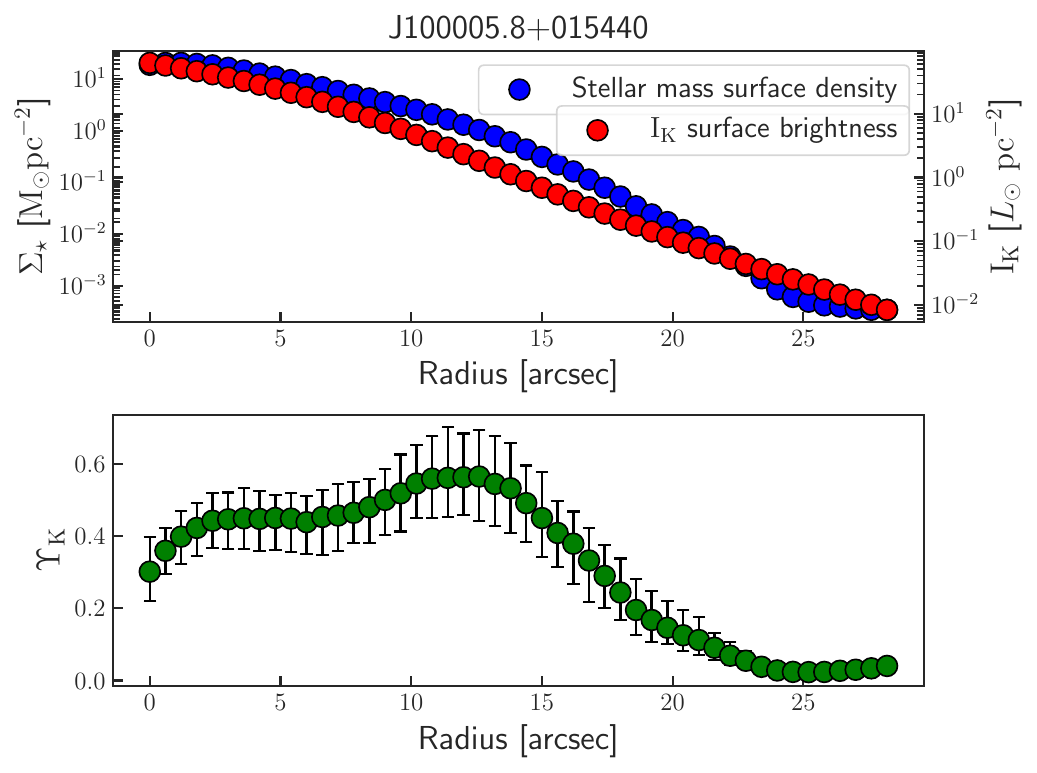}
    \end{subfigure}
    \caption{J100005.8+015440}
\end{figure*}

\begin{figure*}
    \begin{subfigure}[b]{0.48\textwidth}
        \centering
        \includegraphics[height=7cm, keepaspectratio]{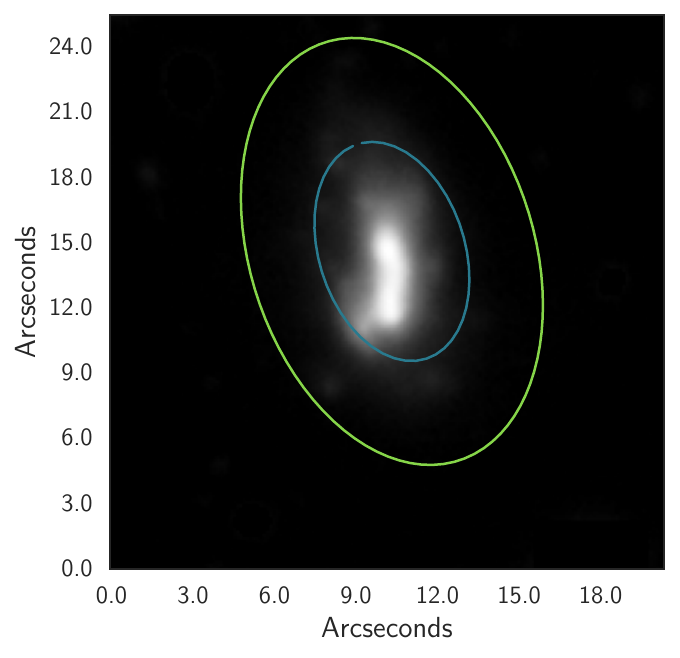}
    \end{subfigure}
    \hfill
    \begin{subfigure}[b]{0.48\textwidth}
        \centering
        \includegraphics[height=7cm, keepaspectratio]{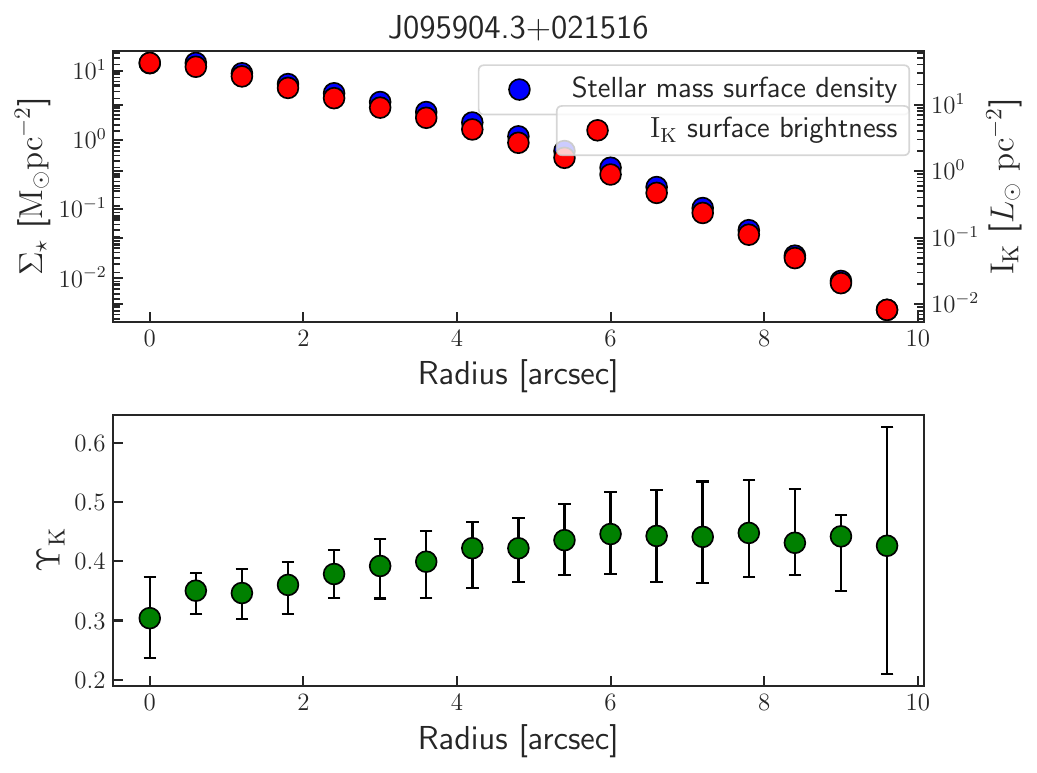}
    \end{subfigure}
    \caption{J095904.3+021516}
\end{figure*}

\begin{figure*}
    \begin{subfigure}[b]{0.48\textwidth}
        \centering
        \includegraphics[height=7cm, keepaspectratio]{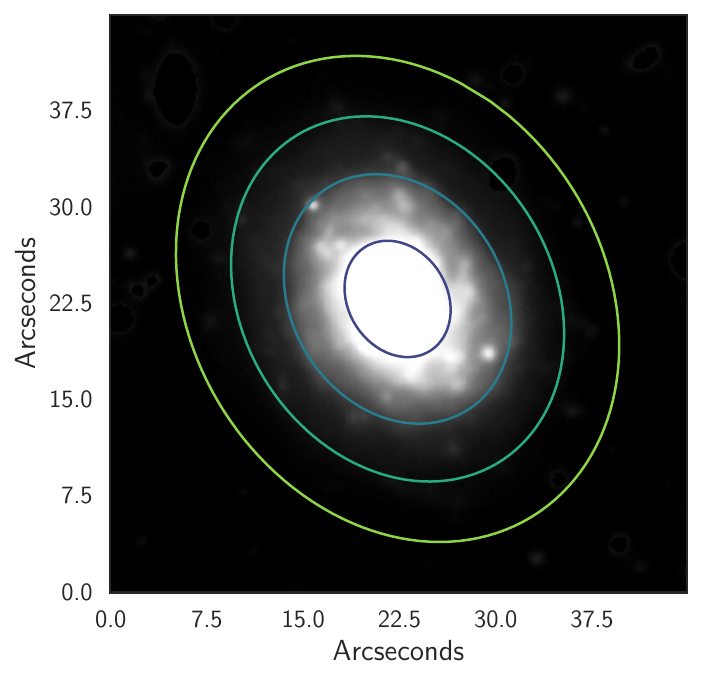}
    \end{subfigure}
    \begin{subfigure}[b]{0.48\textwidth}
        \centering
        \includegraphics[height=7cm, keepaspectratio]{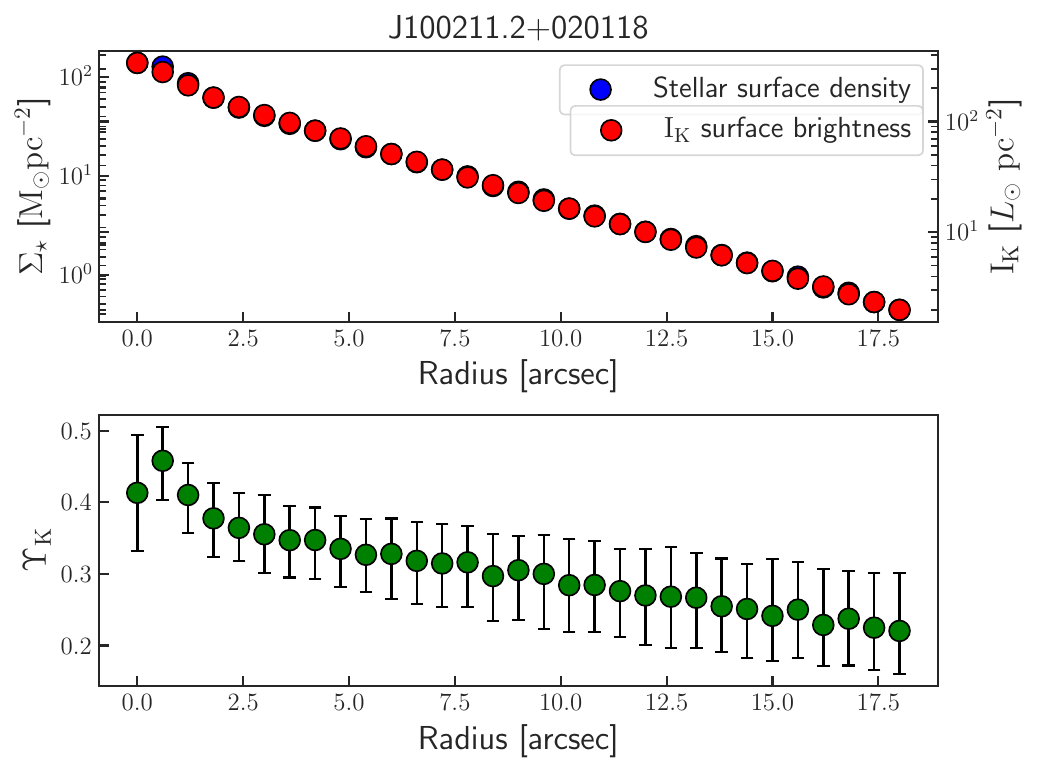}
    \end{subfigure}
    \caption{J100211.2+020118}
\end{figure*}

\begin{figure*}
    \centering
    \begin{minipage}[t]{0.48\textwidth}
        \centering
        \includegraphics[height=7cm, keepaspectratio]{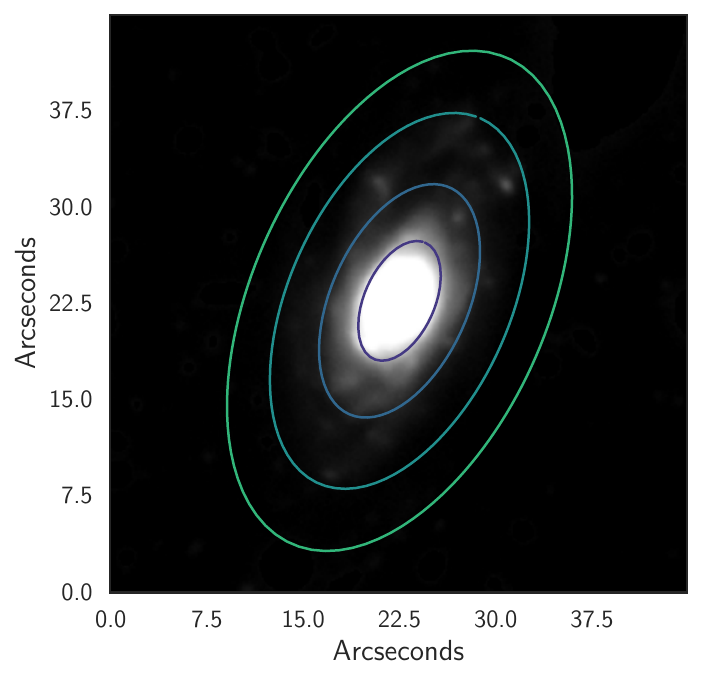}
    \end{minipage}
    \hspace{0.5cm} 
    \begin{minipage}[t]{0.48\textwidth}
        \centering
        \includegraphics[height=7cm, keepaspectratio]{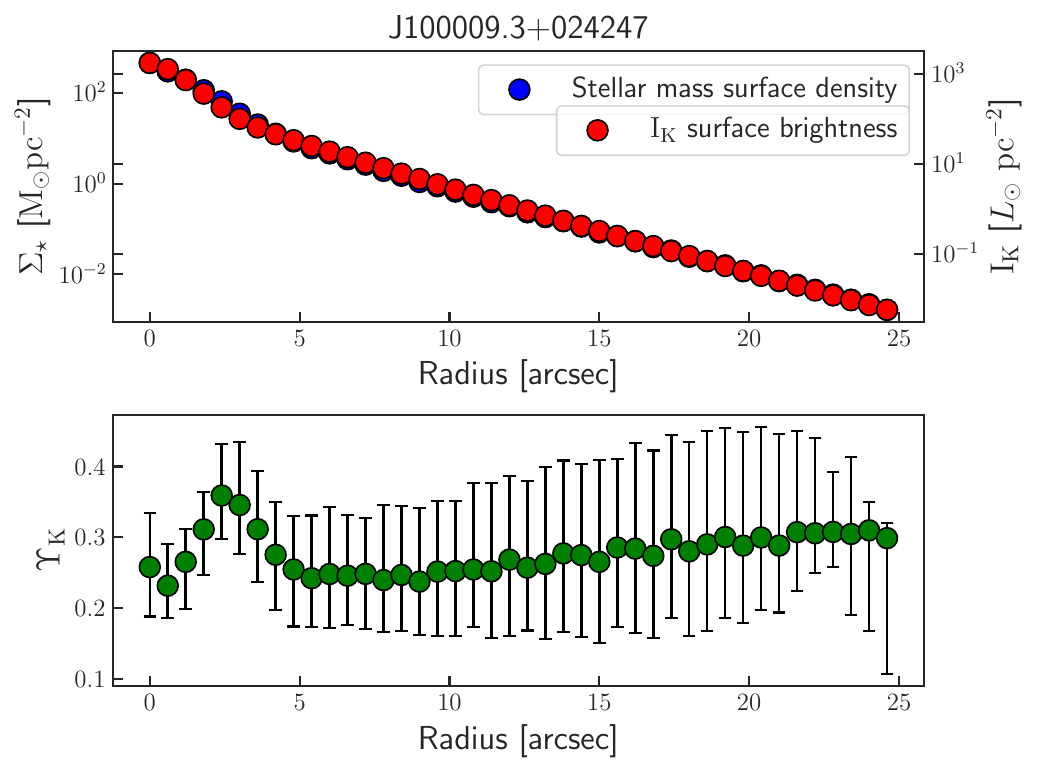}
    \end{minipage}
    \caption{J100009.3+024247}
\end{figure*}

\begin{figure*}
    \begin{subfigure}[b]{0.48\textwidth}
        \centering
        \includegraphics[height=7cm, keepaspectratio]{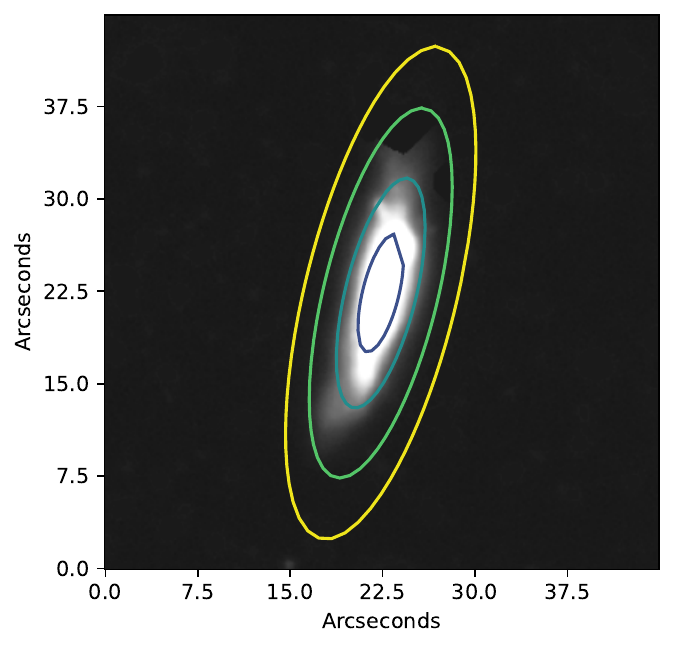}
    \end{subfigure}
    \hfill
    \begin{subfigure}[b]{0.48\textwidth}
        \centering
        \includegraphics[height=7cm, keepaspectratio]{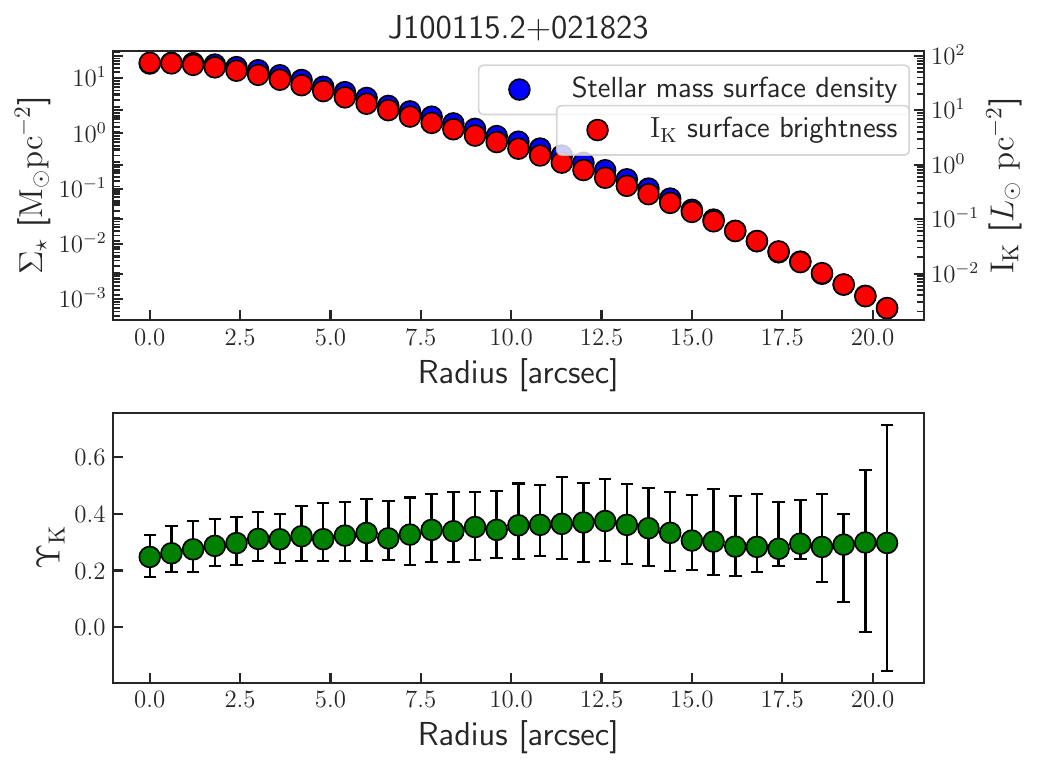}
    \end{subfigure}
    \caption{J100115.2+021823}
\end{figure*}

\begin{figure*}
    \begin{subfigure}[b]{0.48\textwidth}
        \centering
        \includegraphics[height=7cm, keepaspectratio]{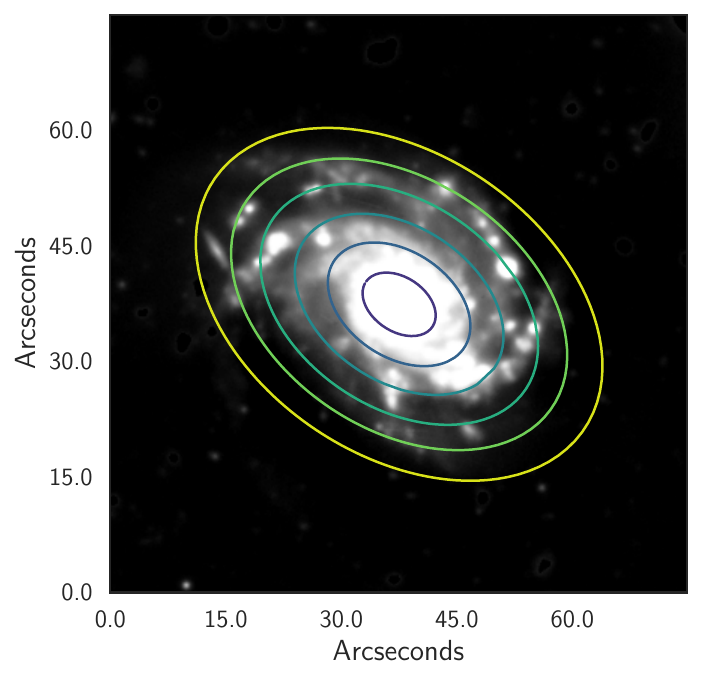}
    \end{subfigure}
    \hfill
    \begin{subfigure}[b]{0.48\textwidth}
        \centering
        \includegraphics[height=7cm, keepaspectratio]{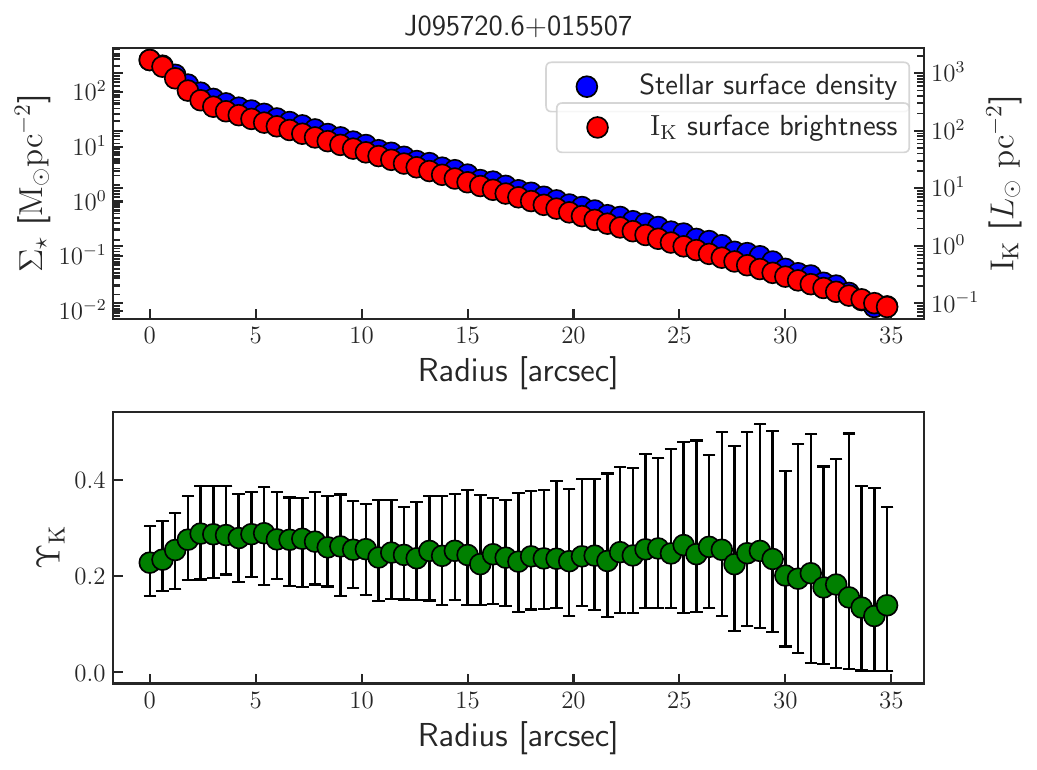}
    \end{subfigure}
    \caption{J095720.6+015507}
\end{figure*}

\begin{figure*}
    \begin{subfigure}[b]{0.48\textwidth}
        \centering
        \includegraphics[height=7cm, keepaspectratio]{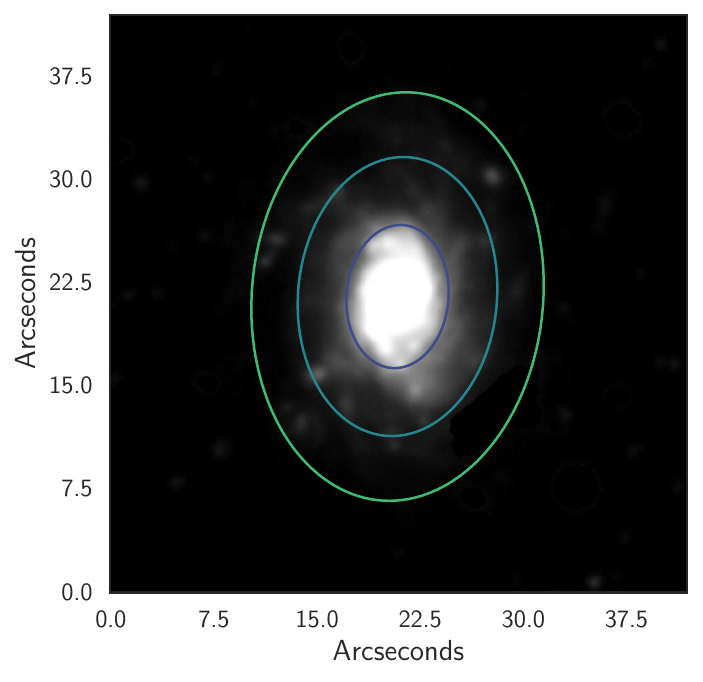}
    \end{subfigure}
    \hfill
    \begin{subfigure}[b]{0.48\textwidth}
        \centering
        \includegraphics[height=7cm, keepaspectratio]{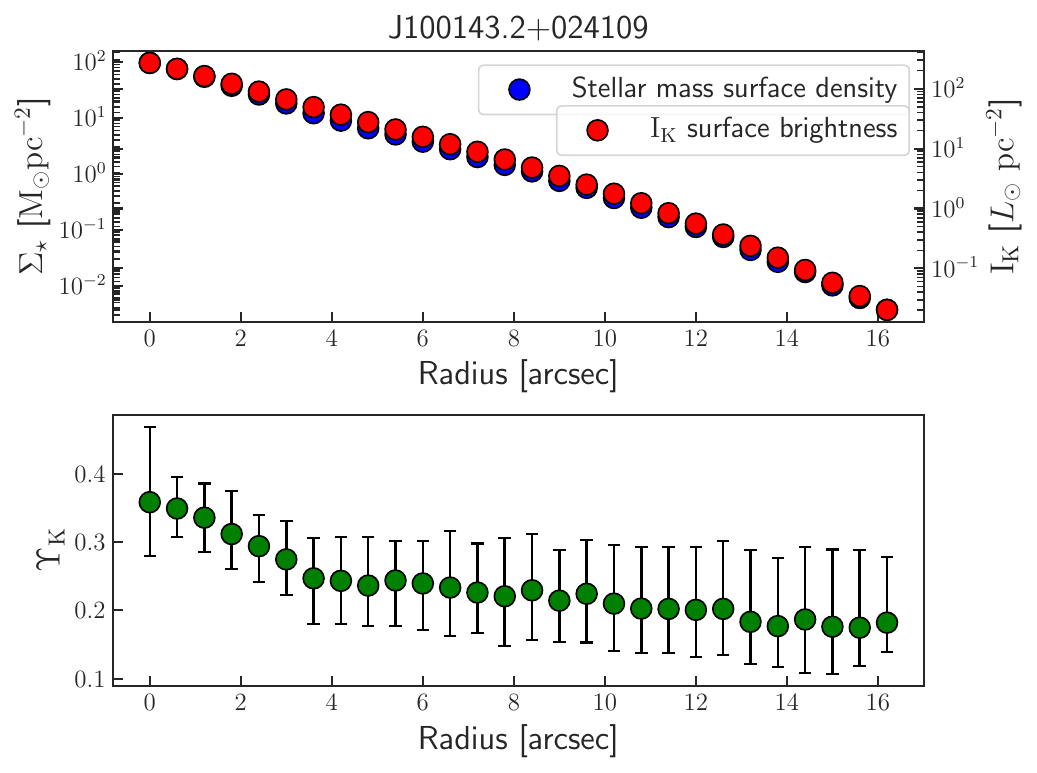}
    \end{subfigure}
    \caption{J100143.2+024109}
\end{figure*}

\begin{figure*}
    \begin{subfigure}[b]{0.48\textwidth}
        \centering
        \includegraphics[height=7cm, keepaspectratio]{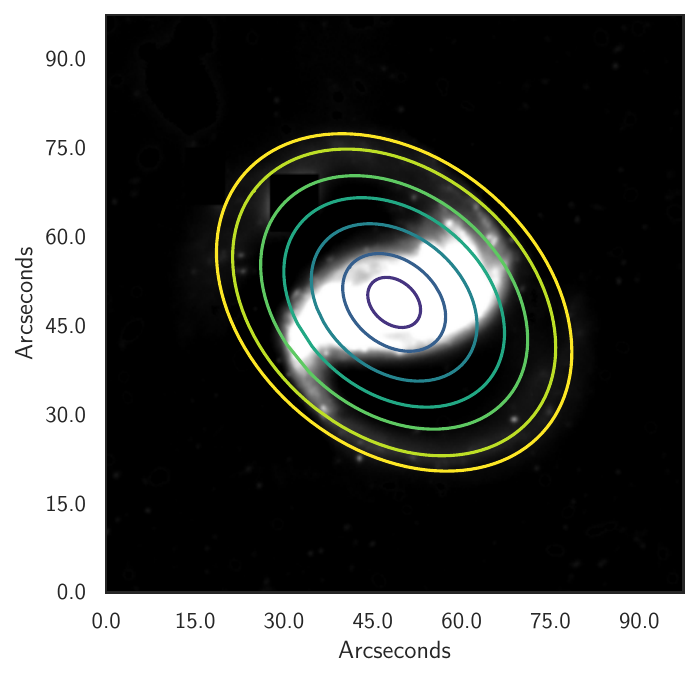}
    \end{subfigure}
    \hfill
    \begin{subfigure}[b]{0.48\textwidth}
        \centering
        \includegraphics[height=7cm, keepaspectratio]{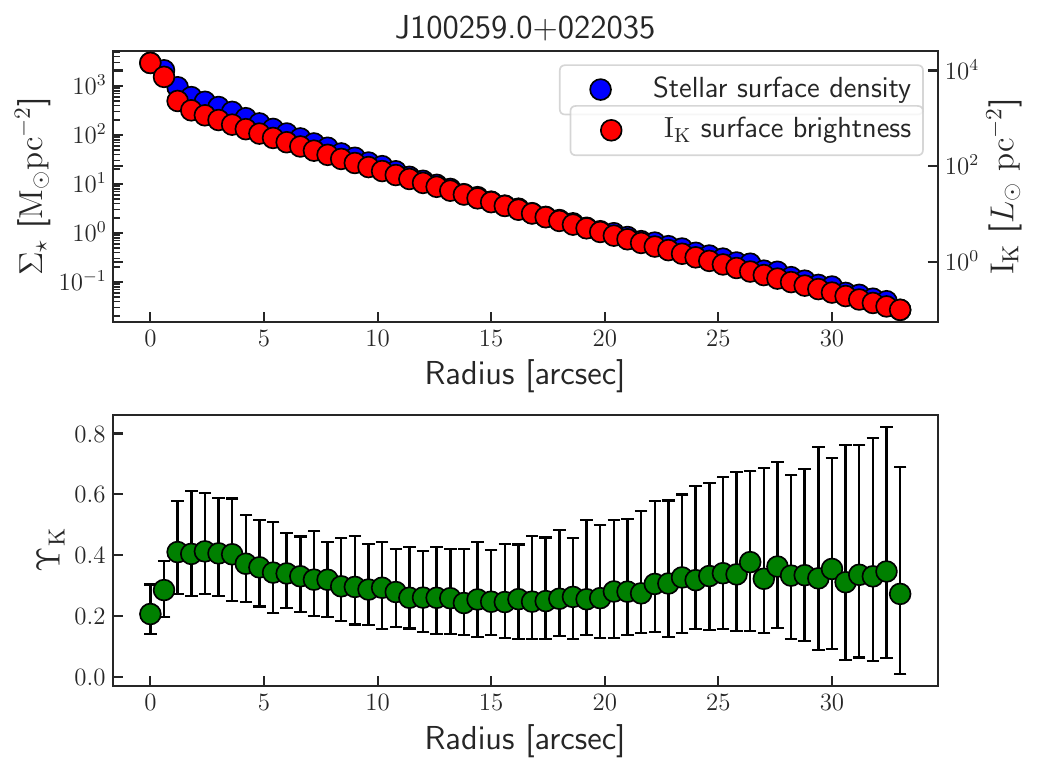}
    \end{subfigure}
    \caption{J100259.0+022035}
\end{figure*}

\begin{figure*}
    \begin{subfigure}[b]{0.45\textwidth}
        \centering
        \includegraphics[height=7cm, keepaspectratio]{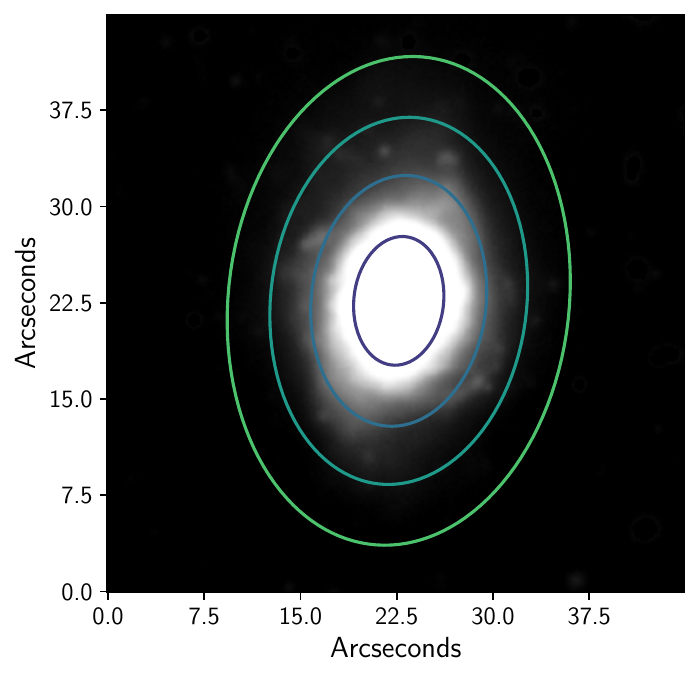}
    \end{subfigure}
    \hfill
    \begin{subfigure}[b]{0.45\textwidth}
        \centering
        \includegraphics[height=7cm, keepaspectratio]{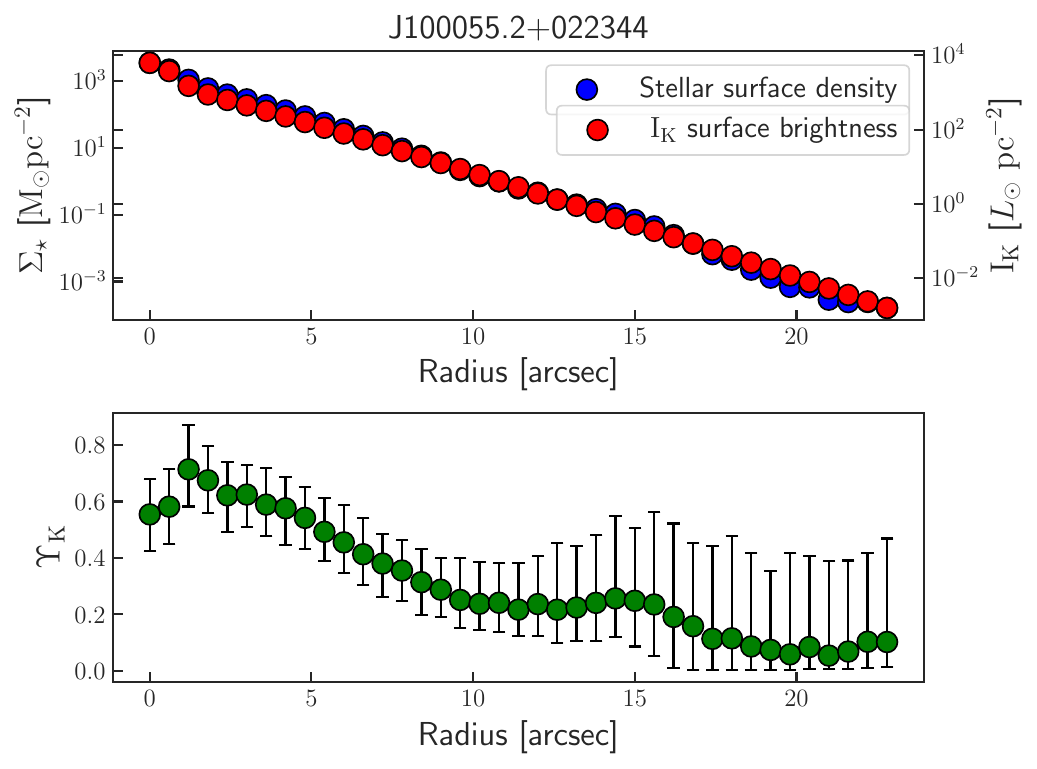}
    \end{subfigure}
    \caption{J100055.2+022344}
\end{figure*}

\begin{figure*}
    \begin{subfigure}[b]{0.48\textwidth}
        \centering
        \includegraphics[height=7cm, keepaspectratio]{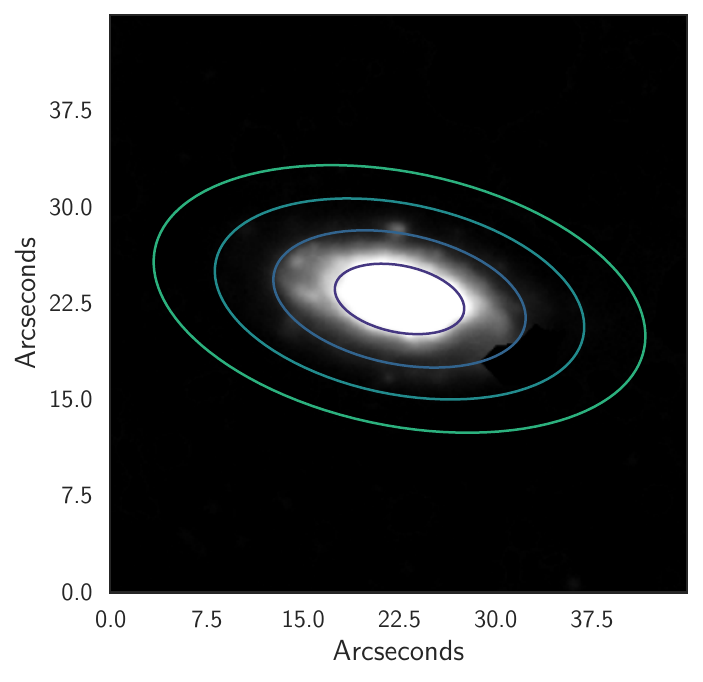}
    \end{subfigure}
    \hfill
    \begin{subfigure}[b]{0.48\textwidth}
        \centering
        \includegraphics[height=7cm, keepaspectratio]{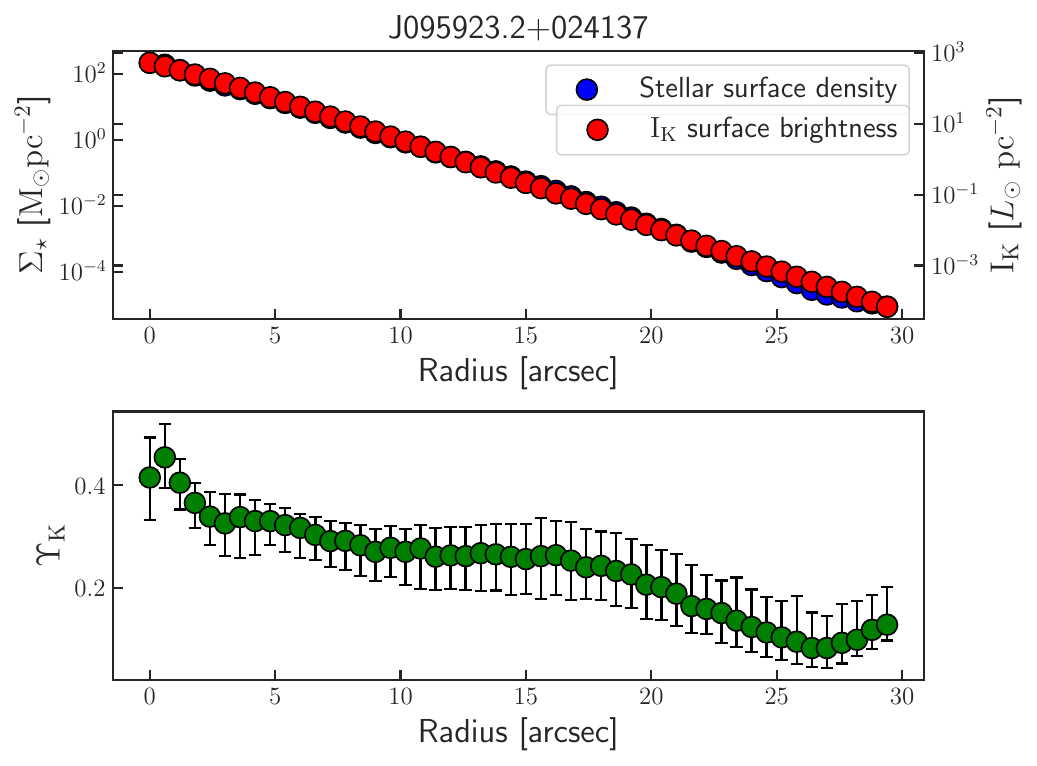}
    \end{subfigure}
    \caption{J095923.2+024137}
\end{figure*}

\begin{figure*}
    \begin{subfigure}[b]{0.48\textwidth}
        \centering
        \includegraphics[height=7cm, keepaspectratio]{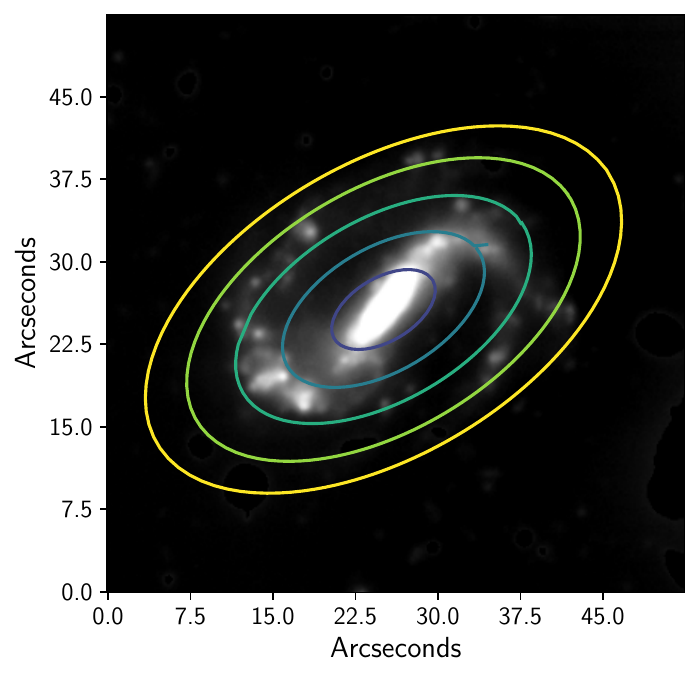}
    \end{subfigure}
    \hfill
    \begin{subfigure}[b]{0.48\textwidth}
        \centering
        \includegraphics[height=7cm, keepaspectratio]{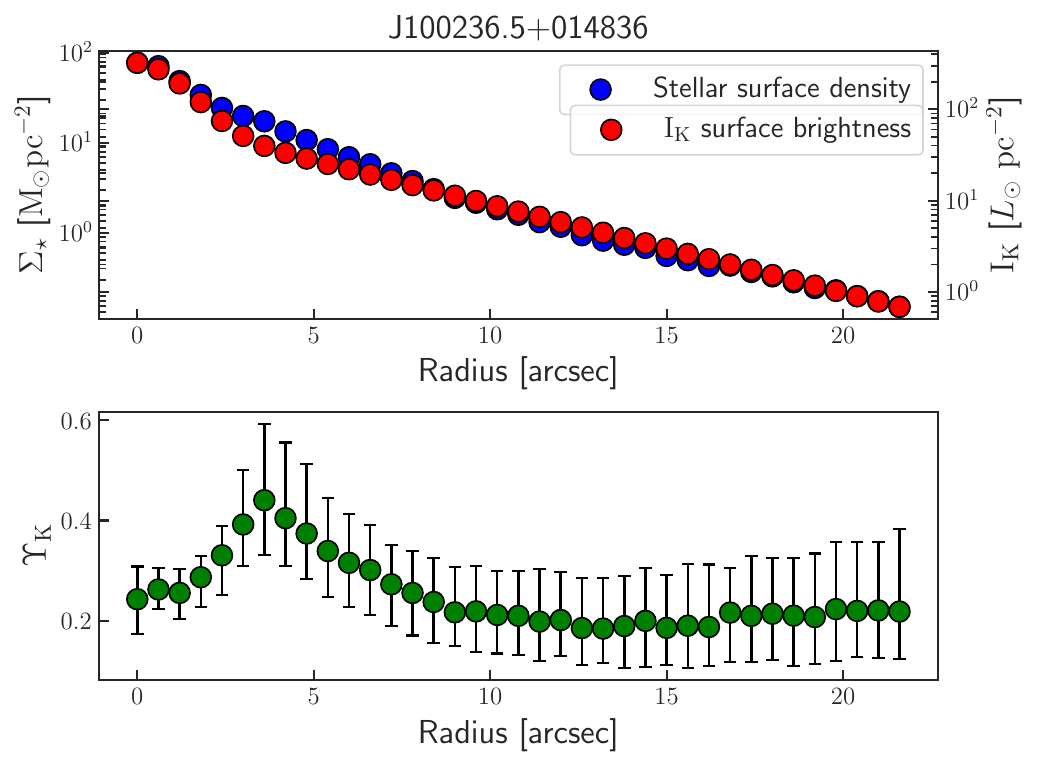}
    \end{subfigure}% need to check this one
    \caption{J100236.5+014836}
\end{figure*}

\begin{figure*}
    \begin{subfigure}[b]{0.48\textwidth}
        \centering
        \includegraphics[height=7cm, keepaspectratio]{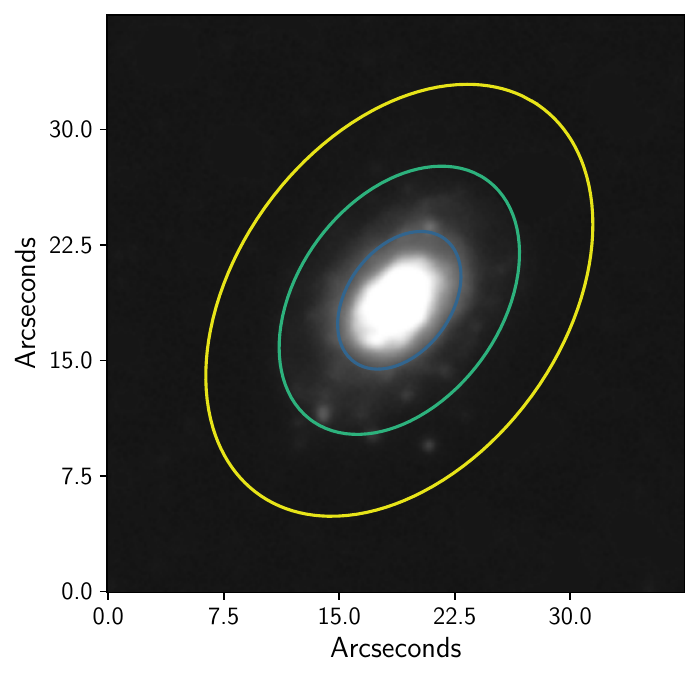}
    \end{subfigure}
    \hfill
    \begin{subfigure}[b]{0.48\textwidth}
        \centering
        \includegraphics[height=7cm, keepaspectratio]{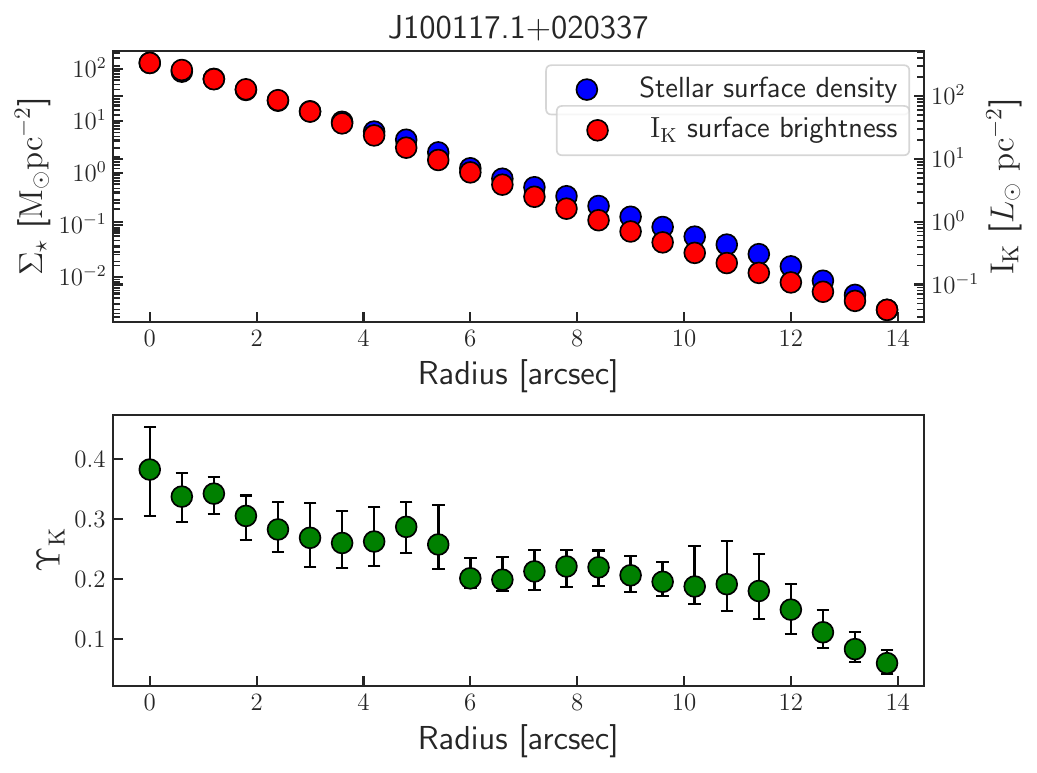}
    \end{subfigure}% need to check this one
    \caption{J100117.1+020337}
\end{figure*}

\begin{figure*}
    \begin{subfigure}[b]{0.48\textwidth}
        \centering
        \includegraphics[height=7cm, keepaspectratio]{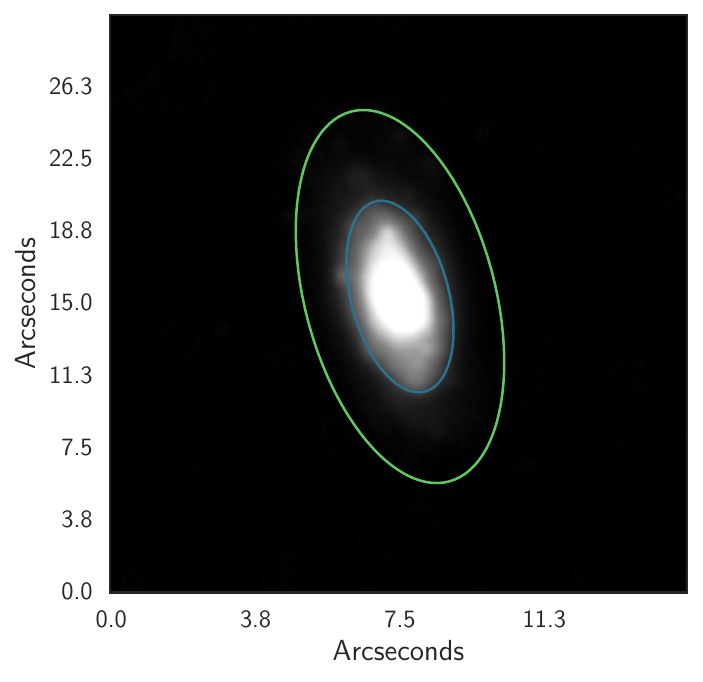}
    \end{subfigure}
    \hfill
    \begin{subfigure}[b]{0.48\textwidth}
        \centering
        \includegraphics[height=7cm, keepaspectratio]{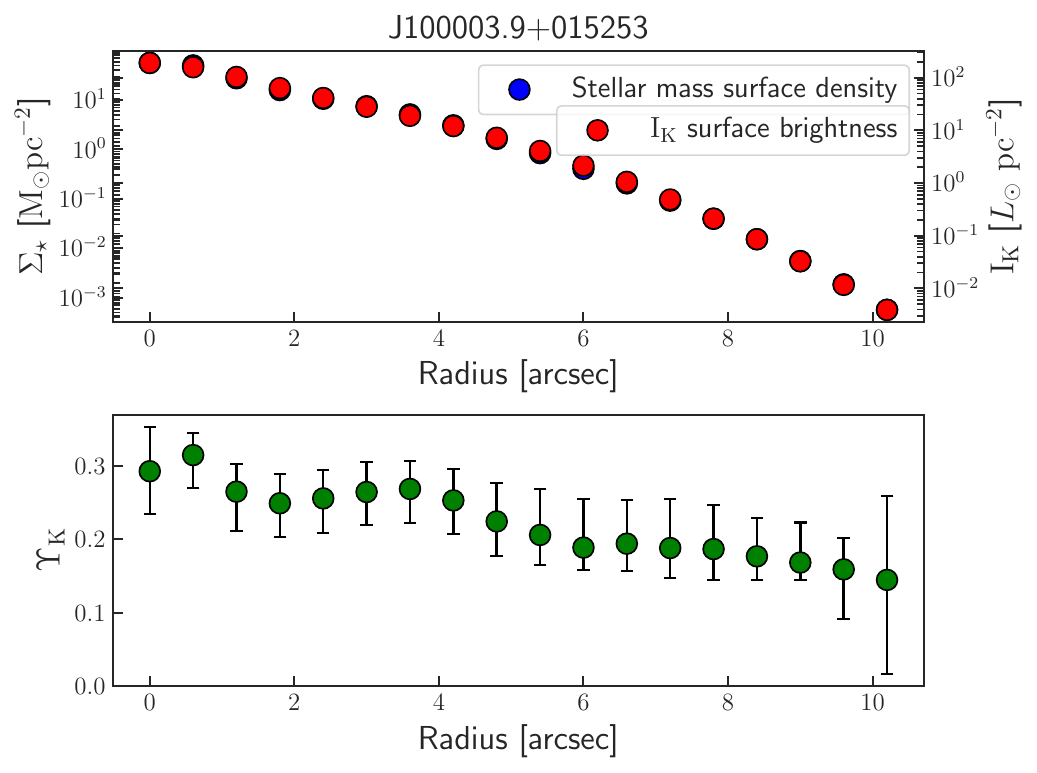}
    \end{subfigure}
    \caption{J100003.9+015253}
\end{figure*}

\begin{figure*}
    \begin{subfigure}[b]{0.48\textwidth}
        \centering
        \includegraphics[height=7cm, keepaspectratio]{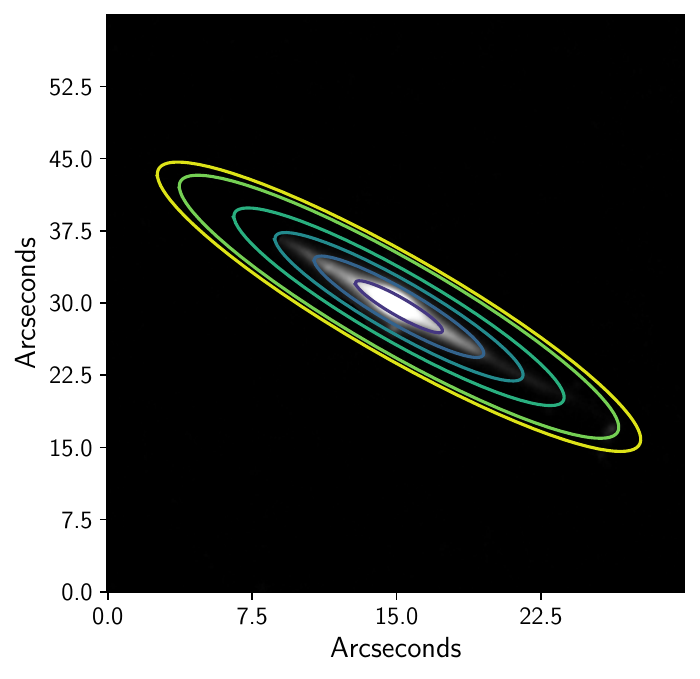}
    \end{subfigure}
    \hfill
    \begin{subfigure}[b]{0.48\textwidth}
        \centering
        \includegraphics[height=7cm, keepaspectratio]{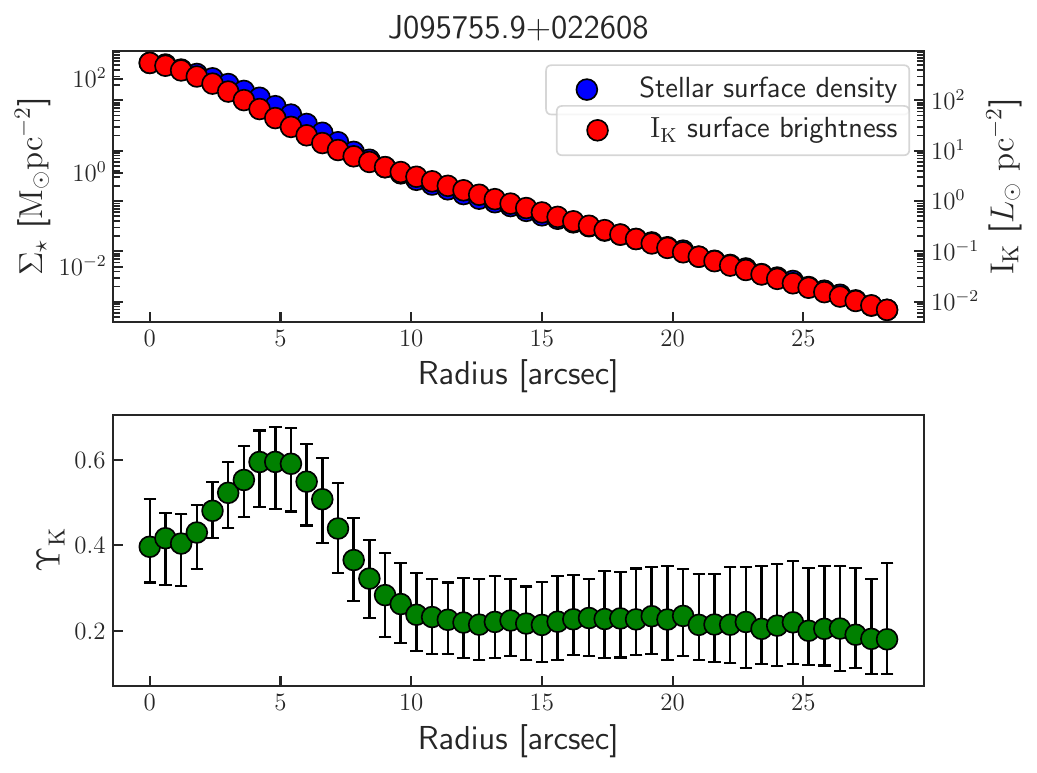}
    \end{subfigure}% mean 0.3 from resolved SED 
    \caption{J095755.9+022608}
\end{figure*}

\begin{figure*}
    \begin{subfigure}[b]{0.48\textwidth}
        \centering
        \includegraphics[height=7cm, keepaspectratio]{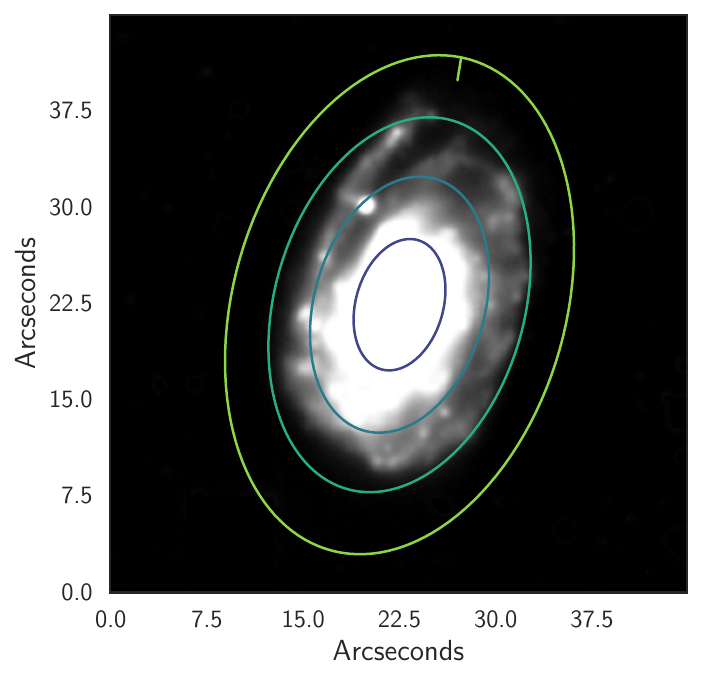}
    \end{subfigure}
    \hspace{0.02\textwidth}
    \begin{subfigure}[b]{0.48\textwidth}
        \centering
        \includegraphics[height=7cm, keepaspectratio]{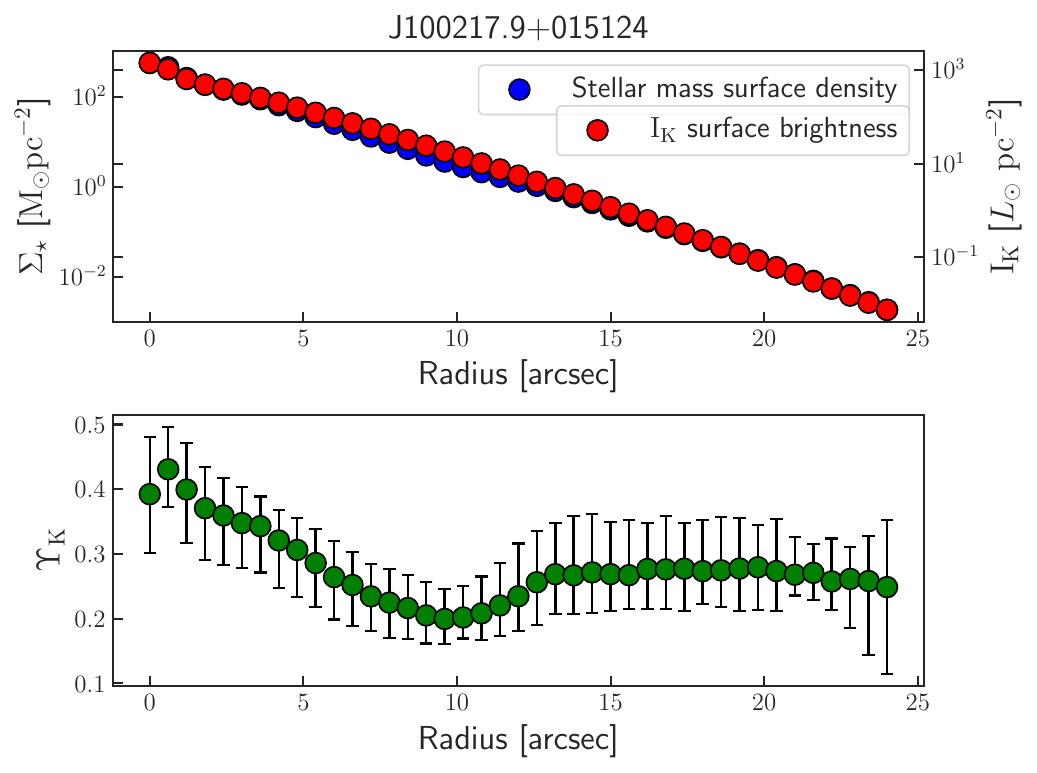}
    \end{subfigure}
    \caption{J100217.9+015124}
\end{figure*}

\begin{figure*}
    \begin{subfigure}[b]{0.48\textwidth}
        \centering
        \includegraphics[height=7cm, keepaspectratio]{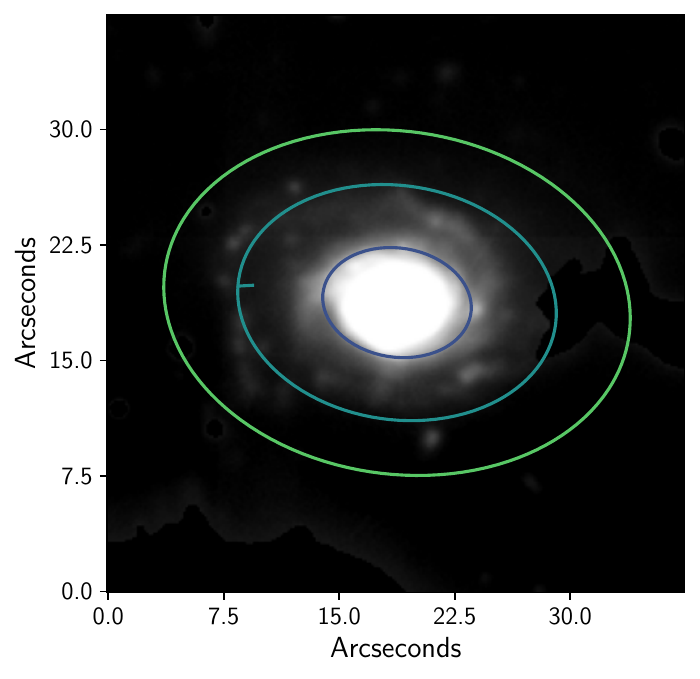}
    \end{subfigure}
    \hfill
    \begin{subfigure}[b]{0.48\textwidth}
        \centering
        \includegraphics[height=7cm, keepaspectratio]{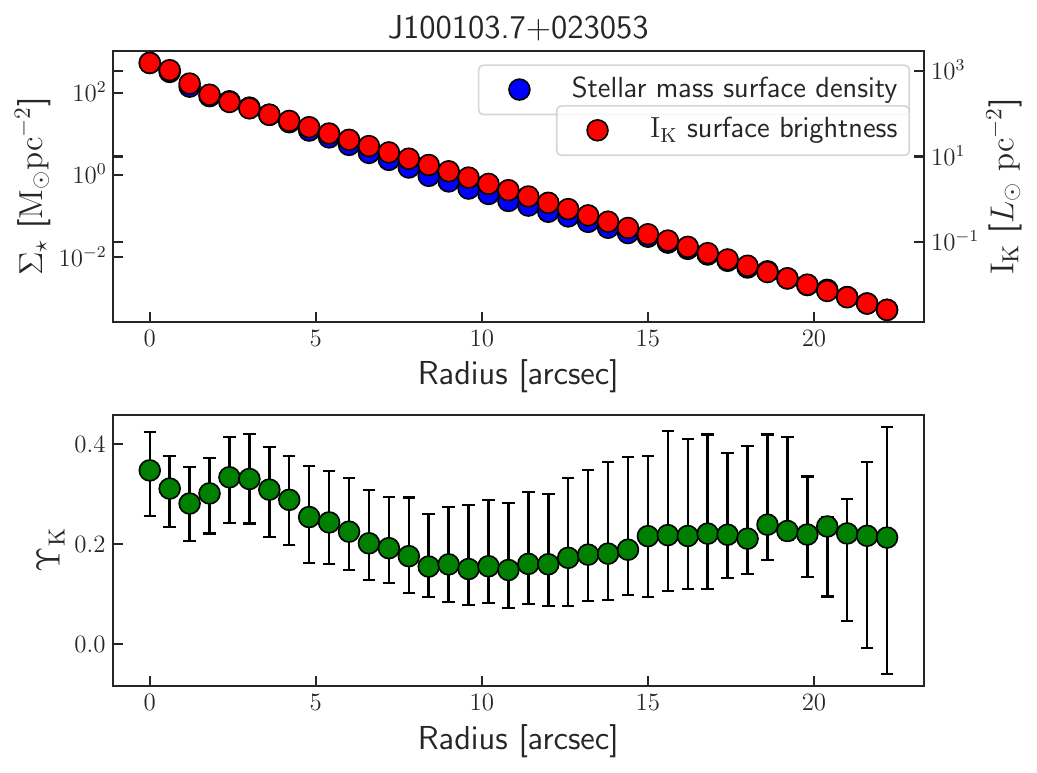}
    \end{subfigure}
    \caption{J100103.7+023053}
\end{figure*}

\begin{figure*}
    \begin{subfigure}[b]{0.48\textwidth}
        \centering
        \includegraphics[height=7cm, keepaspectratio]{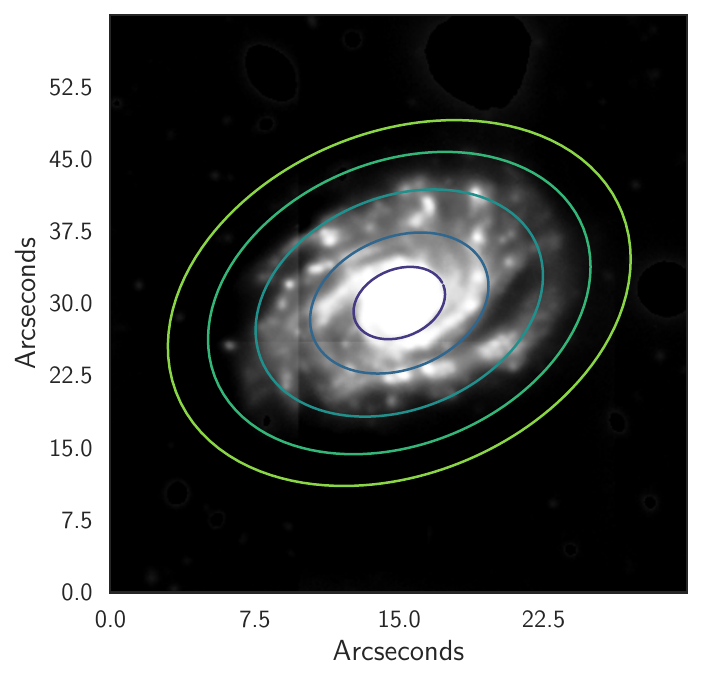}
    \end{subfigure}
    \hfill
    \begin{subfigure}[b]{0.48\textwidth}
        \centering
        \includegraphics[height=7cm, keepaspectratio]{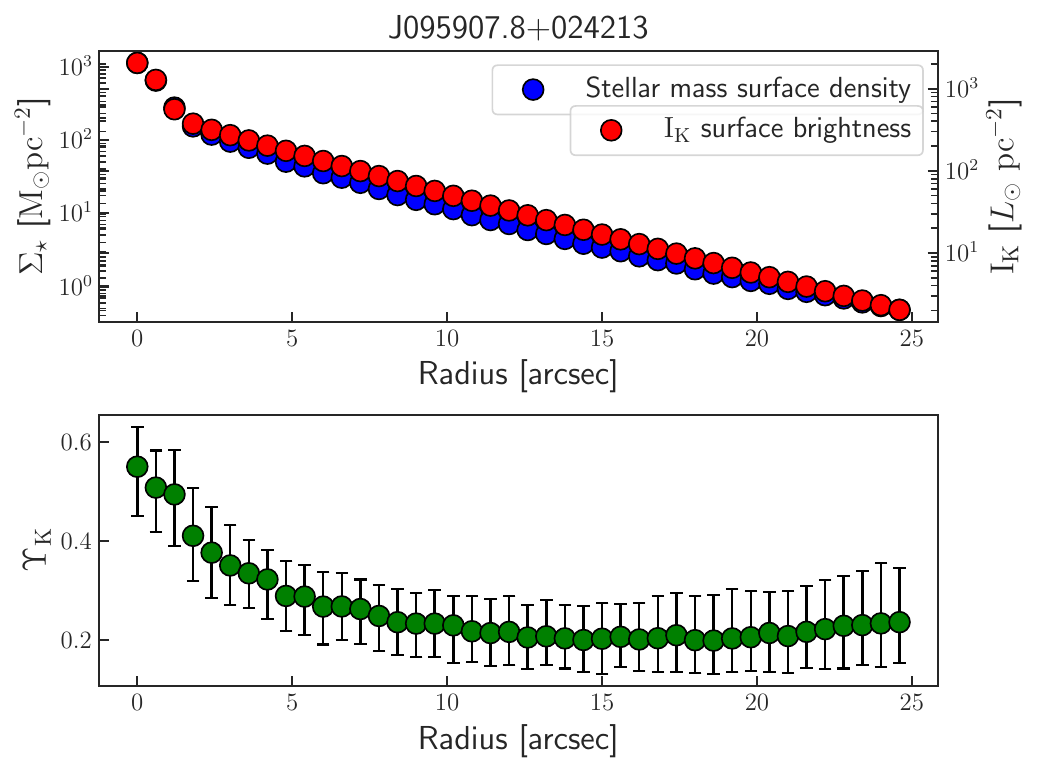}
    \end{subfigure}
    \caption{J095907.8+024213}
\end{figure*}

\section{RAR fits}
\label{appendix:rar_fits}

\begin{figure*}
\centering
\includegraphics[width = \textwidth]{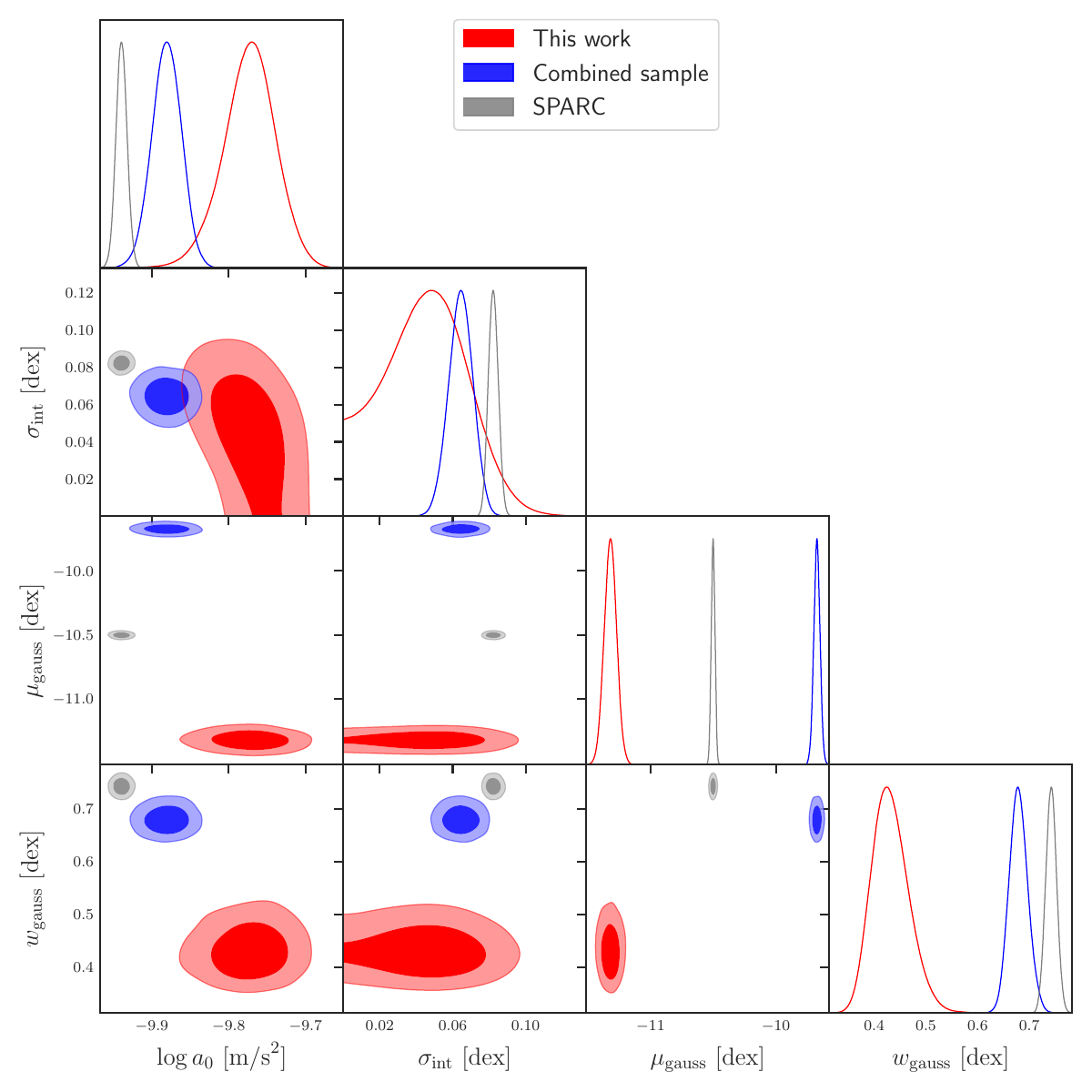}
\caption{The corner plot showing the posterior distribution of the parameters from the RAR sample fit for our sample in red, for the combined sample (this work + high acceleration portion of SPARC) in blue and all the SPARC data in grey. 
$\mu$ and $w$ represent the mean and standard deviation of the Gaussian prior on the true $\log(g_\text{bar})$ values as implemented in the Marginalised Normal Regression method of {\sc Roxy}. 
% $\mathrm{\mu_{gauss}}$ has units of log acceleration, $\mathrm{\sigma_\text{int}}$ and $w_\mathrm{gauss}$ units of dex.
}\label{fig:RAR_fit}
\end{figure*}

\begin{figure*}
        \centering
        \includegraphics[width=\textwidth]{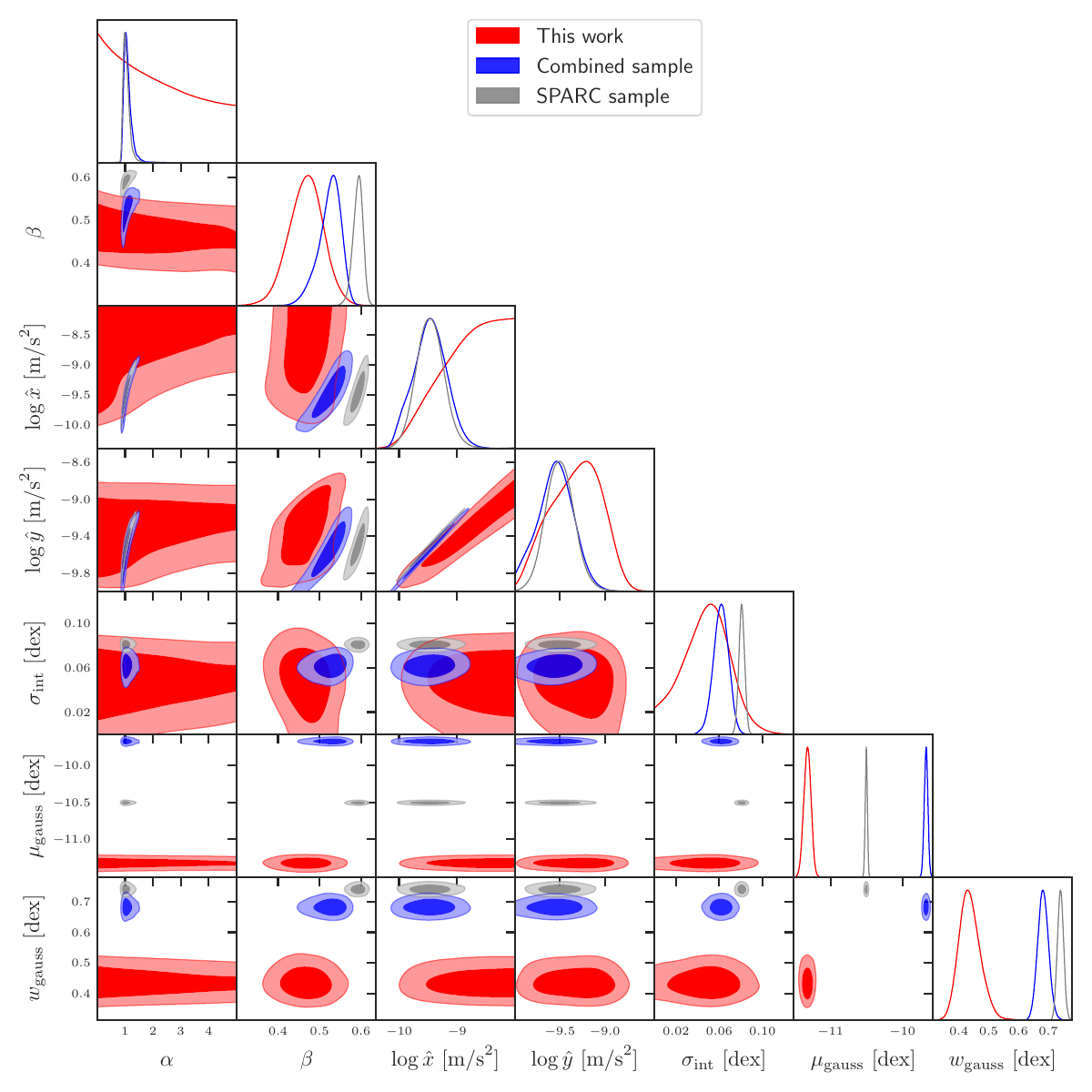}
        %RAR_combined_corner_dblplaw.pdf}
    \caption{The corner plot showing the posterior distribution of the parameters for the double power law for our sample, the combined sample (this work + high acceleration portion of SPARC) and all the SPARC data.} %$\mu$ and $w$ represent the mean and the standard deviation of the Gaussian prior on the true x values, under the assumption of an uncorrelated Gaussian likelihood.}
    \label{fig:corner_dblplaw}
\end{figure*}

We show in Figs.~\ref{fig:RAR_fit} and~\ref{fig:corner_dblplaw} the constraints on the parameters of the RAR and double power-law fits respectively.
%Figures~\ref{fig:RAR_n_IF_corner} and ~\ref{fig:RAR_gamma_IF_corner} show the constraints from the $n$ and $\gamma$ families of interpolating functions, with the corresponding values of the parameters listed in Table~\ref{tab:n-gamma-constraints-table}.

    %\begin{subfigure}[b]{0.48\textwidth}
       % \centering
       % \includegraphics[width=\linewidth]{Figures/RAR_posterior_predictive_best_fit.pdf}
   % \end{subfigure}
   % \hfill  
   %Left: The posterior predictive plot for RAR at 1, 2 and 3 $\sigma$ confidence from our RAR sample fit, with the best fit line in blue from the combined datasets. Right panel: 

%above our higher $\mathrm{log_{g_{bar}}}$ acceleration point

%\begin{subfigure}[b]{0.48\textwidth}
%\centering 
%\includegraphics[width=\linewidth]{Figures/RAR_corner.pdf}
%\caption{The corner plot showing the posterior distribution of the parameters from our RAR sample fit. $\mu$ and $w$ represent the mean and standard deviation of the Gaussian prior on the true $\log(g_\text{bar})$ values as implemented in the Marginalised Normal Regression method of {\sc Roxy}.   $\mathrm{\mu_{gauss}}$ has units of log acceleration, $\mathrm{\sigma_\text{int}}$ and $w_\mathrm{gauss}$ of dex.}
%\label{fig:rar_roxy}
%\end{subfigure}

%\begin{subfigure}[b]{0.48\textwidth}
   % \centering
   % \includegraphics[width=\linewidth]{Figures/RAR_combined_corner.pdf}   
   % \caption{Corner plot for the combined sample (this work RAR + SPARC) for the simpler fit. The intrinsic scatter $\sigma_\text{int}$ is now 0.083 dex.}
   % \label{fig:RAR-combined-corner}
%\end{subfigure}

\begin{table*}
    \centering
    \begin{tabular}{lcccccc}
    \hline
        IF family & Sample & MIGHTEE $\mathrm{\Upsilon_{\star}}$ & SPARC $\mathrm{\Upsilon_{\star}}$ & $a_0$ & Shape & $\sigma_\text{int}$ \\
        \hline
        $n$ & MIGHTEE & Varying $\Upsilon_{K}$ & - & 2.02 $\pm$ 0.15 & 6.84 $\pm$ 3.05 & 0.040 $\pm$ 0.022 \\
        $n$ & MIGHTEE & $\Upsilon_{K} = 0.6 $ &-& 1.48 $\pm$ 0.095 & 9.19 $\pm$ 3.31 &  0.032$\pm$ 0.019 \\
        \hline 
        
        $n$ & SPARC & -& $\Upsilon_{\rm{\star, disc}}$ = 0.5, $\Upsilon_{\rm{\star, bulge}}$ = 0.7 & 1.04 $\pm$ 0.04 & 0.92 $\pm$ 0.03 & 0.081 $\pm$ 0.003 \\
        $n$ & Low acceleration SPARC & - & $\Upsilon_{\rm{\star, disc}}$ = 0.5, $\Upsilon_{\rm{\star, bulge}}$ = 0.7  &  0.91 $\pm$ 0.05 & 0.81 $\pm$ 0.03 & 0.084 $\pm$ 0.003 \\
        
        \hline
         $n$ & MIGHTEE + high acceleration SPARC & Varying $\Upsilon_{\rm{K}}$ & $\Upsilon_{\rm{\star, disc}}$ = 0.5, $\Upsilon_{\rm{\star, bulge}}$ = 0.7 & 1.82 $\pm$ 0.15 & 1.41 $\pm$ 0.11 & 0.061 $\pm$ 0.007 \\         
         $n$ & MIGHTEE + high acceleration SPARC &$\Upsilon_{K} = 0.6$ & $\Upsilon_{\rm{\star, disc}}$ = 0.5, $\Upsilon_{\rm{\star, bulge}}$ = 0.7 & 1.33 $\pm$ 0.12 & 1.26 $\pm$ 0.09 & 0.047 $\pm$ 0.011 \\
        $n$ & MIGHTEE + high acceleration SPARC discs & Varying $\Upsilon_{K}$ & $\Upsilon_{\rm{\star}}$ = 0.27 & 0.96 $\pm$ 0.21 &0.54  $\pm$ 0.05 & 0.028 $\pm$ 0.017 \\
         
        \hline
        \hline

        $\gamma$ & MIGHTEE & Varying $\Upsilon_{K}$ & - & 1.65 $\pm$ 0.15 & 2.61 $\pm$ 1.42 & 0.045 $\pm$ 0.022 \\
         $\gamma$ & MIGHTEE & $\Upsilon_{K} = 0.6 $ & - & 1.13 $\pm$ 0.11 & 2.24 $\pm$ 0.98 & 0.051 $\pm$ 0.022  \\
         \hline
         
        $\gamma$ & SPARC &-& $\Upsilon_{\rm{\star, disc}}$ = 0.5, $\Upsilon_{\rm{\star, bulge}}$ = 0.7 & 0.97 $\pm$ 0.06 & 0.76 $\pm$ 0.05 & 0.081 $\pm$ 0.003 \\
        $\gamma$ & Low acceleration SPARC & -& $\Upsilon_{\rm{\star, disc}}$ = 0.5, $\Upsilon_{\rm{\star, bulge}}$ = 0.7 & 0.78 $\pm$ 0.07 & 0.61 $\pm$ 0.04 & 0.084 $\pm 0.003$ \\
        \hline
        
         $\gamma$ & MIGHTEE + high acceleration SPARC & Varying $\Upsilon_{K}$ & $\Upsilon_{\rm{\star, disc}}$ = 0.5, $\Upsilon_{\rm{\star, bulge}}$ = 0.7 & 1.57 $\pm$ 0.09 & 1.41 $\pm$ 0.16 & 0.063 $\pm$ 0.006 \\
         $\gamma$ & MIGHTEE + high acceleration SPARC & $\Upsilon_{\star} = 0.6$ & $\Upsilon_{\rm{\star, disc}}$ = 0.50, $\Upsilon_{\rm{\star, bulge}}$ = 0.7 & 1.19 $\pm$ 0.10 & 1.12 $\pm$ 0.11 & 0.051 $\pm$ 0.009\\
           $\gamma$ & MIGHTEE + high acceleration SPARC discs & Varying $\Upsilon_{K}$ & $\Upsilon_{\rm{\star}}$ = 0.27 & 1.19 $\pm$ 0.10 & 1.12 $\pm$ 0.11 & 0.051 $\pm$ 0.009\\
        \hline
    \end{tabular}
    \caption{Table~\ref{tab:delta-constraints-table} but for the $n$ and $\gamma$ IF families.}
    \label{tab:n-gamma-constraints-table}
\end{table*}

%\section
%%%%%%%%%%%%%%%%%%%%%%%%%%%%%%%%%%%%%%%%%%%%%%%%%%
% Don't change these lines
\bsp	% typesetting comment
\label{lastpage}
\end{document}